# Beam-Induced Damage Mechanisms and their Calculation

*A. Bertarelli*
CERN, Geneva, Switzerland

**Abstract**
The rapid interaction of highly energetic particle beams with matter induces dynamic responses in the impacted component. If the beam pulse is sufficiently intense, extreme conditions can be reached, such as very high pressures, changes of material density, phase transitions, intense stress waves, material fragmentation and explosions. Even at lower intensities and longer time-scales, significant effects may be induced, such as vibrations, large oscillations, and permanent deformation of the impacted components. These lectures provide an introduction to the mechanisms that govern the thermomechanical phenomena induced by the interaction between particle beams and solids and to the analytical and numerical methods that are available for assessing the response of impacted components. An overview of the design principles of such devices is also provided, along with descriptions of material selection guidelines and the experimental tests that are required to validate materials and components exposed to interactions with energetic particle beams.

## 1 Introduction to beam-induced damage

When subatomic particles or ions interact with matter, they tend to transfer part of their energy to the medium they traverse [1]; this energy loss is ultimately turned into heat and leads to an increase of the temperature in the impacted target. Depending on the amount and distribution of the deposited energy and the time-scale of the phenomenon, i.e. the density of deposited power, different effects may result.

If the density of the deposited energy is relatively small, of the order of 10 W cm$^{-3}$ or less, and extended over a relatively long period of time (of the order of seconds), the structural response can be reduced to a (quasi-) steady-state or slow transient thermomechanical problem. This class of problems can often be linearized and solved using ordinary simulation methods, either analytical or, more usually, numerical, such as standard *finite element method* tools, which allow one the change of material properties with temperature to be taken into account. Examples of these problems include, for instance, simulation of so-called slow losses of particle accelerators in the collimation regions [2, 3].

Conversely, if the deposited power density is much higher and the duration of the interaction is very short (of the order of a few milliseconds or less), dynamic responses will be induced, principally because the thermal expansion of the impacted material is partly prevented by its inertia [4, 5]. These effects, often referred to as thermal shocks, generate dynamic stresses, which propagate through the material at the velocity of sound, in much the same way as when a structure is struck by another body. Depending on the amount of the deposited energy and the melting point of the impacted material, the temperature increase induced by the impact may lead to the formation of shock waves, changes of phase, or the ejection of molten material.

The nature and intensity of the dynamic responses depend on several parameters, mainly the total amount of energy deposited, its distribution, the duration of the impact, the thermophysical and mechanical properties of the impacted material, and the form and dimensions of the device interacting with the beam.

Figure 1 provides an overview of the different types of dynamic response that might be induced in a structure as a function of the density of the deposited power and the duration of the interaction. One can easily observe that the severity of the response is broadly proportional to the deposited power density and inversely proportional to the duration of the interaction; in other words, dynamic response depends on the specific energy deposited on the material.

In spite of the large influence of material properties, some approximate general figures can be extrapolated and used with caution to predict the type of response, regardless of the actual impacted material and boundary conditions. If the deposited energy is below 100 J cm$^{-3}$, the dynamic response will probably remain within the *elastic dynamic regime*, meaning that the induced vibrations and stress waves will not exceed the elastic limit of the material and that the structure will return to its initial undeformed state at the end of the process. Between roughly 100 J cm$^{-3}$ and 10 kJ cm$^{-3}$, we may expect that the stress waves will locally exceed the elastic limit of the material, inducing permanent plastic deformations that cannot be recovered once the dynamic response has waned out (the *plastic dynamic regime*). Above 10 kJ cm$^{-3}$, the stress waves will be strong enough to generate major changes of density and extensive damage to the material, such as fragmentation or explosion. If the impacted material is metal, phase transitions with the formation of liquid, gas, or plasma usually occur: this is usually referred to as the *shock wave regime*. If a significant reduction in density has occurred in the impacted material while particle bunches are still hitting it, the beam will penetrate more and more deeply, given that fewer atoms are available to interact with the incoming particles: this effect is sometimes called *hydrodynamic tunnelling* and may arise within the shock wave regime when the duration of the impact is sufficiently long to allow changes of phase to develop (several hundred nanoseconds or more).

The shock wave regime is also sometimes referred to as the *hydrodynamic regime*, implying that impacted materials start behaving as fluids, losing their mechanical strength (see Section 2.4.3). Since extensive shock-induced damage, such as fragmentation or mechanical spalling, may occur long before the material completely loses its strength, the term 'shock wave regime' appears more comprehensive.

These phenomena, with the possible exclusion of those belonging to the elastic regime, may severely affect the integrity and the functionality of the impacted equipment. A correct understanding and prediction of beam-induced damage is therefore extremely important in the design of any component exposed to direct interaction with intense and energetic particle beams, such as collimators, absorbers, dumps, scrapers, or windows. The same damage mechanisms apply to any device accidentally and rapidly interacting with energetic beams, such as vacuum chambers, magnet components, RF cavities, or beam instrumentation devices.

These lectures will address these topics, particularly focusing on dynamic events that have the potential to generate structural damage. Other energy release mechanisms (e.g. of stored magnetic energy or RF impedance-induced heating) are not explicitly covered here, although their effects might be dealt with in a similar way if the time-scale of the phenomena are relatively short.

Longer-term phenomena, such as radiation damage, which do also play a fundamental role in the design of devices interacting with particle beams, are treated elsewhere (see, for instance, Ref. [6]).

We will first provide a few cases of notable accidents and tests that have occurred in various accelerators in the world over the past few decades, to exemplify the effects of beam impacts on matter.

To explore the mechanisms of beam interaction with matter leading to the various responses and damage states previously described, we will start by introducing the concepts of linear thermo-elasticity, in particular, the case of beam impacts on circular discs and cylinders inducing responses in the elastic domain of the material: these relatively simple cases, although not leading to permanent damage, can be treated analytically and are very useful to gain some insights and physical understanding of the mechanisms of thermally induced stresses and dynamic responses.

For the sake of simplicity, the analysis will mostly deal with isotropic, homogeneous materials. However, these principles and methods can be extended to anisotropic or non-homogeneous materials with some additional mathematical complexity.

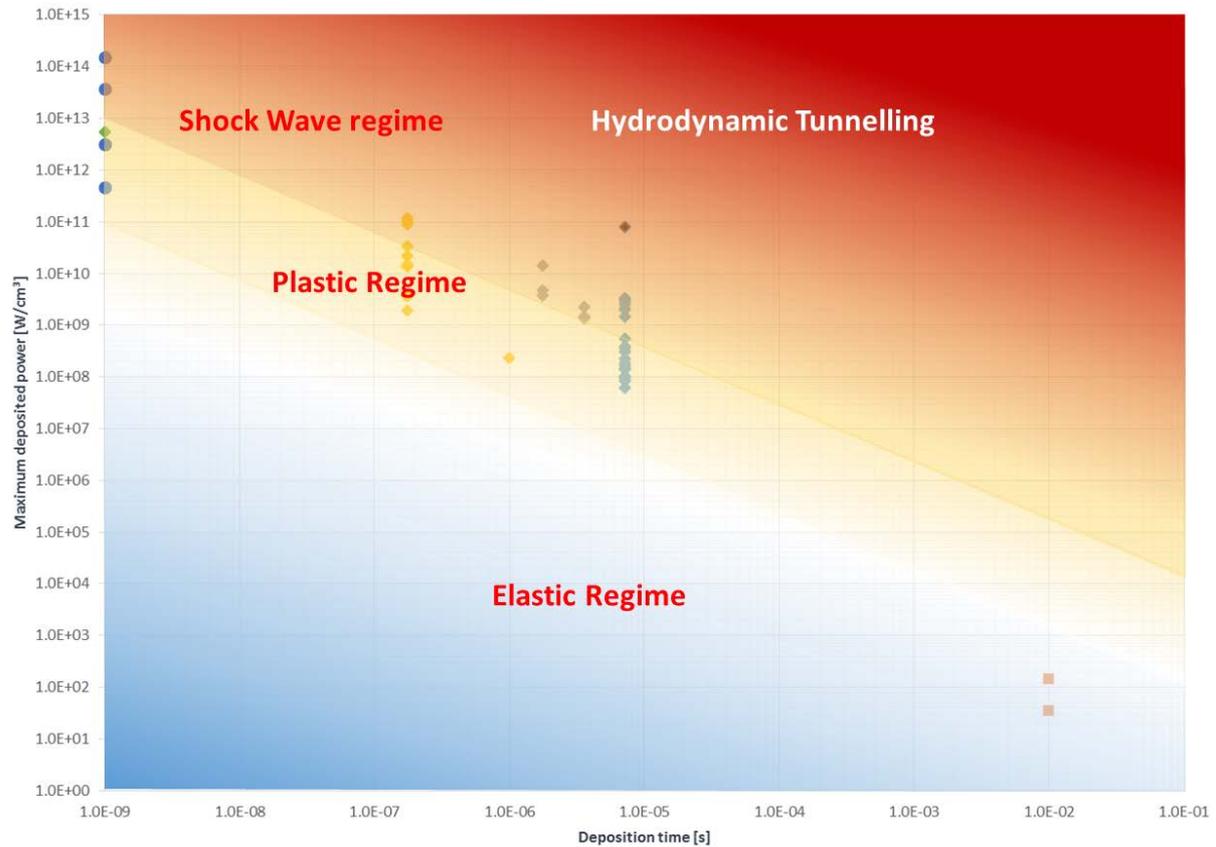

**Fig. 1:** Plot of maximum deposited power versus duration of deposition, showing the different dynamic responses that can be induced in matter by interaction with particle beams. Points represent cases of beam impacts (real or simulated).

We will then introduce non-linear dynamic responses, which require, to be correctly treated, the use of numerical tools ranging from standard finite element methods to sophisticated highly non-linear wave propagation codes, making use of complex material constitutive models.

Next, the principles that should guide the design of equipment subjected to beam-induced accidents will be briefly set out, introducing relevant figures of merit for such cases, and finally we will describe the experimental tests that are essential to validate the numerical simulations and qualify the design of such components.

## 1.1 Examples of beam-induced accidents

Figure 2 shows thin rods that were part of the first neutrino target station installed in the Super Proton Synchrotron (SPS) at CERN [7]; they were impacted by a beam of $1 \times 10^{13}$ protons at 300 GeV/$c$, with a cross-section of roughly 2 mm, with a certain offset with respect to the centre. These rods were made of beryllium and are 100 mm long, 3 mm in diameter; the pulse duration was roughly 23 µs. The accident took place in the early 1970s.

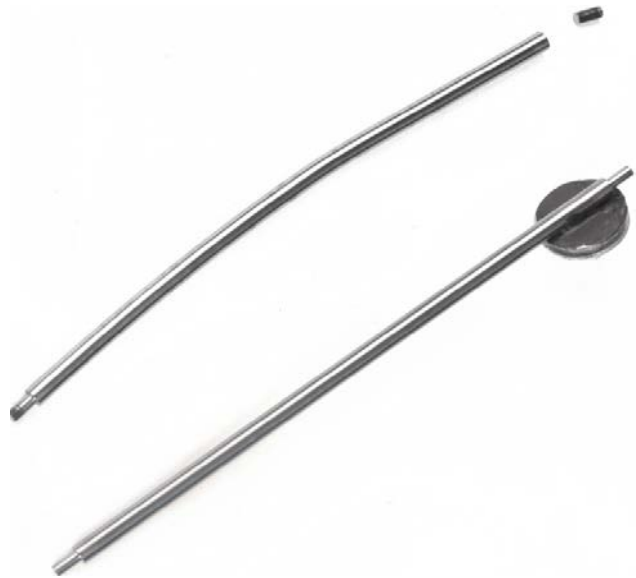

**Fig. 2:** Beryllium rods for first neutrino target installed at CERN-SPS. The unit on top was accidentally hit by an off-axis beam, leading to permanent bending and failure of the rod.

Figure 3 depicts a 30 cm long copper block onto which a beam damage test was performed in 1971 at SLAC [8]; the block was meant to simulate a collimator. An 18 GeV/$c$ e$^-$ beam of ≈2 mm diameter, with a power of roughly 500 kW, impinged the edge of the block. The impact lasted roughly 1.3 s and led to extensive melting in the impacted area. Although this was a relatively slow accident, the type of damage it generated can be compared to those created by faster events.

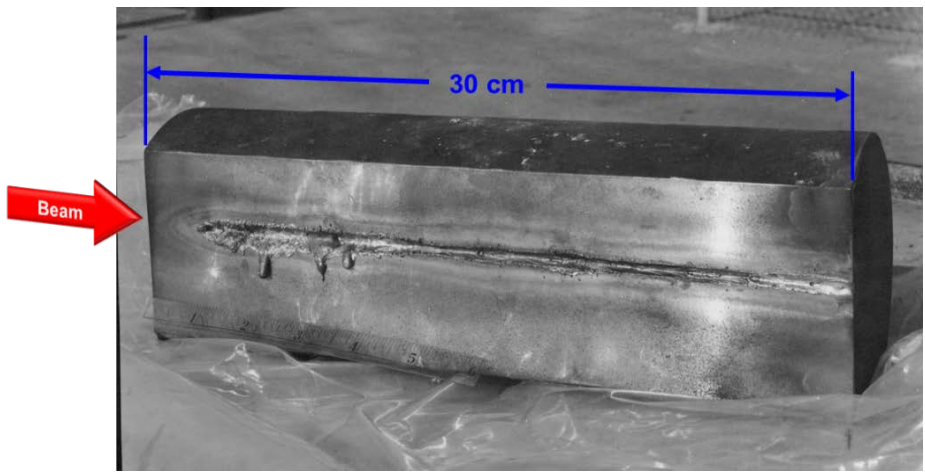

**Fig. 2:** Copper block used for damage test at SLAC in 1971

Figure 4 illustrates the effects of an accident that took place at Tevatron (FNAL). In 2003, a Roman pot accidentally moved towards the Tevatron beam, generating an intense cascade of secondary particles, which led to fast quenches in several superconducting magnets [9]. This in turn caused a drift of the 980 GeV/$c$ proton beam, which eventually hit a primary collimator, made of tungsten alloy, and a secondary collimator, made of stainless steel, resulting in a 2.5–3 mm hole in the 5 mm thick tungsten unit and an extended groove several centimetres long on the stainless steel collimator. The peak density of deposited energy on tungsten alloy is in the range of 1 kJ g$^{-1}$ (≈20 kJ cm$^{-3}$).

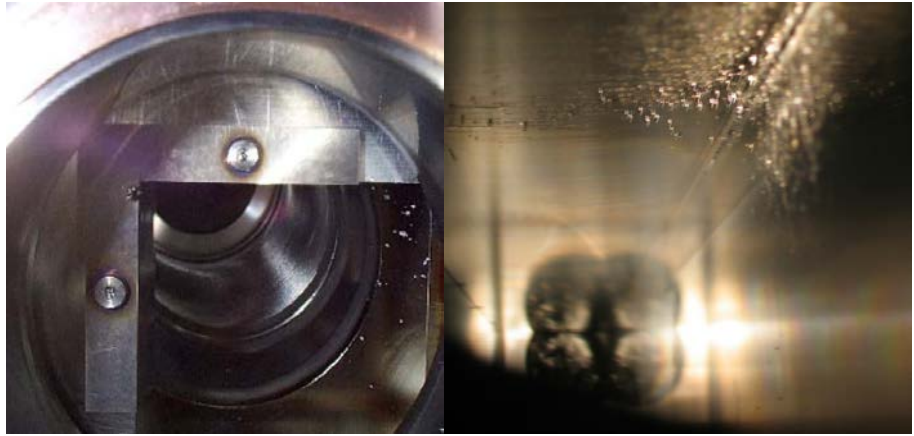

**Fig. 4:** Tevatron collimators accidentally hit by proton beam: left, tungsten alloy primary collimator; right, stainless steel secondary collimator.

During high-intensity extraction from SPS (CERN) in 2004, an incident occurred in which a stainless steel vacuum chamber of a magnet in TT40 transfer line was badly damaged. The beam was a 450 GeV full Large Hadron Collider (LHC) injection batch of $3.4 \times 10^{13}$ p$^+$ in 288 bunches extracted from SPS with a wrong trajectory. The beam drift was induced by the switch-off of a septum magnet [10]. This provoked a 110 cm long groove and a cut of 25 cm on the side of the impact with projection of molten steel on the opposite side (Fig. 5). Both the vacuum chamber and the magnet had to be replaced.

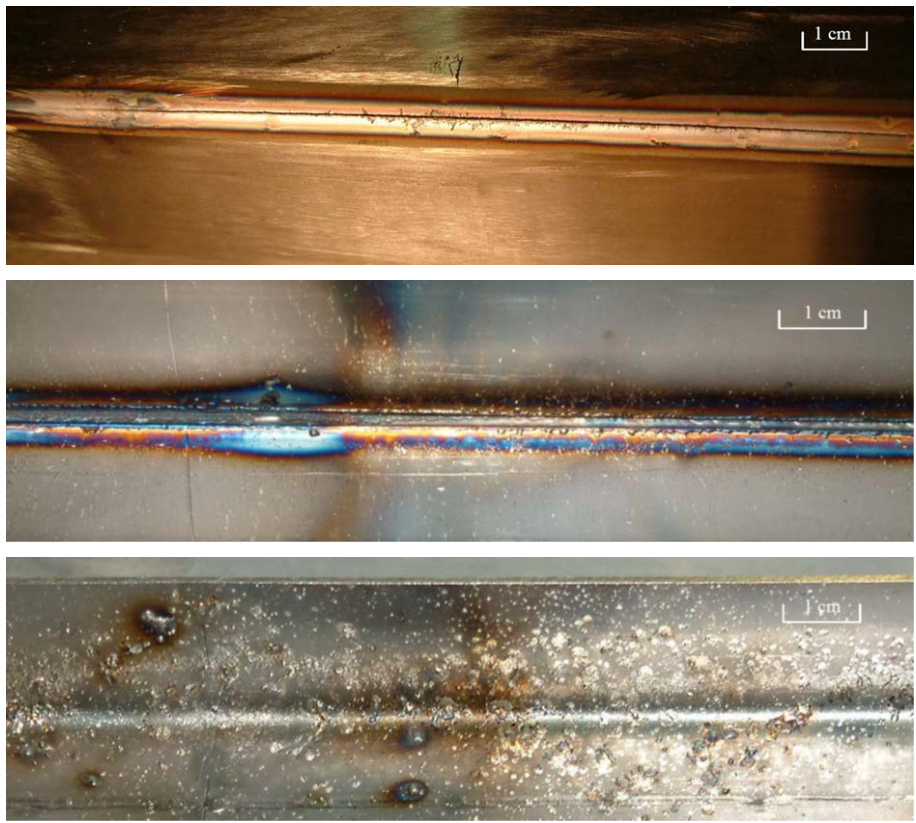

**Fig. 5:** Damages on a vacuum chamber of TT40 magnet (SPS-LHC transfer line). Top: Outside of the vacuum chamber. Centre: Inside of the vacuum chamber, beam impact side. Bottom: Inside of the vacuum chamber, side opposite to the beam impact

The accident presented in Fig. 5 occurred while the beam was being prepared for a series of tests, including one which aimed at determining the damage threshold of several materials under the impact of a 450 GeV proton beam from the SPS [11]. The target consisted of a series of tightly packed plates made of metals commonly used in accelerators, such as copper, stainless steel, and Inconel™ Ni–Cr superalloy. Pulses at four intensities, ranging from $1.32 \times 10^{12}$ p$^+$ to $7.92 \times 10^{12}$ p$^+$ with an average root-mean-square beam size of 0.85 mm ($\sigma_x$, 1.1 mm, $\sigma_y$, 0.6 mm). Effects of the impacts on copper are presented in Fig. 6: letters correspond to different intensities. At $1.32 \times 10^{12}$ (A), no signs of damage are visible, at $2.64 \times 10^{12}$ (B), coloration starts to appear, at $5.28 \times 10^{12}$ (C) and $7.92 \times 10^{12}$ (D) melting becomes clearly visible. These results are essentially in accordance with the value of the calculated peak energy deposition (Fig. 7): for a beam size of 1 mm at 450 GeV, melting in copper is expected to begin at an intensity of $\approx 2.5 \times 10^{12}$ protons (see Section 2.1 for calculation method).

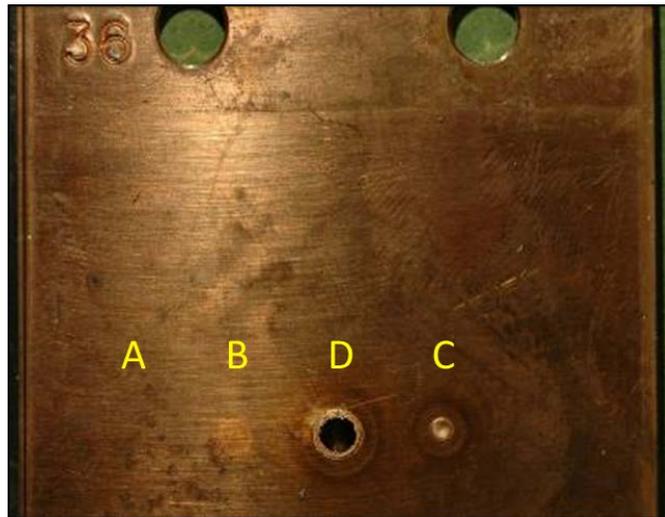

**Fig. 6**: Effect of beam impacts at different intensities on copper target

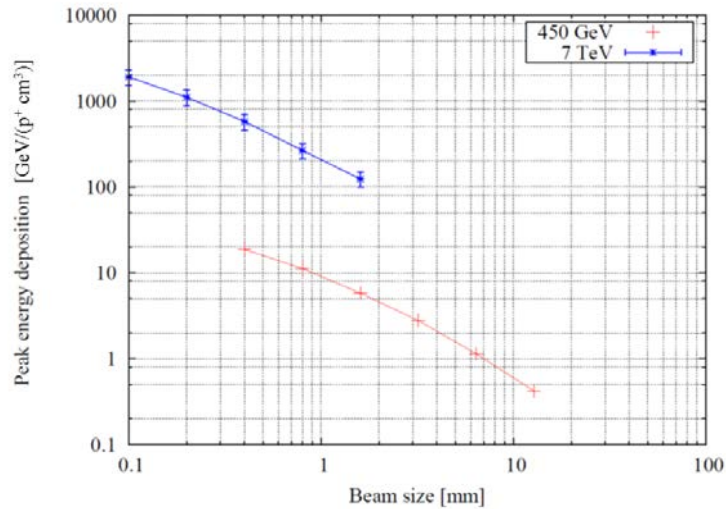

**Fig. 7:** Peak energy deposition in a copper target at 0.45 and 7 TeV as a function of beam size [11]

In 2012, an experiment was carried out at CERN, in the recently commissioned HiRadMat facility [12] to test the behaviour of a LHC tertiary collimator in case of direct beam impact [13]. The beam was extracted from SPS at an energy of 450 GeV/*c*.

Figure 8 shows a collimator jaw, whose active part is made of five blocks of tungsten heavy alloy (Inermet® IT180 from Plansee, Austria), after three distinct tests performed at various beam intensities.

In test 1, 24 SPS bunches with a total intensity of $3.36 \times 10^{12}$ p$^+$ hit the Inermet blocks: a groove several centimetres long and a few millimetres high was produced. This impact was intended to produce an energy distribution similar to that induced by one full LHC bunch ($1.15 \times 10^{11}$ p$^+$) at 7 TeV. In test 2, the impacting beam intensity was $1.04 \times 10^{12}$ p$^+$: according to simulations this was the threshold value at which plastic deformations were first induced, without material fragmentation. In test 3, a train of 72 SPS bunches impacted the jaw with a total intensity of $9.34 \times 10^{12}$ p$^+$. It can easily be inferred that a small fraction of one LHC bunch is sufficient to generate extensive damage, with changes of phase, material ejection, and fragmentation, on high-Z materials, such as tungsten alloys.

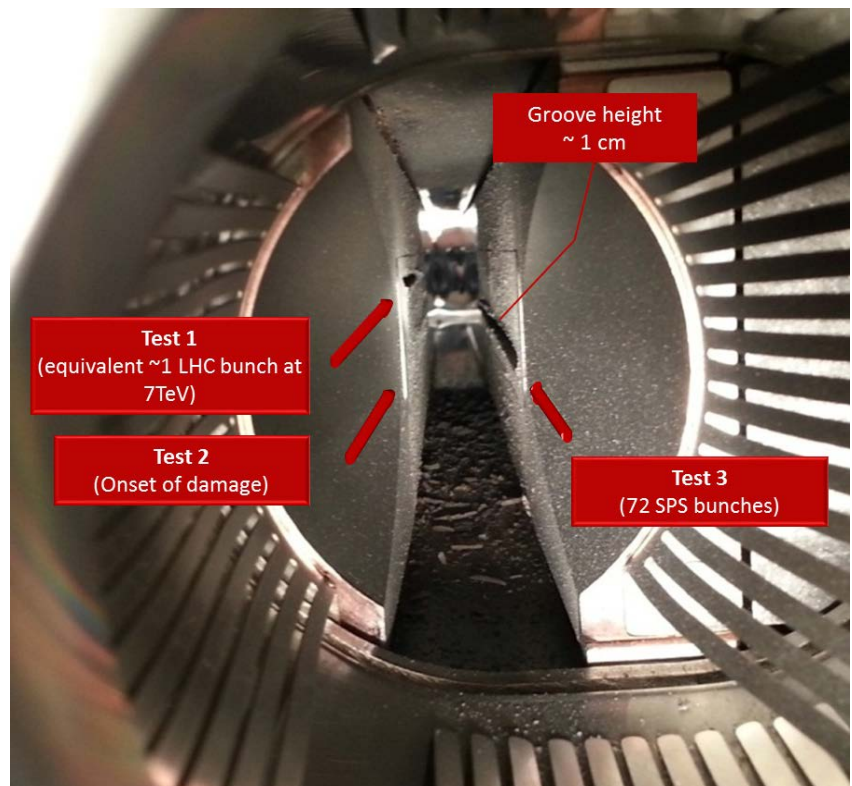

**Fig. 8:** Effects of three impact tests on a LHC tertiary collimator jaw at various beam intensities

## 2  Analysis of beam interaction with matter

As the preceding examples prove, damage phenomena induced by high-energy, high-intensity particle beams bring matter to extreme states, where practical experience and material knowledge is very limited. Hence, the accurate prediction of the structural response to such events becomes very complex. The analysis of these phenomena must rely on methodologies that integrate and couple several fields of science and engineering, numerical tools and experimental verification in a multidisciplinary approach.

From an engineering perspective, these problems can be attacked by dividing the procedure into three successive steps.

– The physical problem. The main goal of this step is to determine how much energy, and where, was deposited onto the relevant body.

– The thermal problem. The objective is to determine which temperature distribution, at which moment in time, was induced in the body by the deposited energy computed in the first step.

– The thermomechanical problem (which may be linear or non-linear). Given the temperature field, the goal is to determine which strains, stresses, deformations, dynamic responses, and phase transitions were generated in the body.

## 2.1 The physical problem

Particles interact with matter through various mechanisms, which typically depend on particle species and energies, and on the density, atomic number, and atomic mass of the impacted material.

Since, as mentioned already, we are not interested in long-term damage mechanism induced by particle irradiation and the changes they induce in material properties, what only matters, for the problems we are dealing with, is that the part of beam energy that is lost in the target during particle-matter interaction is ultimately transformed into heat. These interactions occur during extremely short time-scales, of the order of $10^{-11}$ s [14]: this is sufficiently short to consider the heat generation by each particle as an instantaneous process.

Monte Carlo interaction and transport codes, such as FLUKA, MARS or Geant4, are typically used to simulate these phenomena and predict the distribution of energy deposited per interacting particle [1].

The energy deposition process lasts as long as the particles interact with matter. This depends on the bunch length, the number of interacting bunches and their time spacing. The total deposited heat can be simply calculated by multiplying the single particle energy distribution by the total number of particles.

It is interesting to note that the linear scaling of the deposited energy with the number of particles holds, provided that the material density does not change during the interaction. Substantial changes of density do occur if the temperature increase leads to phase transitions (melting, vaporization, plasma generation, etc.) or if severe shock waves physically displace matter in the region of the impact. In this case, particular caution must be taken, since, if the interaction with the beam is still ongoing as density varies, the energy deposition distribution will be modified by the change of density, typically reducing energy peaks in the upstream part and extending the interaction along longer portions of the target (see Section 2.4.4 for details).

### 2.1.1 *Temperature distribution*

Once the energy deposition distribution is available, the quasi-instantaneous temperature distribution can be easily calculated by taking into account the material specific heat,

$$q_V(x_i) = \int_{T_i}^{T_f} \rho \cdot c_p(T) \cdot \mathrm{d}T(x_i) \ , \tag{1}$$

where $q_V$ is the *deposited energy per unit volume* (J cm$^{-3}$) at the location $x_i$ ($i$ ranges from 1 to 3), $\rho$ is the *density* of impacted material (g cm$^{-3}$), $c_p$ is its *specific heat capacity* at constant pressure (J g$^{-1}$ K$^{-1}$) and $T(x_i)$ is the local *temperature* (K), with $T_i$ and $T_f$ being the temperatures at the beginning and the end of the energy deposition process under analysis.

Attention must be paid when using Eq. (1) to determine the temperature distribution $T(x_i)$; in fact, this relationship implicitly assumes that no heat diffusion occurs while heat generation occurs, i.e. the *duration of the energy deposition* process $\tau$ is much shorter than the *thermal diffusion time* (see next sections for details).

In general, $c_p$ is a function of *temperature*, $T$. When this dependence is not too strong, to a first approximation, an average value, $\bar{c}_p$, can be taken to obtain the quasi-instantaneous temperature increase. In such a case, the temperature increase can be derived explicitly; if, for convenience, we set the initial temperature to zero, we obtain

$$T(x_i) \cong \frac{q_V(x_i)}{\rho \cdot \bar{c}_p} \ . \tag{1a}$$

## 2.2 The thermal problem

To assess the stress state in a body submitted to thermal shocks, it is fundamental to determine the initial temperature distribution and its evolution over time. The temperature evolution is governed by a diffusion process, the *heat equation* (also known as *Fourier's equation*):

$$\rho c_p \frac{\partial T}{\partial t} = \nabla \cdot (\lambda \nabla T) + \dot{q}_V ~, \qquad (2)$$

where $\dot{q}_V$ is the energy deposition rate or *heat generation rate* (W m$^{-3}$), and $\lambda$ is the *thermal conductivity* (W m$^{-1}$ K$^{-1}$).

If we assume that the material is *homogeneous* and *isotropic*, so that physical properties do not change from point to point and with material orientation, we get

$$\frac{\partial T}{\partial t} = a \nabla^2 T + \frac{\dot{q}_V}{\rho c_p} ~, \qquad (2a)$$

where $a = \frac{k}{\rho c_p}$ is the thermal diffusivity (m$^2$ s$^{-1}$).

It is interesting to note that Eq. (2) fails to predict heat transfer phenomena for very short timescales, given that it implies infinite speed of heat signal propagation. This is usually not relevant in the problems we are dealing with, but can play a role in ultra-short phenomena, such as high-frequency laser pulsed heating lasting of the order of femtoseconds.

If thermophysical properties are also constant with temperature, Eq. (2a) becomes a partial differential equation with constant coefficients, which, in some cases, can be solved analytically.

Once energy deposition is completed ($t \geq \tau$, that is, the time $t$ is much greater than $\tau$, the duration of the energy deposition), Eq. (2a) becomes a homogeneous linear partial differential equation.

$$\frac{\partial T}{\partial t} = a \nabla^2 T ~. \qquad (2b)$$

If we assume that the initial temperature distribution is known and that, at least for the short timescales we are interested in, the system is adiabatic regardless of the actual boundary conditions (e.g. active cooling), analytical solutions to Eq. (2b) become available for simple geometries, usually involving Fourier series, Bessel series, Laplace transforms, etc. [15].

A useful case is that of a circular cylinder or disc with an axially symmetrical energy distribution that is constant along the axis: this may well approximate impacts at the centre of thin circular windows or cylindrical targets. In this case, the solution, in cylindrical coordinates, to Eq. (2b) with an initial temperature distribution obtained from Eq. (1a) and adiabatic boundary conditions is given by

$$T(r,t) = \sum_i C_i J_0\left(\frac{\beta_i}{R} r\right) \cdot e^{-\frac{a}{R^2} \beta_i^2 \cdot t} ~, \qquad (3)$$

where $R$ is the outer radius of the disc or cylinder, $J_0\left(\frac{\beta_i}{R} r\right)$ is a Bessel function of the first kind of order zero with $\frac{\beta_i}{R}$ being the eigenvalues of the problem obtained by imposing the adiabatic boundary condition and $C_i$ are numerical coefficients derived from the initial temperature distribution [16].

Analytical solutions can also be derived for more complicated cases, such as beams with rectangular cross-section or discs and cylinders with off-axis energy deposition [15].

Assuming that the energy deposition profiles can be approximated with an axially symmetrical normal Gaussian distribution, as is often the case, the initial temperature field in a disc or circular cylinder takes the form

$$T(r,\tau) = T_0(r) = T_{\max} e^{-\frac{r^2}{2\sigma_b^2}}, \tag{3a}$$

where $\sigma_b$ is the standard deviation of the distribution and $T_{\max}$ is the maximum initial temperature, obtained from Eq. (1a).

As an example, the temperature distribution obtained from solving Eq. (3) for a thin graphite disc with outer radius $R = 5$ mm is shown in Fig. 9, assuming a Gaussian round beam with $\sigma_b = 0.05R$.

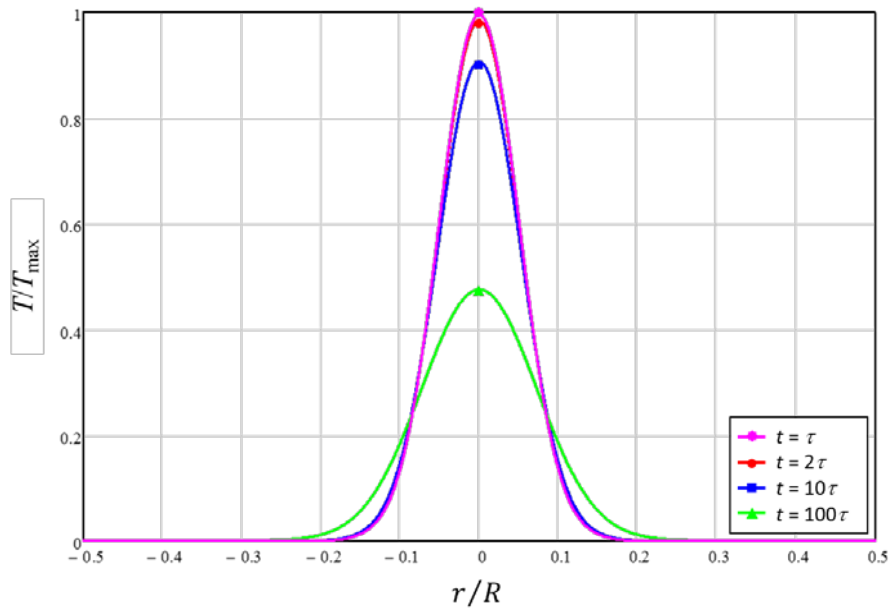

**Fig. 9:** Temperature distribution (normalized to maximum initial temperature $T_{\max}$) in the central area of a graphite circular disc impacted at its centre by a Gaussian round beam with $\sigma_b = 0.05R$ at different instants in time ($\tau$ is the duration of the energy deposition).

It is very important to note that in practically all analytical solutions, regardless of their mathematical complexity, a characteristic time, called the *thermal diffusion time*, $t_d$ can be identified:

$$t_d = \frac{B^2}{a}. \tag{3b}$$

This parameter is related to the time required to reach, by heat diffusion processes, a uniform temperature distribution in a region whose relevant dimension is $B$ (e.g. the radius of a disc).

For the typical dimensions of interest (several millimetres or more), the thermal diffusion time lasts from several to many milliseconds, which is usually much longer than the duration of beam impacts we are concerned with (of the order of microseconds or less): this is why the assumption of instantaneous heat deposition is generally acceptable. Heat diffusion times for materials of interest for accelerator components exposed to interaction with the beam are given in Table 1.

Even if at the impacted component global scale the impact can be considered instantaneous when the diffusion time is much longer than the impact duration, it is important to note that very sharp and narrow temperature peaks may start to flatten even during the energy deposition process: this is because at the sub-millimetric scale the diffusion time becomes much shorter and comparable to the duration of the impact.

Observing Fig. 9, it can be seen that the temperature at the centre of the beam spot tends to decrease relatively slowly: at $t \cong 2\tau$, i.e. after a time equal to the duration of the energy deposition, the maximum temperature has decreased by roughly 1%. In such a case, the assumption of instantaneous energy deposition seems acceptable.

However, for smaller beam sizes, temperature, at the centre of the disc, drops at a much faster rate: as shown in Fig. 10, for $\sigma_b = 0.01R$ (with $R = 5$ mm) the temperature at the centre has already fallen by more than 20% when $t = 2\tau$. In such a case (*a fortiori* with smaller beam sizes), neglecting the heat diffusion processes occurring during the impact is not appropriate and the initial temperature obtained through Eq. (1a) would be largely overestimated.

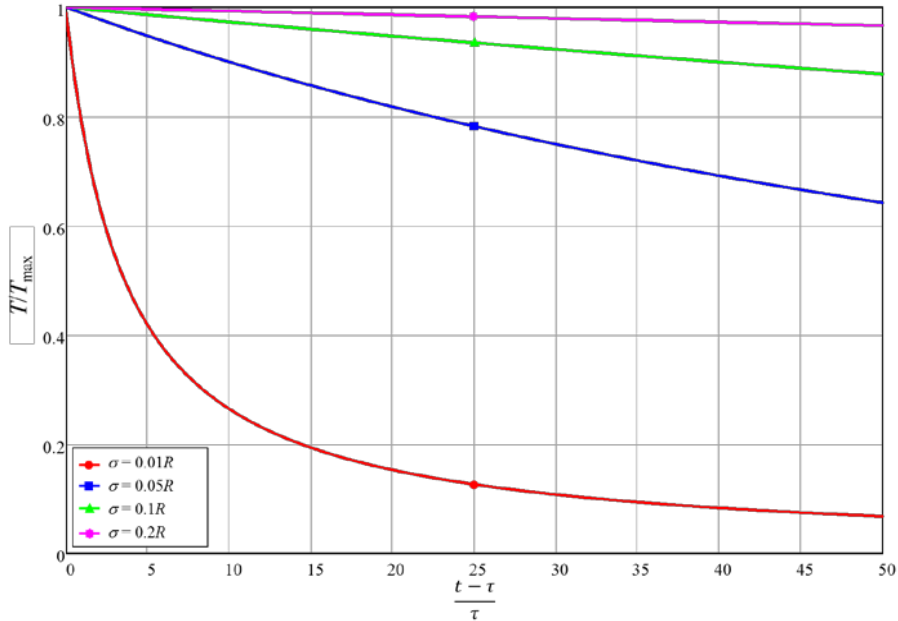

**Fig. 10:** Temperature at the centre of the disc (normalized to $T_{max}$) as a function of time (normalized to impact duration $\tau$) for different beam sizes ($R = 5$ mm).

**Table 1:** Thermal diffusion times on several length scales for materials of interest

| Material | Properties at room temperature | | | | Thermal diffusion time [ms] | | |
|---|---|---|---|---|---|---|---|
| | Density [kg m$^{-3}$] | Specific heat [J kg$^{-1}$·K$^{-1}$] | Thermal conductivity [W m$^{-1}$·K$^{-1}$] | Thermal diffusivity [mm$^2$ s$^{-1}$] | $B$ = 0.1 mm | $B$ = 1 mm | $B$ = 1 cm |
| **Copper (Glidcop)** | 8 900 | 391 | 365 | 104.9 | 0.10 | 9.5 | 953 |
| **Tungsten alloy (Inermet180)** | 18 000 | 150 | 90.5 | 33.5 | 0.30 | 29.8 | 2983 |
| **Molybdenum** | 10 220 | 251 | 138 | 53.8 | 0.19 | 18.6 | 1859 |
| **Titanium alloy (Ti6Al4V)** | 4 420 | 560 | 7.2 | 2.9 | 3.44 | 343.8 | 34378 |
| **Aluminium alloys** | 2 700 | 896 | 170 | 70.3 | 0.14 | 14.2 | 1423 |
| **Molybdenum–graphite (MG-3110)** | 2 500 | 740 | 770 | 416.2 | 0.02 | 2.40 | 240 |
| **Graphite** | 1 850 | 780 | 70 | 48.5 | 0.21 | 20.6 | 2061 |
| **Beryllium** | 1 844 | 1925 | 216 | 60.9 | 0.16 | 16.4 | 1643 |

## 2.3 The linear thermomechanical problem

### 2.3.1 Stresses and strains

Any body submitted to a *mechanical stress*, defined as the limit of the ratio between a force (vector) and the surface it is acting upon, responds by deforming. The ratio of stress-induced deformation to the initial dimension is called *mechanical strain*. For a slender body loaded along its axis, the mechanical (*normal* or *axial*) strain is defined as the change in length, $\delta$ per unit of the original length, $L$, of the body: $\varepsilon_M = \delta/L$. The normal strain is positive if the material 'fibres' are stretched and negative if they are compressed.

More generally, on a given plane of an infinitesimal volume of a body, the stress vector can be decomposed into a component perpendicular to the plane and two orthogonal in-plane components (Fig. 11). The component normal to the surface is called the *normal stress* and the components that act in-plane are the *shear stresses*. These three components, in combination with the three main planes ($x$, $y$ and $z$ or 1, 2, 3), form the nine components of the *stress tensor*. Strain components of the *strain tensor* are defined in a similar way.

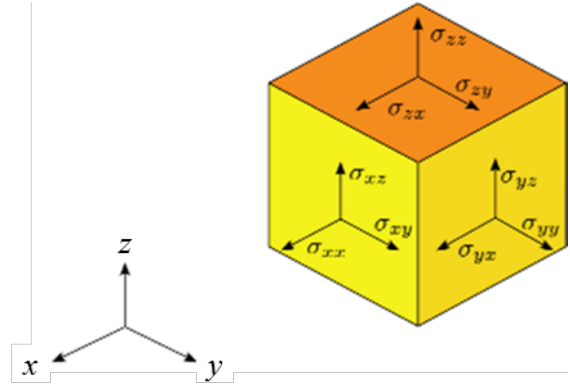

**Fig. 11:** Stress components acting on an infinitesimal volume

### 2.3.2 Linear elasticity

In *linear elasticity* it is postulated that a linear relationships exists between stresses and strains. Mathematically, this is expressed by *Hooke's law*, which constitutes an idealization of the behaviour of most materials submitted to low or moderate stresses.

In indicial notation, for an isotropic body, this relationship takes the expression

$$\varepsilon_{ij}^M = \frac{1}{E}\left[(1+\nu)\sigma_{ij} - \nu\delta_{ij}\sigma_{kk}\right] \quad, \tag{4}$$

where $E$ is the *Young modulus* (Pa), $\nu$ is the *Poisson ratio*, and $\delta_{ij}$ is the Kronecker delta.

The Poisson ratio expresses the tendency of the material to expand in the two directions perpendicular to the direction of compression. Conversely, if the material is stretched, it usually contracts in the directions transverse to the direction of stretching; $\nu$ is the negative ratio of transverse to axial strain.

If only one component of normal stress is acting, Eq. (4) reduces to the well-known linear stress–strain relationship:

$$\varepsilon_{11} = \frac{\sigma_{11}}{E} \quad. \tag{4a}$$

## 2.3.3 Linear thermo-elasticity

It is well known that an unrestrained body submitted to a change of temperature undergoes a dimensional change called *thermal deformation*.

Strains caused by thermal deformation on unrestrained bodies heated from an initial reference temperature (usually uniform and equal to ambient temperature), are called *free thermal strains*, $\varepsilon^T$.

The rate of linear change of dimension per unit temperature variation is called the *linear coefficient of thermal expansion* (CTE), $\alpha$.

$$\alpha(T) = \frac{dL}{L\,dT}\,. \tag{5}$$

The CTE has units of $K^{-1}$ and is, in general, a function of temperature (Fig. 12); at very low temperatures (usually below 80 K), $\alpha$ tends to zero. However, over limited temperature ranges above or around room temperature (RT), it can be averaged to a constant value.

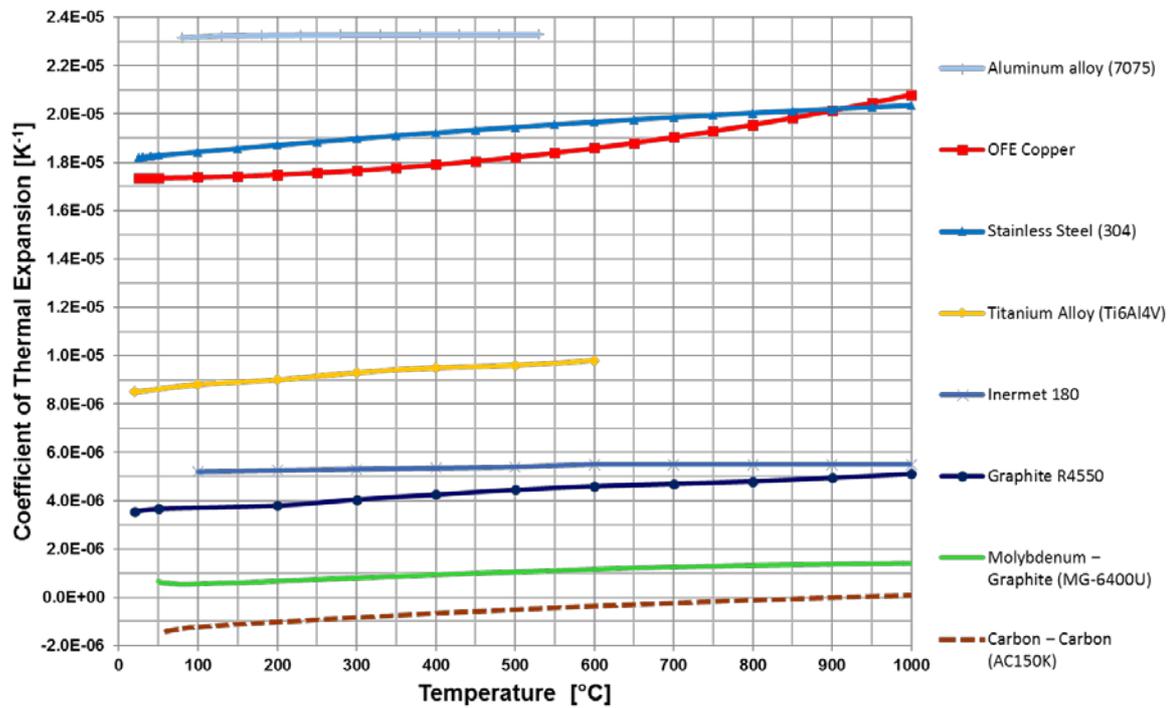

**Fig. 12:** Linear CTE for selected materials as a function of temperature

The linear CTE is related to the *volumetric coefficient of thermal expansion*, $\beta$, by the expression $\alpha = 1/3\,\beta$.

In the nineteenth century, Hooke's law was extended by Duhamel and Neumann to include first-order thermal effects (*linear thermo-elasticity*). They assumed that the *total strain*, $\varepsilon$, at a point consists of two components: the *mechanical strain*, $\varepsilon^M$, and *free thermal expansion*, $\varepsilon^T$.

We then have

$$\varepsilon_{ij} = \varepsilon_{ij}^M + \varepsilon_{ij}^T \tag{6a}$$

In indicial notation, the strain caused by free thermal expansion in an *isotropic* and *homogeneous* body is expressed as

$$\varepsilon_{ij}^{\text{T}} = \alpha \delta_{ij} T \ , \tag{6b}$$

with the initial reference temperature assumed uniform and taken identically equal to zero for convenience.

We note from Eq. (6b) that, for a homogeneous and isotropic body, only normal strain components are affected by a temperature change, that is, the deformation is only volumetric and, if the temperature change is uniform throughout, the shape of the body is maintained.

The *Duhamel–Neumann law* can then be expressed as

$$\varepsilon_{ij} = \frac{1}{E}\left[(1+\nu)\sigma_{ij} - \nu \delta_{ij}\sigma_{kk}\right] + \alpha \delta_{ij} T \ . \tag{7a}$$

It is important to note that, in an isotropic body, shear strains are never induced by free thermal expansion.

In general, stresses may be caused by mechanical loads, temperature gradients creating internal constraints to uniform expansion, or geometric restraints preventing free thermal expansion (*hyperstatic* design).

When such stresses remain well below the *yield strength*[1] of the material, defined as the conventional stress at which the material begins to deform plastically, the material is said to be in an *elastic regime*.

It can be observed that the smaller the CTE, the smaller the thermal strains and, hence, the total stresses: this is a fundamental concept in the design of devices directly interacting with the beam, since, for this type of equipment, mechanical loads are typically negligible and the design is usually isostatic, allowing free thermal expansion. The main (often single) source of stress is non-uniform temperature distribution (or non-homogeneity of CTE, e.g. in composite structures).

In the limiting case where CTE is zero everywhere, no thermal stresses are induced, regardless of the temperature increase!

Inverting Eq. (7a), *quasi-static total stresses* can be obtained:

$$\sigma'_{ij} = \frac{E}{(1+\nu)(1-2\nu)}\left[(1-2\nu)\varepsilon_{ij} + \nu \delta_{ij}\varepsilon_{kk}\right] - \delta_{ij}\frac{E\alpha T}{1-2\nu} \ . \tag{7b}$$

Although Eq. (7b) is time-dependent, since the temperature distribution obtained from Eq. (2) is a function of time, the stress distribution obtained can be considered quasi-static, given that the mass inertia effects are not yet taken into account (this is why the stress tensor components are primed).

Quasi-static stresses can be computed by combining Eq. (7b) with the equations of equilibrium and compatibility and the boundary conditions.

Some cases of special interest for isotropic bodies can be easily derived in the case of particular boundary conditions.

If no deformation is allowed, i.e. all total strain components are zero, as is the case for a fully constrained massive body, we observe that all shear stresses are zero while normal stresses in any element of the body are compressive and given by

$$\sigma'_{ii} = -\frac{E\alpha T}{1-2\nu} \ . \tag{7c}$$

---

[1] Although the terms *elastic limit*, *yield strength* and *flow stress* possess, strictly speaking, slightly different meanings, we will use them interchangeably in these lectures.

For a two-dimensional body, such as a thin, large plate, for which one can reasonably assume that in the through-thickness direction ($z$ or 3) normal stress is zero and deformation is free, taking all other directions as constrained, it can be shown that Eq. (7b) reduces to

$$\sigma'_{11} = \sigma'_{22} = -\frac{E\alpha T}{1-\nu} \ . \tag{7d}$$

Finally, if only one direction is constrained, the other being free to expand, as in longitudinally clamped thin, long beams and rods, it can be shown that only the axial normal stress is non-zero; this is given by

$$\sigma'_{11} = -E\alpha T \ . \tag{7e}$$

In general, for more complex structures and boundary conditions, the study of the elastic thermomechanical response relies on numerical methods (typically implicit, linear *finite element analysis*); however, analytical solutions to Eq. (7) exist for simple geometries: among the most important solutions are those available for long circular cylinders and thin discs.

### 2.3.4   *Thermo-elastic stresses in thin discs and long circular cylinders*

For discs and cylinders, assuming an axially symmetrical, $z$-independent thermal distribution $T(r,t)$ with adiabatic boundary conditions, we get, in a cylindrical reference system, for the radial and circumferential stresses

$$\sigma'_r(r,t) = \frac{E\alpha}{\zeta}\left[\frac{1}{R^2}\int_0^R T(r,t)r\,\mathrm{d}r - \frac{1}{r^2}\int_0^r T(r,t)r\,\mathrm{d}r\right], \tag{8a}$$

$$\sigma'_\theta(r,t) = \frac{E\alpha}{\zeta}\left[\frac{1}{R^2}\int_0^R T(r,t)r\,\mathrm{d}r + \frac{1}{r^2}\int_0^r T(r,t)r\,\mathrm{d}r - T(r,t)\right], \tag{8b}$$

where $\zeta = 1$ for discs and $\zeta = 1 - \nu$ for cylinders.

We note that at $r = 0$, radial and circumferential (or hoop) stresses are identical (compressive) and equal to:

$$\sigma'_r(0,t) = \sigma'_\theta(0,t) = \frac{E\alpha}{\zeta}\left[\frac{1}{R^2}\int_0^R T(r,t)r\,\mathrm{d}r - \frac{1}{2}T_0(t)\right]. \tag{8c}$$

This can be easily verified by observing that

$$\lim_{r\to 0}\frac{1}{r^2}\int_0^r T(r,t)r\,\mathrm{d}r = \frac{1}{2}T_0(t) \ .$$

Since the body is free to expand radially, at $r = R$, radial stress is zero, while hoop stresses are always larger or equal to zero.

Making use of Eq. (1a), one can observe that the first term in Eq. (8a) is proportional to the *total deposited energy* (per unit length) $Q_d$, which, for an adiabatic problem, once the impact is concluded, remains constant over time and is therefore proportional to the uniform *final temperature*, $T_F$. Stresses at the centre and outer rim can then be easily computed once the maximum temperature (which is proportional to the peak energy) is known.

In particular,

$$\sigma'_r(0,t) = \sigma'_\theta(0,t) = \frac{E\alpha}{\zeta}\left[\frac{Q_d}{2\pi R^2 \rho \bar{c}_p} - \frac{1}{2}T_0(t)\right] = \frac{E\alpha}{2\zeta}\left[T_F - T_0(t)\right], \text{ for } t \geq \tau \qquad (8d)$$

and

$$\sigma'_\theta(R,t) = \frac{E\alpha}{\zeta}\left[\frac{Q_d}{\pi R^2 \rho \bar{c}_p} - T_0(t)\right] = \frac{E\alpha}{\zeta}\left[T_F - T_0(t)\right], \text{ for } t \geq \tau. \qquad (8e)$$

Figure 13 shows radial and circumferential stresses for a typical axially symmetrical energy distribution at various times. At times larger than $t_d$, when the temperature becomes uniform and equal to $T_F$, radial and hoop stresses go to zero everywhere.

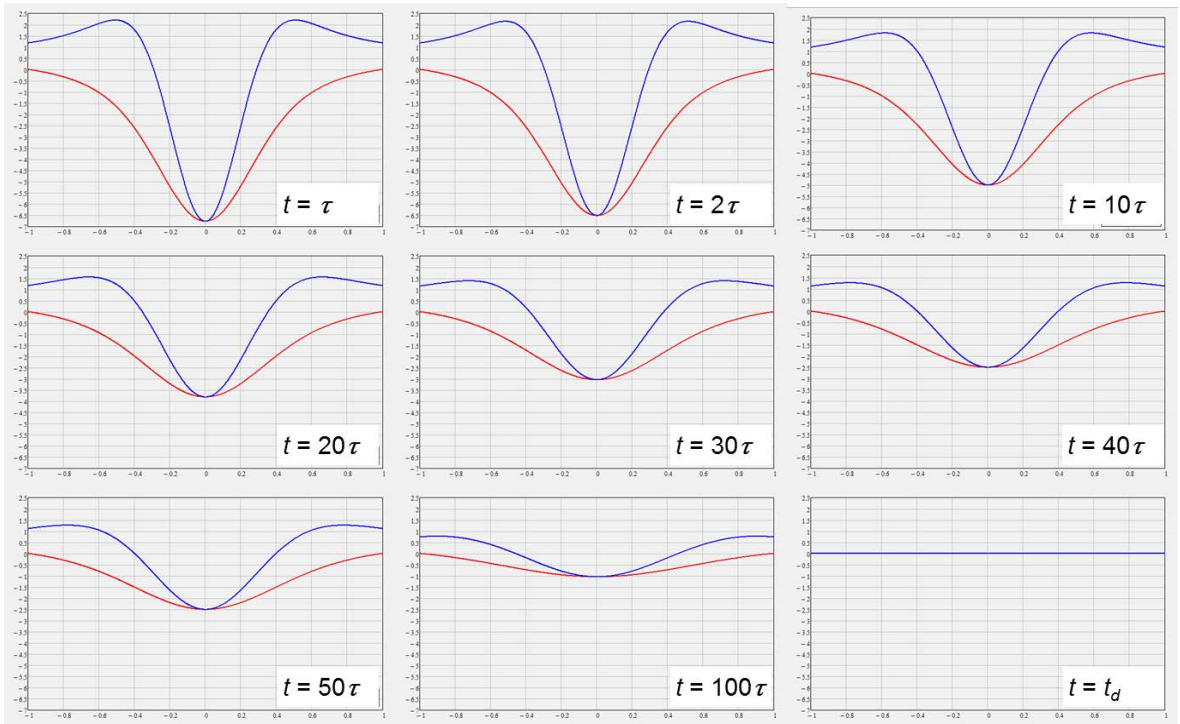

**Fig. 13:** Radial and circumferential stresses in a circular cylinder as a function of radial coordinates for an axially symmetrical normal distribution at various times.

For $t < \tau$, assuming that no thermal diffusion has occurred yet, one can simply scale linearly with time to the stress values at $t = \tau$.

The axial stress component is usually disregarded for thin discs, since through-thickness stresses are negligible; conversely, in the case of long, slender structures, such as rods, bars, or beams, axial stresses become very important and are the main cause of the dynamic response.

For such structures, to compute axial stresses, it is initially assumed that the body is restrained at its ends, i.e. the axial strain is zero throughout ($\varepsilon_z = 0$). In this hypothesis (known as the method of strain suppression [17]), quasi-static axial stresses can easily be derived from Eq. (7b).

For a long cylinder, using cylindrical coordinates, we get

$$\sigma'_z = \nu(\sigma'_r + \sigma'_\theta) - E\alpha T, \qquad (9)$$

using radial and circumferential components obtained from Eqs. (8a) and (8b).

The distribution of axial stresses integrated over the cross-section results in a net compressive force $R(t)$ (remember that we are blocking axial expansion, for convenience), which is a function of time and is equal to

$$R(t) = 2\pi \int_0^R \sigma'_z(r,t)\,\mathrm{d}r = -E\alpha \frac{Q_\mathrm{d}}{\rho c_p} = -E\alpha T_\mathrm{F} \pi R^2 = -F_{\mathrm{ref}}\ ,\ \text{for } t \geq \tau\ . \qquad (10)$$

We observe that the resultant axial force $R(t)$ is proportional to the *total deposited energy* (per unit length) $Q_\mathrm{d}$.

Very importantly, since $Q_\mathrm{d}$ is conserved after the impact (we are assuming an adiabatic problem), for $t \geq \tau$, $R(t)$ remains constant and proportional to the final uniform temperature $T_\mathrm{F}$, regardless of the actual deposited energy distribution.

For $t < \tau$, $R(t)$ increases, following the trend of the deposited energy, so, disregarding the actual bunched structure of the beam, to a first approximation, it can be assumed that $R(t)$ increases linearly from zero to a constant value, so that $R(t) = -F_{\mathrm{ref}}\,g(t)$, $g(t)$ being the unit function shown in Fig. 14.

If the structure is simply supported and free to expand (as is usually the case for an isostatic structure), to restore the free-end boundary condition and allow thermal expansion $\delta_\mathrm{T}$, a traction force opposed to the compressive resultant ($F(t) = -R(t)$) can be superposed at the two ends of the rod (Fig. 15).

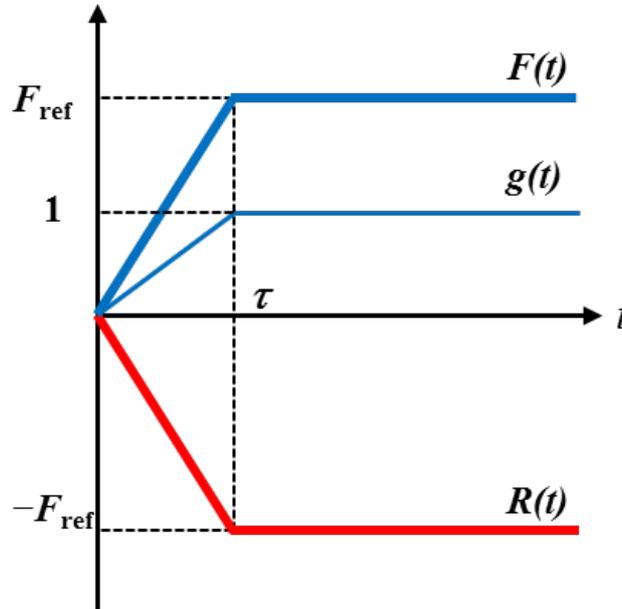

**Fig. 14:** Time history of the compressive force $R(t)$ acting on a rod clamped at its ends and impacted by a beam (impact duration $\tau$) and of the traction force $F(t)$ to be superposed to restore the free-end boundary conditions. The unit function $g(t)$ having the same trend of $F(t)$ is also shown.

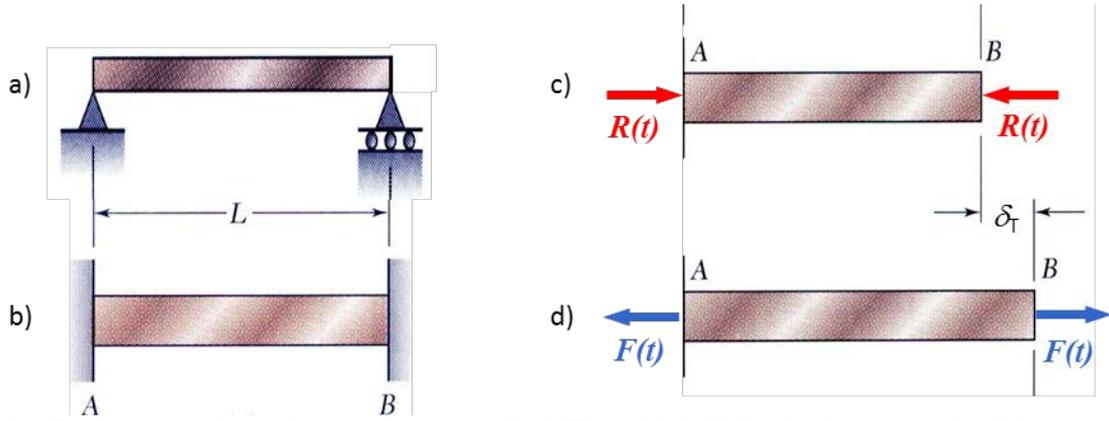

**Fig. 15:** Strain suppression and restoration approach. a) The rod is initially simply supported and free to expand. b) For convenience, the axial strain is suppressed, as if the rod were clamped. c) This leads to the generation of two compressive forces at the ends. d) The simply supported boundary condition is restored by superposing a traction force at the two ends.

Thanks to this approach, we have reduced the problem of a slender structure submitted to rapid heating to the well-known mechanical problem of the dynamic response of a beam to a pulsed axial excitation $F(t)$ with rise time $\tau$ (Fig. 12) applied at its ends. This response generates dynamic axial stresses $\sigma_{z_d}$ (uniform on the rod cross-section) that can be superposed on the quasi-static stresses given by Eq. (9). The total axial stress is therefore given by

$$\sigma_z = \sigma'_z + \sigma_{z_d} . \tag{11}$$

Dynamic stresses appear because of the coupling between the rapidly applied load (thermal or mechanical) and the inertia of the material: since the heating occurs in a very short time $\tau$, during this process, thermal expansion in the bulk material is partly prevented by its mass inertia. However, at the two free ends, expansion is allowed to occur from the very beginning since nothing prevents particle displacement in the material. Expansion starts from the two rod ends, propagating towards the centre of the structure at the *speed of sound*, $C_0 = \sqrt{E/\rho}$. In this way, two elastic stress waves are generated; to some extent, this is equivalent to an axial spring that is rapidly moved outwards at its end (Fig. 16). Since these are expansion waves, dynamic tensile stresses propagate and are superposed on the compressive axial stresses, which are due to the initially 'clamped' state.

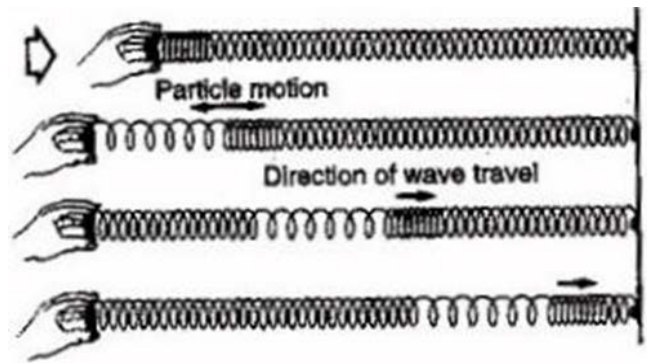

**Fig. 16:** Analogy of elastic wave propagation in a rod: rapid displacement at the free end generates a wave, which is propagated towards the centre.

The mechanical response of a simply supported cylindrical rod to a pulsed force with respect to time is a well-known problem in the theory of vibrations. It can be solved by resorting to such procedures as the *mode-summation method*. The *axial displacement*, $u(z,t)$, is expanded in terms of longitudinal *natural modes*, $\phi_z(z)$, and *generalized coordinates*, $q_z(t)$:

$$u(z,t) = \sum_i \phi_{z_i}(z) q_{z_i}(t) . \tag{12}$$

The solution can be obtained by means of *Lagrange's equation* for each independent mode:

$$\frac{d^2 q_{z_i}}{dt^2} + \omega_i^2 q_{z_i} = F_{z_i} . \tag{13}$$

The *natural modes* and the *natural (circular) frequencies* of longitudinal vibration for a simply supported beam of length $L$ are given, respectively, by

$$\varphi_{z_i}(z) = \sqrt{2} \cos\left(i\pi \frac{z}{L}\right), \tag{13a}$$

$$\omega_{z_i} = \frac{i\pi}{L} \sqrt{\frac{E}{\rho}} . \tag{13b}$$

The generalized forces $F_{z_i}$ are given by

$$F_{z_i}(t) = \frac{F_{\text{ref}} \sqrt{2}}{m}\left[1-(-1)^i\right] \cdot g(t) . \tag{13c}$$

Generalized coordinates $q_z(t)$ are obtained from the response of a system with a single degree of freedom excited by a ramp function given by Eq. (13c):

$$q_{z_i}(t \geq \tau) = \frac{F_{z_i}}{(\omega_{z_i})^2} \cdot \left(1 - \frac{\sin(\omega_{z_i} t)}{\omega_{z_i} \tau} + \frac{\sin[\omega_{z_i}(t-\tau)]}{\omega_{z_i} \tau}\right),$$

$$q_{z_i}(t < \tau) = \frac{F_{z_i}}{(\omega_{z_i})^2} \cdot \left(\frac{t}{\tau} - \frac{\sin(\omega_{z_i} t)}{\omega_{z_i} \tau}\right). \tag{14}$$

Finally, the dynamic longitudinal stress component is calculated as follows:

$$\varphi'_{z_i}(z) = -\sqrt{2} \cdot \frac{i\pi}{L} \cdot \sin\left(i\pi \cdot \frac{z}{L}\right), \tag{15a}$$

$$u'_z(z,t) = \sum_i \phi'_{z_i}(z) \cdot q_{z_i}(t) , \tag{15b}$$

$$\sigma_{z_d}(z,t) = E \cdot u'_z(z,t) , \tag{15c}$$

where $u'_z(z,t)$ denotes the first derivative of $u_z$ with respect to $z$, i.e. the longitudinal strain.

Figure 17 shows the evolution of dynamic axial stresses induced along the rod by the elastic wave at different times.

At the beginning of the impact, a tensile stress wave starts travelling at the speed of sound $C_0$ from both ends towards the centre while the force $F(t)$ (as well as axial stresses) linearly build up, generating a ramped wavefront. At the end of the impact ($t = \tau$), $F(t)$ and the axial stress stop increasing, reaching constant values of $F_{\text{ref}}$ and $\sigma_{\text{ref}}$, respectively; the two wavefronts have now covered a length equal to $\tau \cdot C_0$. It can be shown that the value of the axial stress reached at this moment is simply given by

$$\sigma_{\text{ref}} = \frac{F_{\text{ref}}}{\pi R^2} = E\alpha T_{\text{F}} \ . \tag{16}$$

The head of each elastic wave reaches the rod centre after one quarter of the *wave period* $t_M = 2L/c$; after this time, the two waves start superposing, continuing to increase the axial stress at the centre during a time equal to $\tau$, after this time a maximum stress equal to $2\sigma_{\text{ref}}$ is attained. The waves continue to propagate towards the other end of the rod, where the stress value always remains equal to $\sigma_{\text{ref}}$, as imposed by the boundary force $F_{\text{ref}}$ acting there. At half the wave period, both wave heads reach the opposite ends of the rod and begin to become reflected as a compressive wave, so decreasing the stress value. After one full wave period, the head of the reflected wave has reached the departure end, being reflected again as an expansion wave; after an additional time equal to $\tau$, all the wave ramped front has been reflected and the cycle starts again.

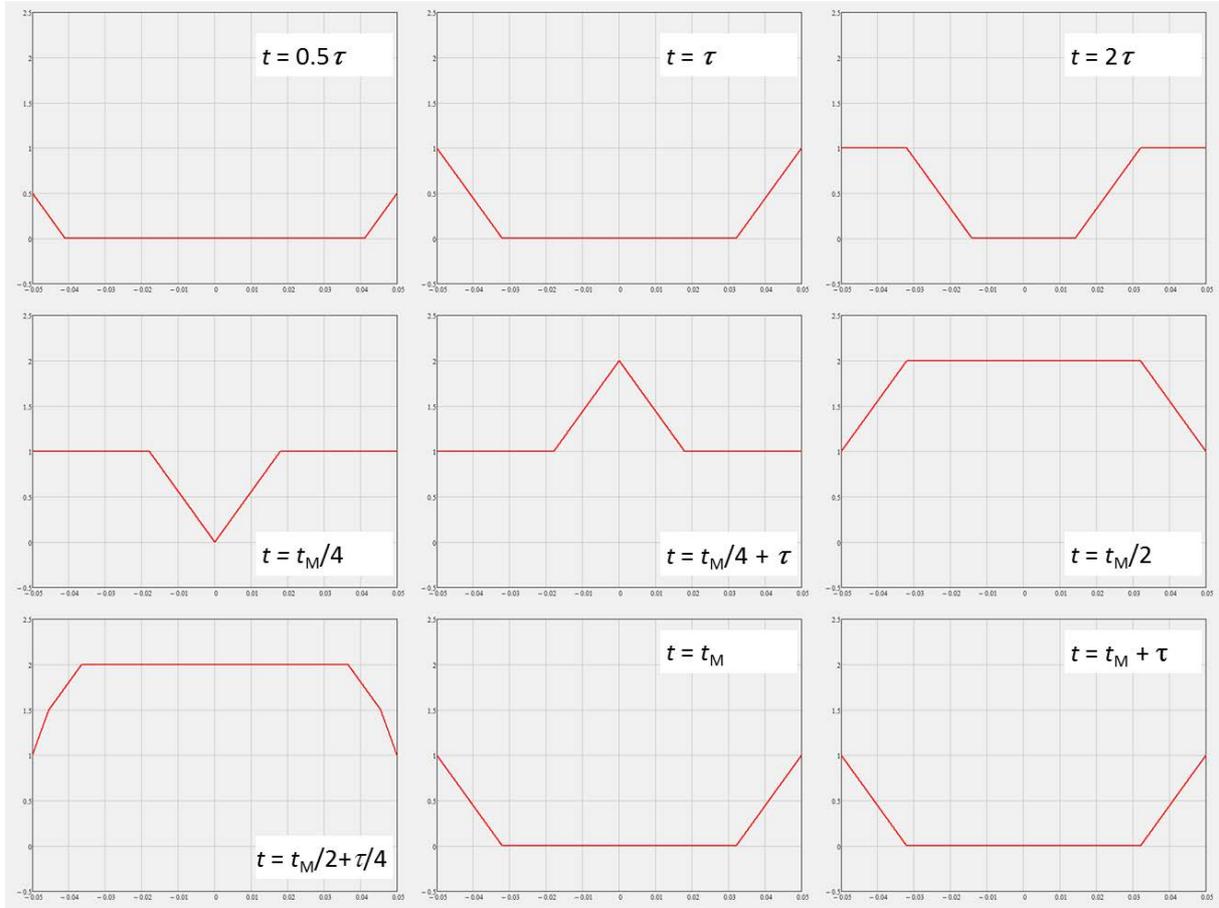

**Fig. 17:** Dynamic axial stress along a rod, scaled to $\sigma_{\text{ref}}$, at various times

Figure 18 shows the evolution over time of the axial dynamic stress.

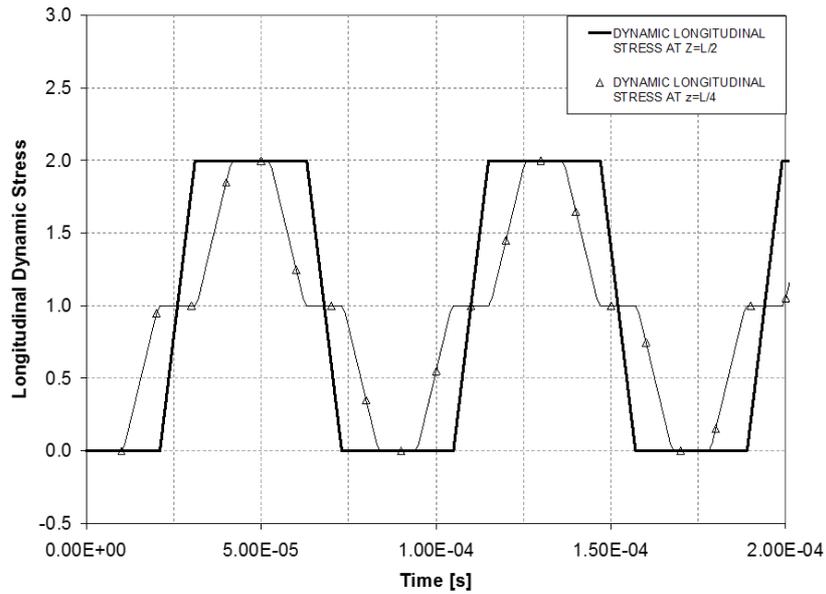

**Fig. 18:** Dynamic axial stress scaled to $\sigma_{\text{ref}}$ as a function of time at the centre and one quarter of the rod [15]

A similar approach, although more cumbersome, can be used for dynamic radial stresses. However, this component is small compared with quasi-static stresses in slender structures and can usually be neglected without affecting the general result [15].

If the beam hits the rod at a certain offset with respect to the centre, the energy deposition is no longer axially symmetrical and the problem becomes mathematically more complex, although it can still be solved analytically [18]. In such a case, dynamic bending stresses do appear, in addition to the dynamic stresses induced by the axial force. This can be explained by the fact that the resulting force at the two ends has an offset with respect to the centre of the beam, generating a bending moment, which varies in time. This effect is the probable cause of the permanent bending of the target rods depicted in Fig. 1: in that case, the beam, which hit the rod off its axis, probably caused dynamic bending stresses that exceeded the yield strength of the material, inducing a permanent bending of the rod.

## 2.4   The non-linear thermomechanical problem

High-energy accelerator components are usually designed to work in the *elastic domain*; however, in the case of highly energetic beam accidents, the dynamic response of the structure can largely exceed this regime and lead, depending on the intensity of the phenomenon, to permanent deformations, very high pressures, changes of material density, phase transitions, intense stress waves, material fragmentation, and explosions [19].

When a fast transient load generates stresses with an amplitude exceeding the elastic limit of the material, the response will decompose into an elastic and a plastic wave. The plastic stress wave usually propagates at velocities lower than the elastic speed of sound ($C_0$). However, if the energy is high enough to provoke stresses and rates of deformation (*strain rate*) exceeding a critical threshold of the order of $10^4$ s$^{-1}$ (Fig. 19), an energetic shock wave is formed, propagating at a velocity higher than $C_0$, and potentially leading to severe damage to the affected component [20].

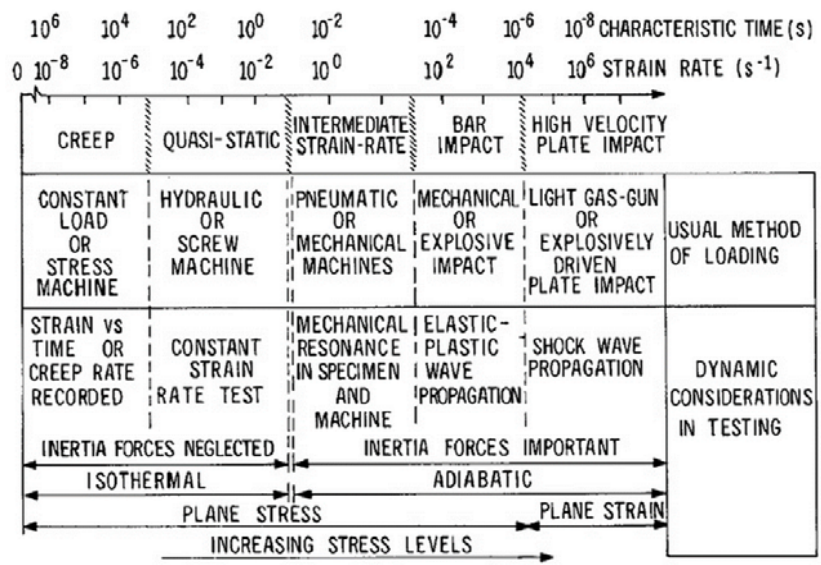

**Fig. 19:** Mechanical behaviour with changing strain rates and load duration [21]

Although an unambiguous identification of the critical threshold depends on the type of shock and on the geometrical conditions and is not always straightforward, as we have seen in Section 1, it is convenient to distinguish the responses according to their severity in a *plastic wave regime* or *plastic dynamic regime* when well below the threshold and in a *shock wave regime* above it. The former is usually associated with a limited permanent deformation induced by the beam impact but not with catastrophic failure, which is typical for the latter regime.

The treatment of both problem classes using pure analytical methods is virtually impossible and one must resort to numerical methods. However, the complexities of the tools required to compute the effects of these two types of regime are usually quite different, the analysis of events implying intense shock waves requiring more sophisticated numerical tools.

### 2.4.1 Numerical methods for beam-induced dynamic phenomena

#### 2.4.1.1 Time integration methods

Two substantially different numerical approaches are available for the study of non-linear dynamic phenomena: when the response is relatively long and slow and the interest lies in the long-term global behaviour of a complex structure (such as oscillations and permanent deformation), rather than in capturing highly dynamic effects at a local scale, finite element tools relying on implicit time-integration schemes are preferred. On the one hand, these algorithms have the advantage of being unconditionally stable, allowing large time steps; on the other hand, they are computationally expensive (a stiffness matrix inversion is required at each step) and affected by numerical damping. All standard finite element codes, such as Ansys, Abaqus, and Nastran, belong to this category.

When large physical variations, such as large changes of density, phase transitions (melting, vaporization, plasma formation), fragmentation, or explosions, occur in a very short time, one must resort to an advanced class of numerical tool called *wave propagation codes* or *hydrocodes*. These are strongly non-linear finite element tools, using explicit time-integration schemes, which are conditionally stable, so that a short time step must be chosen according to the element dimension to ensure scheme stability (Courant–Friedrichs–Lewy condition). However, they are computationally efficient, since no stiffness matrix inversion is required. They are typically employed to study very fast and intense loading on materials and structures (high velocity impacts, explosions, crashes, etc.).

*2.4.1.2  Mesh schemes*

Several types of mesh scheme are available to describe the governing equations and their discretization in highly non-linear structural analyses.

The Lagrangian description is the most widely adopted scheme for structural analysis, both in standard finite element methods and in wave propagation codes: a Lagrangian mesh moves and distorts with the material it models as a result of forces from neighbouring elements; mesh nodes correspond to and move with 'physical' material points. This algorithm is usually very efficient; however, convergence problems can be met when material deformations are very severe, since elements, following the material, become highly distorted. When this occurs, mesh re-zoning is possible, but this is burdensome and introduces errors.

In such cases, alternatives must be found. In an Eulerian description, space is divided into fixed cells through which material flows: it is very well suited for problems involving extreme material movements (hypervelocity impacts, fluid mechanics, gas dynamics). It is computationally intensive, requiring higher element resolution and finer meshes than a Lagrangian scheme; moreover, the treatment of constitutive equations is complicated, owing to the convection of materials through the elements. This method is available in certain hydrocodes and is very extensively adopted for computational fluid dynamics calculations.

A compromise between Euler and Lagrange description is represented by the *arbitrary Lagrange Euler* (ALE) formulation: this hybrid technique tries to capture the advantages of both Lagrangian and Eulerian formulations. Typically, nodes on mesh boundaries and material interfaces move with the material (Lagrangian description), while all other interior nodes may either move with the material (Lagrange) or remain fixed in space (Euler). Most modern hydrocodes allow selection of this formulation, which is typically adopted to treat problems involving fluid/structure interaction.

An additional, relatively new technique for solving computational continuum dynamics problems is so-called *smoothed-particle hydrodynamics* (SPH). This is a mesh-free method ideally suited for certain types of problem with extensive material damage and separation. In this computational method, the material is modelled by a lattice of discrete elements (particles) with a spatial distance of interaction (smoothing length) over which their properties are weighted by a kernel function. Particles are interpolation points from which values of functions and their derivatives can be estimated at discrete points in the continuum. SPH particles can interact with Lagrange, Shell, and ALE meshes.

This method offers the possibility of studying crack propagation inside a body or the motion of expelled material fragments or liquid droplets. It is, therefore, well suited to the study of extreme beam impacts, in which explosion and *mechanical spalling* (ejection of material fragments from a surface of the impacted body) are involved (Fig. 20).

The SPH interaction points (particles) must, generally, be very small and packed to model the material accurately: a compromise must be found between accuracy and computation time.

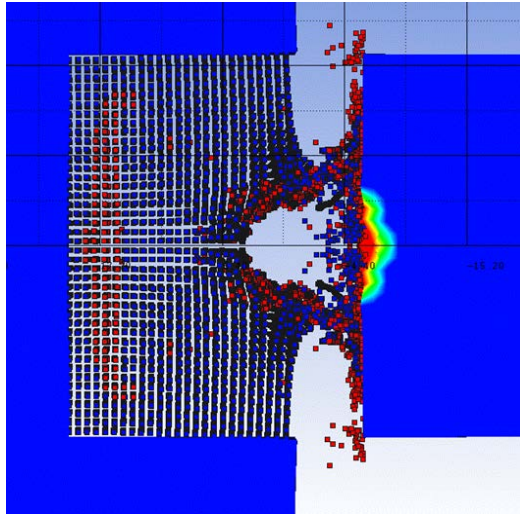

**Fig. 20:** Simulation of the impact of a 7 TeV bunch of $1.3 \times 10^{11}$ protons on a LHC tertiary collimator jaw. The left jaw is partly modelled by a SPH and Lagrangian mesh, while the opposite jaw is purely Lagrangian. Note the interaction between the ejected SPH particles and the Lagrangian mesh.

*2.4.2    The plastic dynamic regime*

When stresses exceed the elastic limit, materials typically undergo irreversible, non-linear plastic deformations, so that, on the removal of loads, the affected component will experience a permanent change of shape. This may occur under both quasi-static and dynamic conditions.

In dynamic conditions, the inelastic behaviour not only depends on the intensity of the applied load, as in quasi-static conditions, but also on the rate at which these loads are applied: this behaviour is called *rate-dependent* plasticity or *viscoplasticity*.

Plastic deformation, particularly for metals and alloys, basically occurs through a slip mechanism linked to the motion of dislocations, and the flow stress is essentially defined by the resistance to dislocation motion. The motion of dislocations inside the lattice is typically prevented by two types of obstacle: long- and short-range barriers.

Short-range barriers are strictly correlated to the material lattice and may include the lattice resistance itself, the resistance due to point defects, such as vacancies and self-interstitials, other dislocations intersecting the slip plane, alloying elements, and solute atoms. These dislocations can surmount short-range barriers partly by the action of shear stress due to externally applied loads and partly by increasing their thermal energy: higher temperatures tend to decrease the force required to move the dislocations. The strain rate has an opposite effect with respect to temperature: the dislocation is given less time to overcome the obstacle, attenuating the effect of the thermal energy and, consequently, increasing the force required to move dislocations. This contribution to the flow stress is called the *thermal* component and is a decreasing function of temperature and an increasing function of strain rate.

The long-range barriers may include grain boundaries, far-field forests of dislocations, and other microstructural defects with far-field influence. The resistance due to long-range barriers is often referred to as the *athermal* component of the flow stress. This type of obstacle cannot be overcome by additional thermal energy. The athermal part increases with increasing accumulated dislocations whose elastic field hinders the motion of mobile dislocations. While this elastic field does not explicitly depend on temperature, it is affected by temperature through the elastic moduli and their dependence on the temperature, and through the effect of the temperature history on the density of far-field dislocation forests. At suitably high temperatures, materials anneal, leading to a reduction in the dislocation density and hence in the corresponding elastic field stress; this is reflected in a reduction of the athermal component.

In its simplest setting the flow stress is expressed as

$$\sigma_y = \sigma_{th}(\varepsilon, \dot{\varepsilon}, T) + \sigma_{ath}(\varepsilon) \ ,  \qquad (17)$$

where $\sigma_{th}$ and $\sigma_{ath}$ are the thermal and athermal components of the resistance to the dislocation motion, respectively.

At moderate strains, the stress–strain curve exhibits non-linear trends like those depicted in Fig. 21; in some cases, the behaviour can be simplified, approximating the material response with a bilinear hardening law:

$$\sigma = E\varepsilon_{el} + E'\varepsilon_{pl} \ . \qquad (18)$$

In Eq. (18), $E'$ is the slope of the plastic linear function, sometimes called the *tangent modulus*, $E$ is the Young (elastic) modulus, and $\varepsilon_{el}$ and $\varepsilon_{pl}$ are the elastic and plastic components of the strain, respectively. If $E' = 0$, the material is said to be *elastic–perfectly plastic*. If more accuracy is sought, the stress–strain curve can be approximated by a multilinear elastic–plastic function.

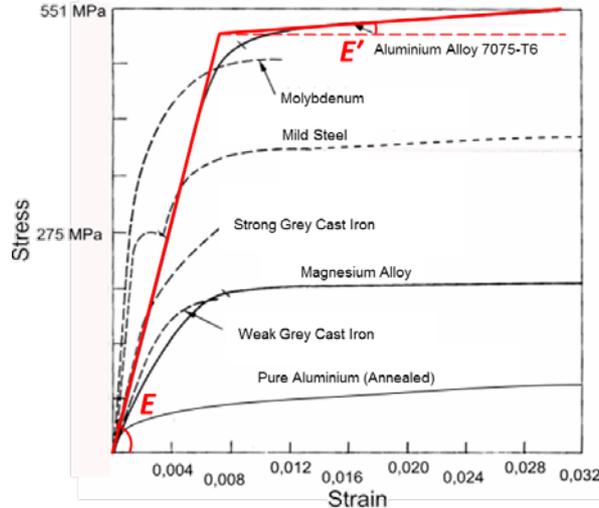

**Fig. 21:** Stress–strain curves beyond elastic region for several relevant materials. A bilinear approximation for an aluminium alloy is shown.

Along with permanent deformations, when large plastic strains occur, material density is also slightly affected: however, changes of density within plastic regime are in general small and can still be considered negligible.

Taking these assumptions into consideration, dynamic responses can be treated at an acceptable degree of approximation with non-linear, implicit *finite element method* codes, such as Ansys.

An example of this type of analysis is provided by a simulation of the effects of a LHC injection error on LHC secondary collimators: this scenario may lead to the impact on the collimator carbon/carbon (C/C) jaw of a full SPS batch of 288 bunches at 450 GeV ($3.2 \times 10^{13}$ protons over 7.2 μs), with transverse impact amplitudes up to 5–6 $\sigma_b$ [22]. In 2004, full-scale robustness tests were performed on collimator prototypes in the SPS extraction line to study such accidental cases. The two jaws of the prototype were submitted to a series of impacts at 450 GeV in two different conditions: (1) with increasing beam intensities at a fixed beam impact depth of 5 mm from the jaw surface and (2) with beam impact depths from 1 mm to 6 mm at a beam intensity of $3.2 \times 10^{13}$ protons. For material comparison, one of the jaw blocks was made of C/C (as for the series production), while the other was made of isotropic graphite (Fig. 22).

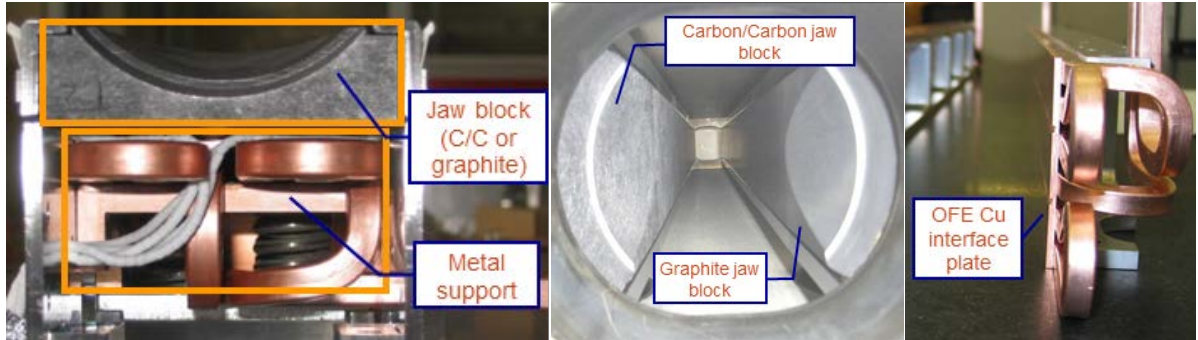

**Fig. 22:** Front view of the jaw assembly of a LHC secondary collimator (left); the two jaws in the collimator tank after completion of robustness tests in 2004 (centre); jaw metal support, showing the thin interface plate and the cooling pipes brazed on it (right).

Visual inspection of the jaw blocks after completion of the tests revealed no sign of mechanical damage; however, measurements performed on jaw assemblies revealed a permanent deformation of the metal support on both jaws of roughly 300 µm, with a well-repeated pattern (the maximum deflection was located towards the downstream end of the support, where the highest temperatures occurred).

An explanation for the residual deformation can be inferred on the basis of the dynamic stresses developing in solids in the case of very fast heating due to material inertia partially preventing free thermal expansion, as explained in previous sections.

The maximum temperature increase expected in the 3 mm thick, oxygen-free electronic copper (OFE Cu) interface plate of the jaw metal support is of the order of 70°C. A simple analytical assessment of the elastic stress can be made by assuming that no in-plane expansion is possible; in this hypothesis, using Eq. (7b) for thin plates and using the thermo-elastic properties of copper, we find, for the in-plane stresses,

$$\sigma_{x_{\max}}^{\lin} = -\frac{E\alpha\Delta T_{\max}}{1-\nu} \cong -210 \text{ MPa}.$$

This value largely exceeds the (compressive) yield strength of annealed OFE copper, which is limited to 50–70 MPa. Hence, residual strains will be present when the thermal load wanes. These compressive plastic strains are eccentric with respect to the neutral axis of the metal support and will lead to a permanent deflection of the metal support with a maximum sag towards the end experiencing higher temperatures.

This analytical assessment allows the permanent deflection and its shape to be justified qualitatively; however, it is practically impossible to estimate the magnitude of the effect quantitatively. To do so, it is necessary to resort to a non-linear, implicit, finite element analysis, including the effects of temperature, contacts, time, and plasticity (fast transient, coupled-field, elastic–plastic analysis).

Plasticity in metal components was modelled with both bilinear and multilinear kinematic hardening; results are shown in Fig. 23. As anticipated by the analytical estimate, the largest residual plastic strains are found on the thin copper plate, their magnitude (up to 0.12%) and extension being compatible with the simplified approach. The calculated permanent deflection (357 µm) of the metal support is close to the measured values and matches the actual deformed shape well.

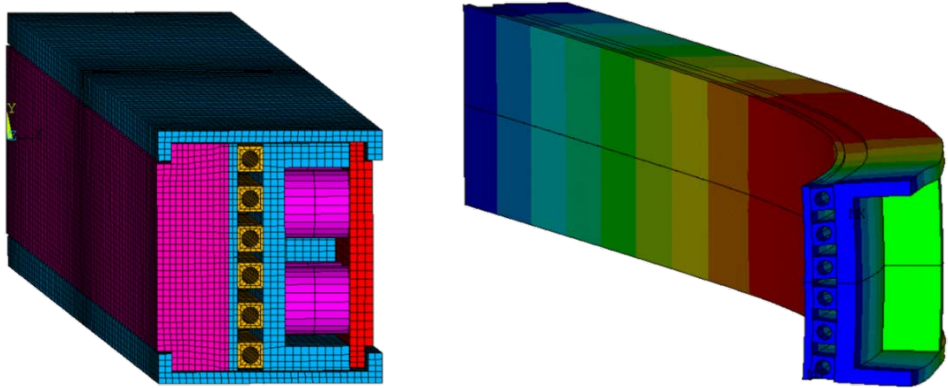

**Fig. 23:** Finite element model of the jaw assembly (left) and computed residual deflection of the jaw metal support (357 µm max) (right).

On the basis of these results, it was decided to modify the jaw assembly series design by changing the thin plate material from OFE copper to Glidcop®, a copper alloy reinforced with a fine dispersion of alumina, which has a much higher yield strength (>200 MPa): analysis of an updated model of the series jaw gave a permanent deflection of 16 µm. This improvement was achieved because plasticity is no longer attained on the thin plate and only occurs in a limited portion of the cooling pipes.

Another interesting outcome of the transient computations is the amplitude of the transverse oscillations occurring during the shock: as shown in Fig. 24, the maximum deflection at the centre of the C/C jaw reaches almost 1.5 mm after ≈12 ms; it is also worth noting that, during the transient part of the shock, the ends of the jaw may depart from the support by as much as 1.3 mm. The flexural frequency of oscillation is about 45 Hz.

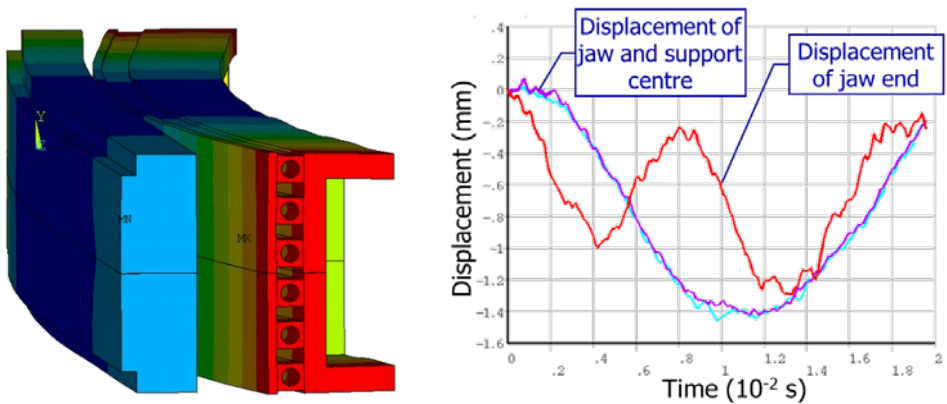

**Fig. 24:** Deflection of the jaw assembly after ≈12 ms (peak value) and time history of lateral displacement for jaw and support.

### 2.4.3 The shock wave regime

As mentioned previously, when the deposited energy is high enough to provoke stresses and strain rates exceeding a critical threshold, an energetic shock wave is formed, propagating at a velocity higher than $C_0$. A shock wave is characterized by a sharp discontinuity in pressure, density, and temperature across its front. This event may be associated with a number of severe effects on matter, including phase transitions, explosions, and spalling (Fig. 25).

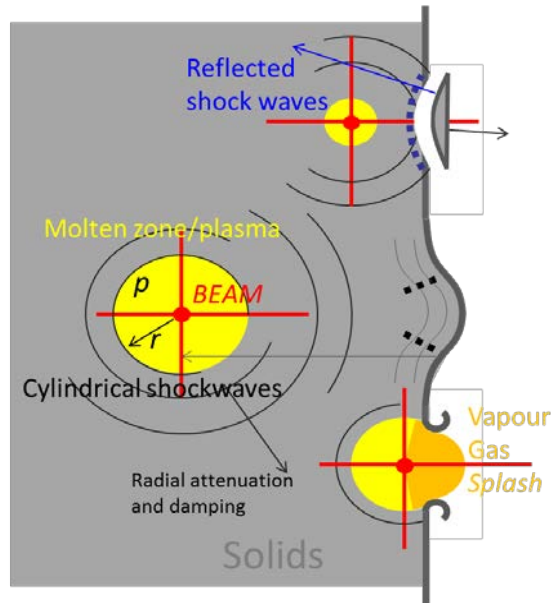

**Fig. 25:** Mechanism and effects that may be induced in solid by high-intensity particle beams at various regions of the impact [23].

In this respect, it is interesting to note that for metallic materials undergoing fast heat deposition, shock waves do not usually appear unless phase changes occur: if one assumes *uniaxial strains*, critical strains required to generate shock waves are in the range of 15% for tungsten and 7.5% for copper, whereas the total deformation at the melting point is in the range of 2% for both metals [19].

Conversely, for graphitic materials or other highly refractory materials, such as ceramics, the shock wave regime can be attained much sooner than the occurrence of extensive phase transitions.

As already mentioned, standard, implicit finite element techniques are not adapted to treat this class of problem and wave propagation codes or hydrocodes must be invoked [24].

In the usual continuum mechanics treatment, the complete stress tensor, which describes the material condition state, is divided into two components: deviatoric and hydrostatic tensors [20]. The name hydrocode stems from the original assumption of purely hydrostatic (fluid-like) behaviour of the impacted solids, which is typically acceptable when achieved stresses greatly exceed the flow strength of the material and the stress tensor can be approximately reduced to its hydrostatic component only; nowadays, the deviatoric component, responsible for material strength, is also taken into account but the original name is still widely used. Examples of codes used extensively to treat thermally induced fast dynamic phenomena are Autodyn, BIG2, and LS-Dyna. BIG2 is a two-dimensional code with a pure hydrodynamic solver which neglects the deviatoric component of the stress tensor. It was developed by Fortov *et al.* [25] for hypervelocity impacts and detonations, based on a Godunov-type numerical scheme. LS-Dyna [26] is a general-purpose transient dynamic finite element program including an implicit and explicit solver with non-linear thermomechanical capabilities. Finally, Autodyn [27] is a commercial explicit analysis tool particularly suitable for modelling the non-linear dynamics of solids, fluids, and gases, and their interactions.

*2.4.3.1  Equations of state*

The hydrostatic response in a hydrocode is governed by the *equation of state* (EOS), which expresses the relation between thermodynamic variables (such as pressure, $P$, internal energy, $E$, entropy, $S$, density, $\rho$, and temperature, $T$). All these variables define the thermodynamic state of the matter. For a thermodynamic system that is in equilibrium, the state of the system is completely defined if two independent and intensive variables are known. Usually, in hydrodynamics, the internal energy replaces temperature as independent variable. In this case, the EOS assumes the general form

$$P = P(\rho, E)$$ . (19)

An EOS represents a set of surfaces, on which it is possible to define one-dimensional paths, which identify isothermal, isobaric, isochoric, isentropic processes (Fig. 26).

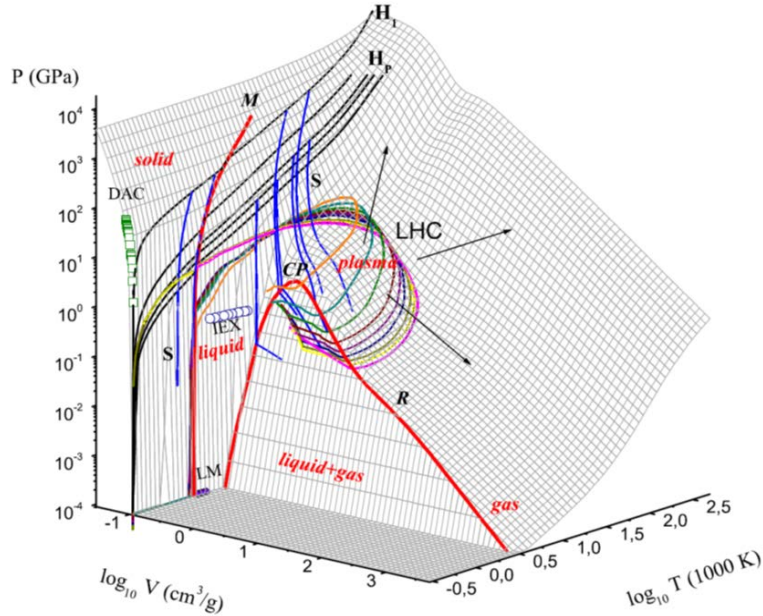

**Fig. 26:** Three-dimensional pressure–volume–temperature surface for copper [28]

The EOS implemented in commercial hydrocodes can be analytical or tabular. Analytical EOSs include, for example, the ideal gas law, linear EOSs (equivalent to linear thermo-elastic stress–strain relationships), polynomial EOSs, and the Mie–Grüneisen, GRAY, PUFF, and Tillotson EOSs: their use is limited, since they can usually describe only a single phase region. A tabular EOS, such as those provided by the SESAME database maintained by the Los Alamos National Laboratory, can be employed to evaluate material behaviour over different phases without loss in precision. Moreover, polynomial EOSs can be interpolated from tabular ones [20].

*2.4.3.2 Strength models*

The deviatoric behaviour of a material is usually expressed by the *strength model*. In very fast and intense phenomena, the mechanical strength of the material is largely affected by the mechanical and thermodynamic variables that contribute to the material deformation process: the material may experience sharp discontinuities in pressure and temperature; the inner volume can melt, losing its yield strength, while the surrounding zone is still solid and subjected to heavy plasticity, generated by the shock wave propagation. The variables governing the material behaviour in the plastic regime are typically deformation (both plastic and elastic), strain rate, temperature, and pressure.

In previous decades, numerous material models in computational plasticity were proposed, for the description of deviatoric behaviour. Models are classified according to their nature in empirical, semi-empirical, and physically based models. Empirical models, such as the one proposed by Johnson and Cook [29], do not possess any physical basis and are phenomenological, obtained by interpolation of experimental data. The physically based models are obtained by starting from the transformation in the material occurring during a deformation process. An example of a semi-empirical model is the Zerilli–Armstrong model [30], which is based on the dislocation mechanics theory and presents a different formulation for body-centred cubic and face-centred cubic materials. An alternative semi-empirical model is the Steinberg–Cochran–Guinan–Lund model [31], which was first developed for the description of high strain rate behaviour and subsequently extended to low strain rates. A completely physical-based and more complex model is the mechanical threshold stress [32]. Most of these material models are usually implemented in commercial FE codes, such as LS-Dyna and Autodyn.

When strain rate starts to play an important role, the Johnson–Cook model is one of the most popularly adopted strength models. This is an empirical multiplicative model, in which the effects of plastic strain, strain rate, and temperature are uncoupled; it is particularly suitable for metals and ductile materials. According to the Johnson–Cook model, the flow stress is defined as

$$\sigma_y = \left(A + B\varepsilon_{pl}^n\right)\left(1 + C \ln \frac{\dot{\varepsilon}_{pl}}{\dot{\varepsilon}_0}\right)\left[1 - \left(\frac{T - T_r}{T_m - T_r}\right)^m\right], \quad (20)$$

where $\varepsilon_{pl}$ is the equivalent plastic strain, $\dot{\varepsilon}_{pl}$ is the plastic strain rate, $A$ is the quasi-static elastic limit, and $B$ and $n$ are the work hardening parameters, which influence the slope and shape of the flow stress in the plastic domain. The parameter $n$ usually assumes values between 0 (for a perfectly plastic model) and 1 (for a piecewise linear model). $C$ expresses the sensitivity to the strain rate, while $\dot{\varepsilon}_0$ is the effective plastic strain rate of the quasi-static test used to determine the yield and hardening parameters $A$, $B$ and $n$ (in the original formulation, it was set equal to 1). The thermal effects are described by the thermal softening coefficient $m$, the actual temperature $T$, the reference temperature $T_r$ used when determining $A$, $B$, and $n$ and the melting temperature $T_m$ at which the material loses its shear strength and starts to behave like a fluid. The thermal parameter $m$ determines the concavity of the temperature function: if $m < 1$, the function is convex, if $m > 1$, it is concave and if $m = 1$, the temperature influence is linear.

These parameters can be obtained through a set of experimental tests, which include Hopkinson bars, Taylor cylinders, and tensile and compression quasi-static tests at different temperatures [20]. Values of JC parameters for various materials are provided in Table 2.

Table 2: Parameters of Johnson–Cook strength model for selected materials

| Material | Melting point [K] | A [MPa] | B [MPa] | n | C | m |
|---|---|---|---|---|---|---|
| OFHC copper | 1356 | 90 | 292 | 0.31 | 0.025 | 1.09 |
| Cartridge brass | 1189 | 112 | 505 | 0.42 | 0.009 | 1.68 |
| Nickel 200 | 1726 | 163 | 648 | 0.33 | 0.006 | 1.44 |
| Armco iron | 1811 | 175 | 380 | 0.32 | 0.060 | 0.55 |
| Electrical iron | 1811 | 290 | 339 | 0.40 | 0.055 | 0.55 |
| 1006 steel | 1811 | 350 | 275 | 0.36 | 0.022 | 1.00 |
| 2024-T351 aluminium | 775 | 265 | 426 | 0.34 | 0.015 | 1.00 |
| 7039 aluminium | 877 | 337 | 343 | 0.41 | 0.010 | 1.00 |
| 4340 steel | 1793 | 792 | 510 | 0.26 | 0.014 | 1.03 |
| S-7 tool steel | 1763 | 1539 | 477 | 0.18 | 0.012 | 1.00 |
| Tungsten alloy | 1723 | 1506 | 177 | 0.12 | 0.016 | 1.00 |

2.4.3.3  *Failure models*

A *dynamic failure model* is typically used in association with a strength model to describe the failure mechanism of a material submitted to rapidly applied loads and determine its structural limits.

The factors that influence dynamic failure are typically the material properties and microstructure, the applied loads and the conditions they induce (stress, strain rate, and temperature), and the ambient environment.

As will be shown in more detail in Section 3.1, depending on the failure mode, the materials are classified as brittle (such as ceramics or glass) or ductile (such as metals or polymers). The former class is characterized by very limited plastic deformation before failure and nearly flat fracture surfaces originating from a single crack propagation. Ductile materials exhibit large plastic deformations, usually with necking phenomena and the typical cup-and-cone shaped failure surface, which is the result of nucleation, growth, and coalescence of voids in the material.

Dynamic failure models can be divided into two categories. In the first category, the material is supposed to fail when locally it overcomes a limit for one or more variables (such as strain to fracture, tensile hydrostatic stress, or maximum principal stress). This type of failure mechanism could be used to describe dynamic brittle failure or phenomena such as spalling. The second category includes failure models that are based on cumulative damage mechanisms: the material starts to be damaged if some property limits are exceeded; damage evolution is then controlled by a damage parameter that can increase until complete failure is achieved. This type of failure mechanism is used to describe dynamic ductile failure.

Examples of dynamic failure models are the maximum plastic strain failure criterion, the minimum hydrostatic pressure failure criterion ($P_{min}$), and the Grady spall model.

As exemplified in Fig. 25, if intense particle beams impact a solid close to a free surface, the compressive shock wave is immediately reflected and turned into a tensile wave, causing bulk failure and material ejection (spalling) if its amplitude is higher than the material hydrostatic strength.

This mechanism is usually reproduced in hydrocodes through the $P_{min}$ model, which broadly corresponds to the maximum normal stress criterion for slower load conditions; material models may also take into account the energy necessary for crack formation, calculated on the basis of the material fracture toughness.

One example of hydrocode computations is the simulation of accidental beam impacts of one or more full bunches on a tertiary collimator for the LHC [33]. The analysis was carried out by making use of Autodyn and simulating the whole collimator jaw assembly (Fig. 27). The jaw section directly interacting with the beam is composed of five Inermet 180 blocks, each 200 mm long, fixed with stainless steel screws to a housing made of OFE Cu. The copper housing is, in turn, brazed to cooling pipes made of copper–nickel alloy (90% Cu, 10% Ni), which are then brazed to a back stiffener made of Glidcop.

Two complementary three-dimensional models were implemented in Autodyn, based respectively on (a) a Lagrangian mesh of the full jaw assembly, to study the shock wave propagation and assess possible damage in each element of the jaw assembly, and (b) a SPH model of the most loaded Inermet block, to study the high-speed ejection of tungsten particles and their impact on the tank and on the opposite jaw.

Table 3 provides details of the constitutive models used for each of the relevant materials. It is worth noting that water in the cooling pipes was also included in the analysis.

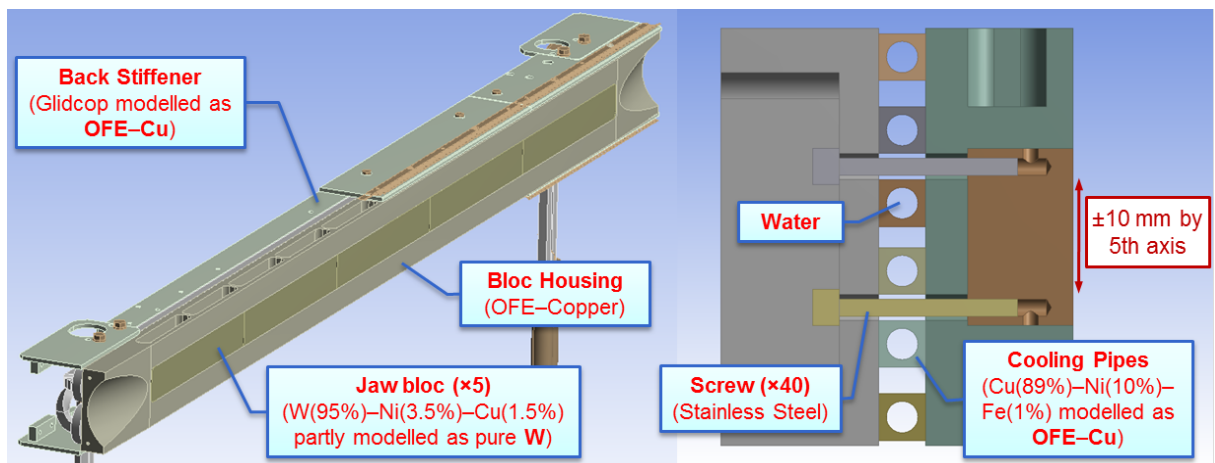

**Fig. 27:** Three-dimensional view and cross-section of the jaw assembly of a LHC tertiary collimator

**Table 3:** Material constitutive models for LHC tertiary collimator analysis

| Material | EOS | Strength model | Failure model |
| --- | --- | --- | --- |
| Inermet/tungsten | Tabular(SESAME) | Johnson–Cook | Max plastic strain and $P_{min}$ |
| OFE copper | Polynomial | Johnson–Cook | Johnson–Cook |
| Stainless steel AISI 316 | Shock | Johnson–Cook | Max plastic strain |
| Water | Shock | – | $P_{min}$ |

Seven accident cases, with different degrees of severity and probability, were identified (Table 4). All of the cases are based on an asynchronous beam abort event, assuming that each bunch has the same impact parameter (2 mm). The beam energy corresponds to the values that were expected for run 1 of the LHC. The impinging proton pulses constitute trains of bunches of $1.3 \times 10^{11}$ particles with energy up to 5 TeV, spaced by 25 ns.

**Table 4:** List of accident cases

| Case | Beam energy [TeV] | Normal emittance [μm rad] | No of impacting bunches | Energy on jaw [kJ] |
| --- | --- | --- | --- | --- |
| 1 | 3.5 | 3.50 | 1 | 38.6 |
| 2 | 5 | 7 | 1 | 56.2 |
| 3 | 5 | 3.5 | 1 | 56.5 |
| 4 | 5 | 1.75 | 1 | 56.6 |
| 5 | 5 | 1.75 | 2 | 111.3 |
| 6 | 5 | 1.75 | 4 | 216.1 |
| 7 | 5 | 1.75 | 8 | 429.8 |

A complete FLUKA model of the collimator was set up and full shower simulations were carried out for each case to provide the deposited energy distribution.

The impact of every bunch with matter leads to a sudden increase in temperature and pressure, which in turn generates an outbound shock wave. The expansion wave that follows may in turn lead to a substantial reduction in density. Hence, one should, in principle, update the FLUKA model for each bunch and re-perform the simulations with the new density map, since changes in densities substantially affect the deposited energy distribution (FLUKA–hydrocode coupling). However, for the simulated accident cases, the change of density during the impact of the bunch train was found to be negligible: therefore, the energy distribution map calculated in the initial state could be used throughout the whole simulation.

To determine consequences on the collimator and LHC operation, three different damage levels were defined:

- Level 1 – Collimator not to be replaced. Limited jaw damage: an intact spare surface can be found relying on the 5th axis movement, which permits a maximum vertical shift of ± 10 mm; negligible permanent jaw deformation.
- Level 2 – Collimator to be replaced. Damage to the jaw incompatible with 5th axis travel; other components (e.g. screws) may also be damaged.
- Level 3 – Long downtime of the LHC. Very severe damage to the collimator, leading to water leakage into beam vacuum.

Results predict that all the single-bunch cases, both at 3.5 and 5 TeV, at all emittances, fall within damage level 1. The primary variable determining damage extent on the jaw is the total energy deposited: the size of the damaged region is already much larger than the beam size so that no sensible difference is found when varying the beam emittance. Even in the less destructive cases, a sizeable plastic deformation is found on the copper support and on the cooling circuit; a groove on the surface of impacted Inermet blocks, with an extension roughly proportional to the bunch energy, well reproducible with the SPH method, is also generated, while Inermet fragments are projected towards the opposite jaw.

It was also found that a key role in determining the damage extension induced by beam impacts on a composite structure is played by the shock impedance matching between adjoining components. Shock impedance in a given material is defined as

$$Z = A\rho_0 U_s \, , \tag{21}$$

where $A$ is the interface surface, $\rho_0$ is the initial density and $U_s$ is the shock velocity.

Owing to the large shock impedance mismatch between tungsten and copper (high $Z_W$ to $Z_{Cu}$ ratio), most of the wave energy is confined inside the Inermet®180 blocks: this limits the damage produced in other critical components, such as the cooling pipes (Fig. 28).

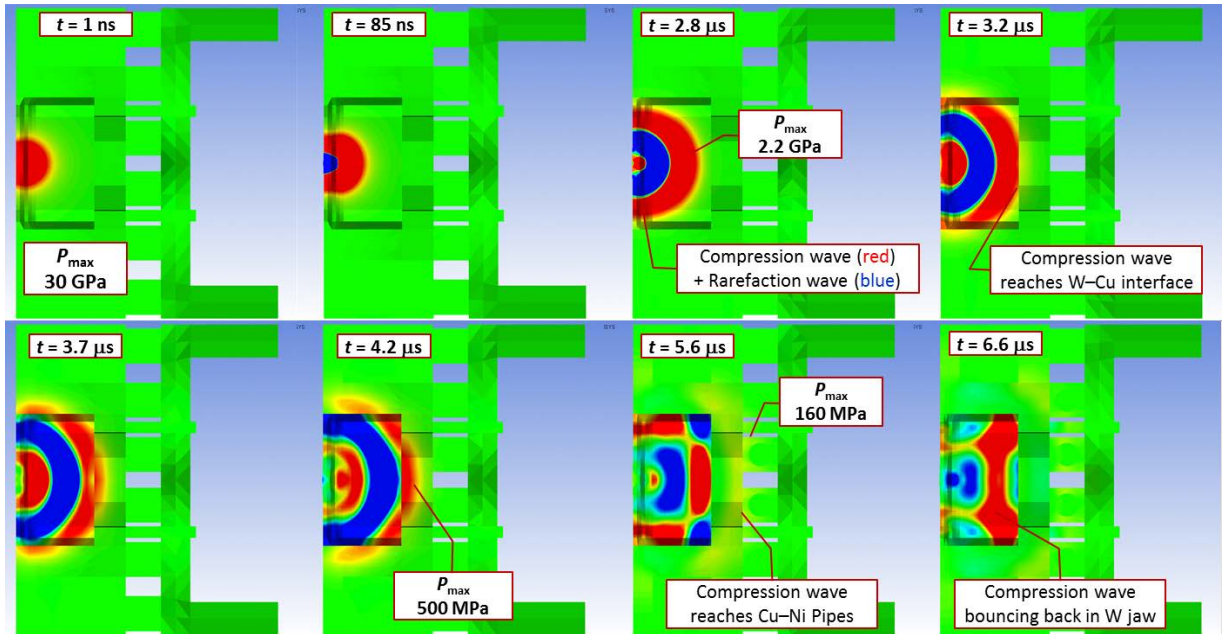

**Fig. 28:** Case 4: propagation of the shock wave in the jaw assembly shown at various instants. Note that the wave is mostly reflected at the W–Cu interface and only partially transmitted to the copper housing.

It can also be observed that the jaw damage extension at 5 TeV (case 4) is at the limit of damage level 2; plastic deformations on cooling pipes and screws remain limited, and tungsten particles are sprayed on a larger area of the opposite jaw (Fig. 29). This jaw is not directly damaged; however, its final flatness may be affected by possible re-solidified droplets stuck on its surface.

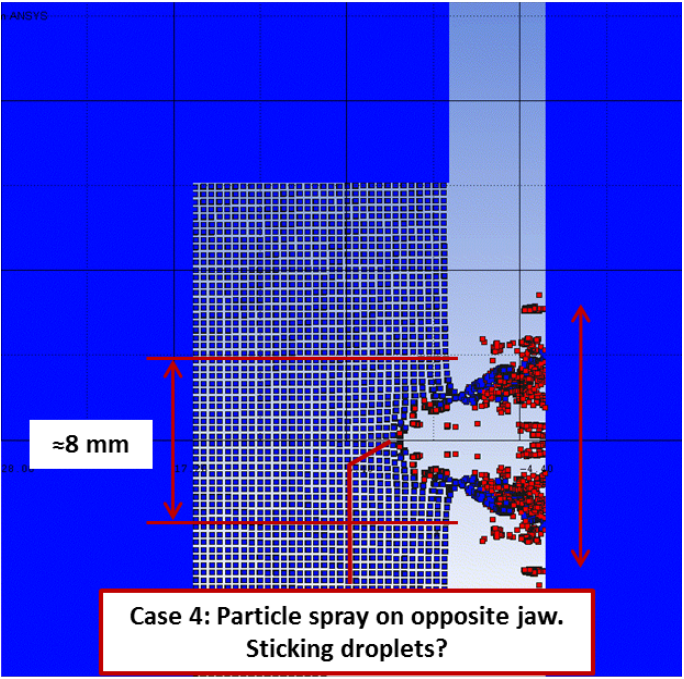

**Fig. 29:** Case 4: damage extension on Inermet block – note particles sprayed on opposite jaw

For cases 5 and 6, the jaw damage cannot be compensated by $5^{th}$-axis travel (damage level 2). Severe plastic deformations can be observed on cooling pipes and screws, although visible failures are not detected. The SPH simulations anticipate permanent damage on the opposite jaw, provoked by tungsten particles impacting at elevated velocity (Fig. 30).

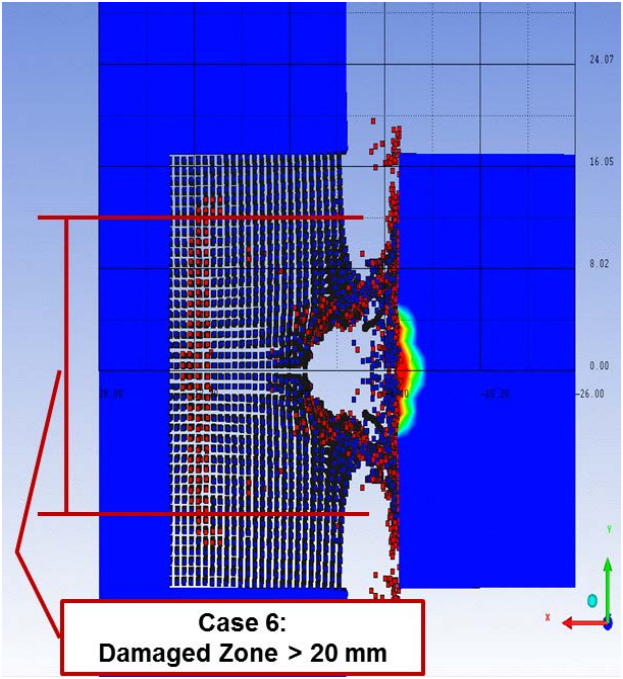

**Fig. 30:** Case 6: high-speed particle spray provoking an extended damage on the opposite jaw

The only case studied leading to damage level 3 is case 7. In this scenario, one may expect: a) high risk of water leakage due to very severe plastic deformation on the pipes (plastic strain up to ≈21%); b) extended eroded and deformed zone on the tungsten jaw; c) projections of hot and fast, solid fragments ($T \approx 2000$ K, $V_{max} \approx 1$ km/s) onto the opposite jaw with slower particles hitting tank covers at velocities just below the ballistic limit; d) risk of permanent bonding between the two jaws due to the projected re-solidified material (Fig. 31).

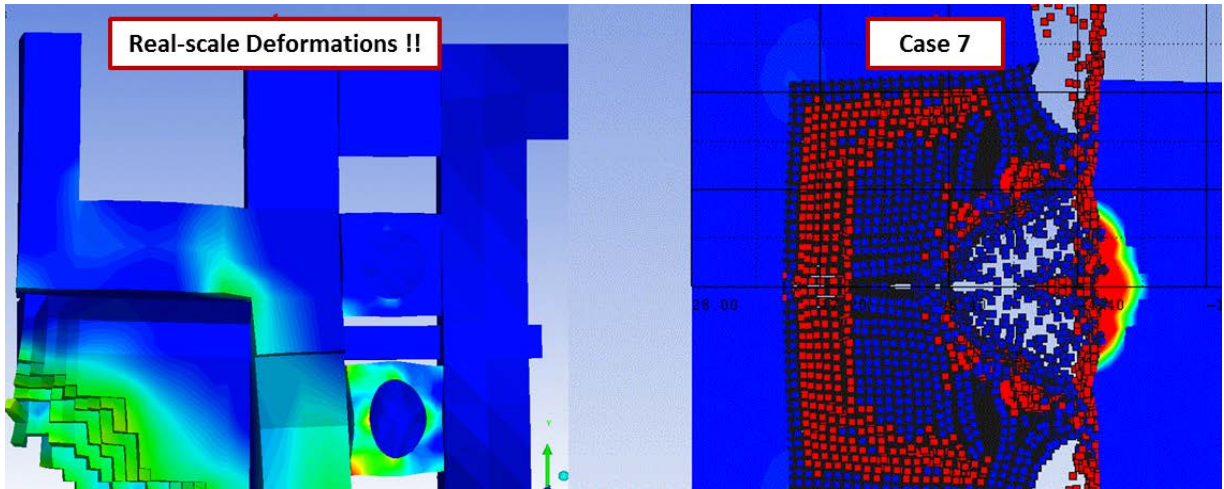

**Fig. 31:** Case 7: plastic strain in Cu and Cu–Ni (left) and damage extension on the two jaws (right)

### 2.4.4 Hydrodynamic tunnelling

In the previous example, the energy deposition calculated for the first bunch on the pristine material was maintained also for subsequent bunches since the change of density induced by the impinging particles was found to be negligible for the duration of the impact. The same approach was followed for similar calculations on other structures [34].

As already discussed, however, the expansion wave, which follows the compression shock wave generated by prolonged intense impacts, when propagating radially away from the impacted region, may displace material outwards, reducing material density. Additionally, material density at the target core abruptly drops because of induced phase transitions (to liquid, gas, and plasma), along with pressure release. If subsequent bunches arrive after a lapse of time sufficiently long to allow the rarefaction wave to develop and pressure to decrease drastically, particles will experience a considerable increase of interaction length (which is density-dependent) and penetrate the matter more deeply, owing to density reduction: this extreme phenomenon is sometimes called the *hydrodynamic tunnelling* effect [28].

In such cases, a coupling between the interaction–transport code and the wave propagation code is necessary, to take into account the change of density during the interaction of the beam with matter.

The following example is that of a tungsten cylindrical target impacted at its centre by 30 LHC full bunches at the energy of 7 TeV [35].

Results of the first FLUKA simulation, performed on pristine material, are uploaded in the LS-DYNA mechanical model. Then, for each bunch, the coupling algorithm performs the following operations:

- Immediately before the impact of bunch *n*, it obtains, from LS-DYNA, the density map induced by the impact of all previous bunches.
- It updates the regions of the FLUKA model that underwent significant density changes (in excess of a few percent).

- It runs a new FLUKA calculation, to be imported in LS-DYNA, simulating the impact of bunch *n*.

Results show that the density variation leads to some reduction of deposited energy at each bunch: as shown in Fig. 32, the energy deposition peak penetrates more deeply at each successive bunch.

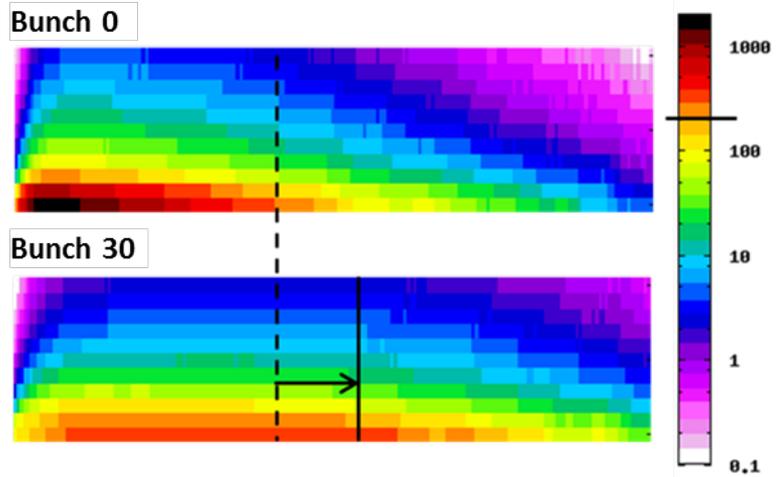

**Fig. 32:** Energy deposition (GeV cm$^{-3}$) in longitudinal section for 1st and 30th bunches

Comparison with the uncoupled solution shows that pressure is also affected: its maximum value, in the beam axis direction, decreases as the shock wave penetrates the material (Fig. 33). Results also confirm that the differences between coupled and uncoupled analyses are significant only when a substantial density reduction occurs: for the studied cases more than 10 bunches are necessary (Fig. 34).

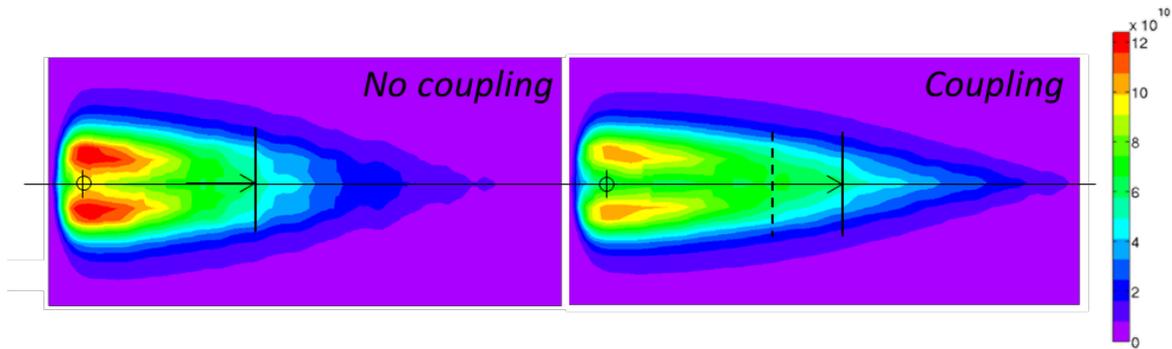

**Fig. 33:** Differences in pressure between FLUKA–hydrocode coupled and uncoupled simulations for a tungsten target impacted by 30 LHC bunches. Pressures are in Pa.

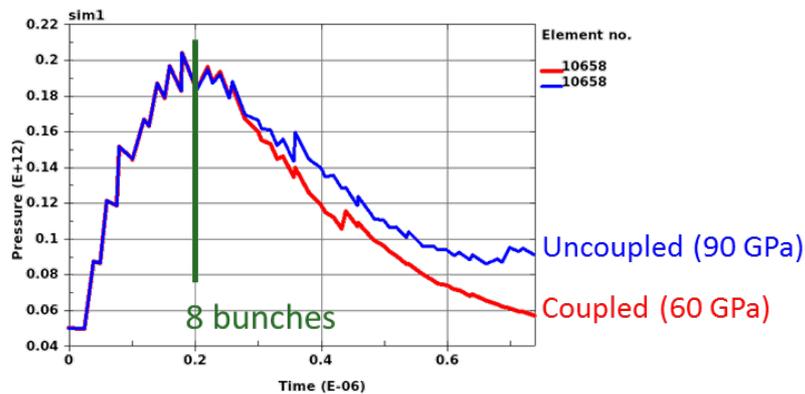

**Fig. 34:** Comparison between coupled and uncoupled solution of maximum pressure in a target element as a function of time. Differences become appreciable after roughly 10 LHC bunches.

# 3 Design principles of beam intercepting devices

## 3.1 Introduction to failure criteria

As we have seen, components directly exposed to interactions with particle beams, such as collimators, absorbers, targets, dumps, or windows, which, in short, we shall call beam intercepting devices (BIDs) are subjected, like any engineering component, to complex loadings in tension, compression, bending, torsion, or pressure, or combinations of these, so that, at a given point in the material, stresses often occur in more than one direction. If sufficiently severe, such combined stresses can act together to cause the material to yield (i.e. to exceed its elastic limit and undergo plastic, irreversible deformation) or fracture (as is more often the case for brittle materials). Predicting the safe limits for use of a component made of a given material under combined stresses in the elastic or elastic–plastic regimes requires the application of a *failure criterion*. Failure criteria in the more extreme shock wave regimes are usually replaced by the dynamic failure models introduced in Section 2.4.3, although the use of 'standard' failure criteria is sometimes also possible in the shock wave regime.

A number of different failure criteria are available, some of which predict the failure by yielding and others by fracture. The former are called *yield criteria* and the latter *fracture criteria*. In general, yielding is considered a form of failure in that the permanently deformed component is expected to no longer meet its design requirements, e.g. because of loss in precision, alignment, or load-carrying ability; in some cases, however, a limited plastic deformation may be tolerated, provided it does not impair component functionality.

Failure criteria are usually based on values of stress, so that their application involves in general calculation of an *equivalent stress* that condenses the complex state of stress into a single value, which is then compared with the yield or fracture strength of the material. To do so, it is always possible to identify a coordinate system in which the complete three-dimensional state of stress can be reduced to one in which only normal stresses are acting; the axes of this particular coordinate system are called *principal directions* and the corresponding normal stresses are called *principal stresses*, conventionally indicated $\sigma_1$, $\sigma_2$, and $\sigma_3$ (Fig. 35). Therefore, the failure criterion usually simplifies to $f(\sigma_1, \sigma_2, \sigma_3) = 0$.

For isotropic materials, it is often useful to rewrite the three principal stresses in terms of the so-called *invariants* of the stress tensor, because they are independent of the orientation of the coordinate system:

$$
\begin{aligned}
I_1 &= \sigma_1 + \sigma_2 + \sigma_3, \\
I_2 &= \sigma_1\sigma_2 + \sigma_2\sigma_3 + \sigma_3\sigma_1, \\
I_3 &= \sigma_1\sigma_2\sigma_3.
\end{aligned}
\tag{22}
$$

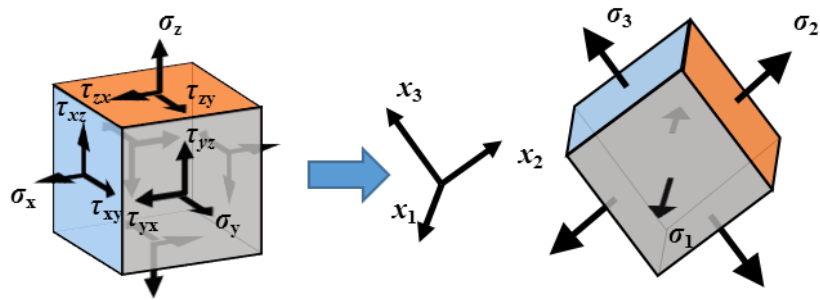

**Fig. 35:** Transformation of a general three-dimensional state of stress into a state of stress with normal stresses only (principal stresses).

In some cases, however, particularly when dealing with materials for which a linear elastic behaviour cannot easily be found, it might be appropriate to base the failure criterion on strains rather than stresses.

Since a given material may fail by either yielding or fracturing, depending on its properties, the state of stress, and the loading conditions (quasi-static or dynamic), no single failure criterion is suitable for every material under any state of stress and for all conditions: the choice of an appropriate failure criterion is therefore a critical step in the design of a structural component and must be carefully considered. An overview of the most adapted criteria for BIDs is given in the next sections.

Safety coefficients are adopted to protect against approximation of failure criteria and uncertainties in the knowledge of the state of stress.

### 3.1.1    Maximum distortion energy theory

The *maximum distortion energy theory*, also known as the *von Mises yield criterion*, *Huber–Hencky–von Mises yield criterion*, or *octahedral shear stress yield criterion* is a failure theory extensively used for ductile materials.

In applying stresses to a structural element, mechanical work is done; for a material within the elastic regime, all of this work is stored as potential energy. This internal *strain energy* can be partitioned into one part associated with *volume change* (caused by hydrostatic stress, $\sigma_{avg} = (\sigma_1 + \sigma_2 + \sigma_3)/3$ and another part associated with *distortion* of the shape of the material element by the remaining portion of the principal stresses, corresponding to the deviatoric components of the stress tensor (Fig. 36).

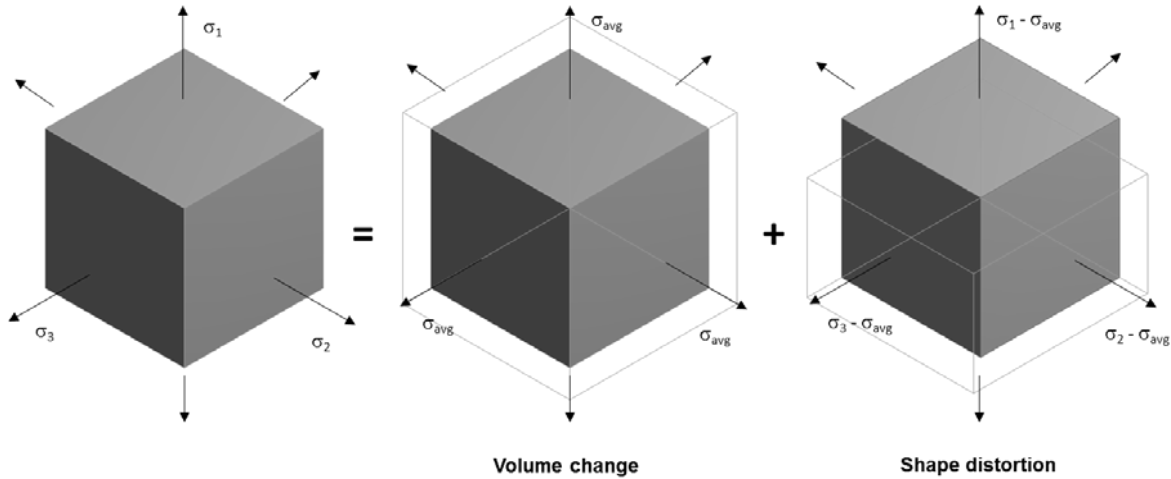

**Fig. 36:** Contributions to the deformation of a material element: volume change is due to hydrostatic stress, shape distortion to the deviatoric portion of principal stresses.

Experimental observations have shown that ductile materials do not yield when subjected to uniform hydrostatic stresses: based on this empirical evidence, Huber proposed that material yielding occurs when the *distortion energy* per unit volume $u_d$ reaches the distortion energy per unit volume of the same material when subjected to yielding in a tension test $(u_d)_Y$.

Mathematically, this reduces to the following relationship between an equivalent stress $\sigma_{eq}$ and the yield strength $\sigma_Y$:

$$\sigma_{eqVM} = \frac{1}{\sqrt{2}}\sqrt{\left(\sigma_1 - \sigma_2\right)^2 + \left(\sigma_2 - \sigma_3\right)^2 + \left(\sigma_3 - \sigma_1\right)^2} = \sigma_Y \ . \tag{23}$$

According to this criterion, yielding is supposed to occur in the material, when locally the equivalent stress (defined previously) reaches or exceeds the yield strength of that material.

It can also be shown that in Eq. (23) failure by yielding is assumed to occur when the octahedral shearing stress in the material reaches a value equal to the maximum octahedral shearing stress in a tension test at yield.

Equation (23) defines a three-dimensional surface in the principal stress space representing a circular cylinder having its axis on the line $\sigma_1 = \sigma_2 = \sigma_3$ (Fig. 37). Any combination of principal stresses falling inside this cylindrical boundary is below the yield stress and hence safe according to the von Mises criterion, while the surface itself represent the geometrical locus of yielding.

A safety factor can be defined as the ratio between the equivalent stress and the yield strength: one can observe that in the case of purely hydrostatic stresses, the principal normal stresses are all equal and the equivalent stress is zero, so the safety factor against yielding is infinite. This state of pure hydrostatic stress is represented graphically by the axis of the circular cylinder.

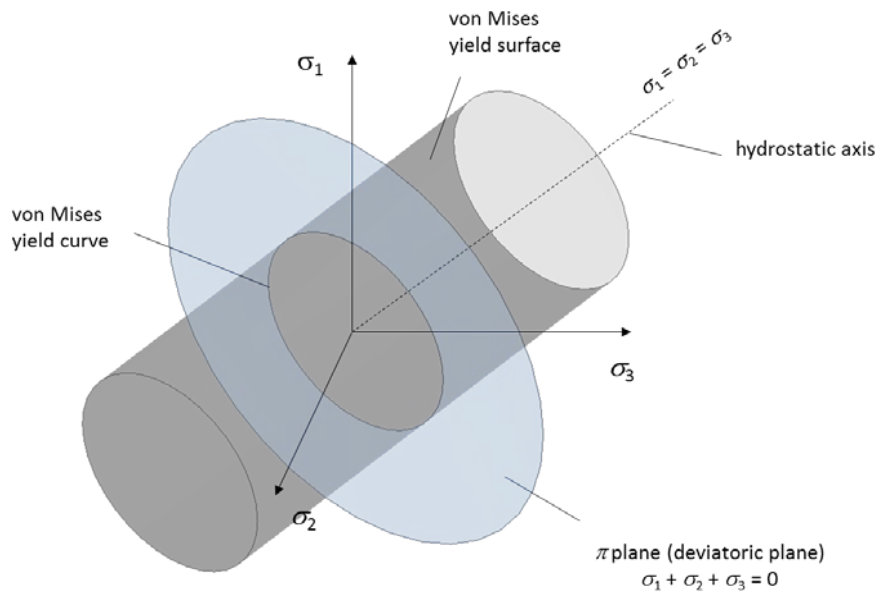

**Fig. 37:** Three-dimensional yield surface of the von Mises yield criterion

If any one of the three principal stresses is zero, the intersection of the yield surface with the plane of the remaining two principal stresses gives an ellipse, as shown in Fig. 38.

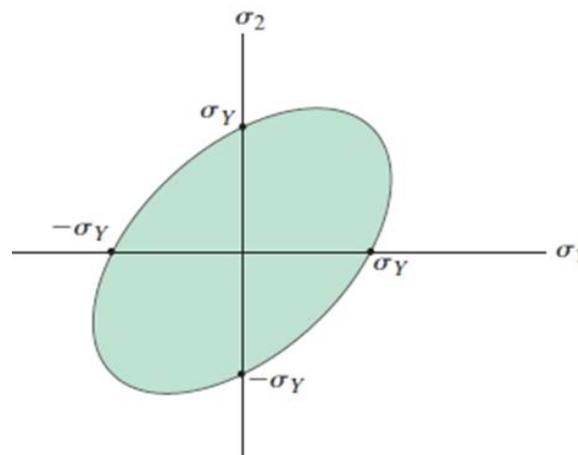

**Fig. 38:** Yielding locus for the von Mises criterion for plane stress ($\sigma_3 = 0$)

### 3.1.2 Maximum shear stress theory

As discussed previously, yielding in ductile materials is associated with the effect of the deviatoric part of the stress tensor: on this basis, Tresca and Guest suggested that failure by yielding occurs when the maximum shear stress in any plane reaches a critical value corresponding to the maximum shear stress that causes the same material to yield when it is subjected only to axial tension.

An equivalent stress can again be identified: failure occurs when this equivalent stress equals the yield strength of the material. Mathematically, the *maximum shear stress theory* or *Tresca–Guest yield criterion* is expressed by

$$\sigma_{eqTG} = \max\left(|\sigma_1 - \sigma_2|, |\sigma_2 - \sigma_3|, |\sigma_3 - \sigma_1|\right) = \sigma_Y . \tag{24}$$

In the principal stress space, the locus of yielding corresponds to the surface of a hexagonal prism with its axis given by the line $\sigma_1 = \sigma_2 = \sigma_3$ (which implies, like the von Mises criterion, that hydrostatic stress does not affect yielding).

For plane stress (one of the principal stresses being zero), the maximum shear stress theory is represented by a distorted hexagon, obtained by the interception of the hexagonal prism with a plane $\sigma_I = 0$.

A comparison between the failure loci provided by the von Mises and Tresca–Guest criteria for plane stress reveals that the latter is slightly more conservative, particularly for pure shear when $\sigma_1 = -\sigma_2$ (Fig. 39).

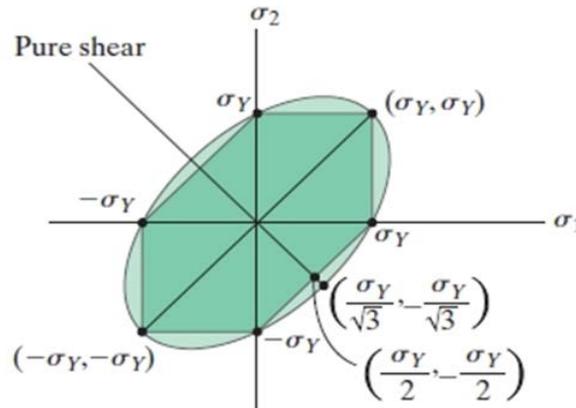

**Fig. 39:** Comparison between von Mises and Tresca–Guest failure loci in case of plane stress

### 3.1.3 Drucker–Prager yield criterion

In most cases, materials possess the same yield strength in tension and compression; however, in certain cases, compressive and tensile yield strength might be significantly different, the former being much larger than the second; these are said to be *uneven materials*. In such cases, an extension of the von Mises criterion, taking into account non-symmetry in stress–strain curves, can be invoked.

The so-called *Drucker–Prager yield criterion* is commonly used to model various pressure-dependent materials. It can be expressed using the principal stresses as

$$\sqrt{\frac{1}{6}\left[(\sigma_1 - \sigma_2)^2 + (\sigma_2 - \sigma_3)^2 + (\sigma_3 - \sigma_1)^2\right]} = A + B(\sigma_1 + \sigma_2 + \sigma_3) . \tag{25}$$

Note that the left-hand side of the equation corresponds (up to a constant factor) to the von Mises equivalent stress, cf. Eq. (23)). $A$ and $B$ are two independent material parameters that can be derived from the tensile and compressive strengths, $\sigma_t$ and $\sigma_c$. The value of $A$ defines the size of the yield locus in principal stress space and thus relates to the overall strength of the material. The parameter *B* describes the dependence on hydrostatic pressure, i.e. on the first invariant of the stress tensor.

As opposed to the infinite cylinder representing the von Mises model, the Drucker–Prager yield locus forms a cone along the hydrostatic axis, which widens in the direction of compressive stress states (Fig. 40).

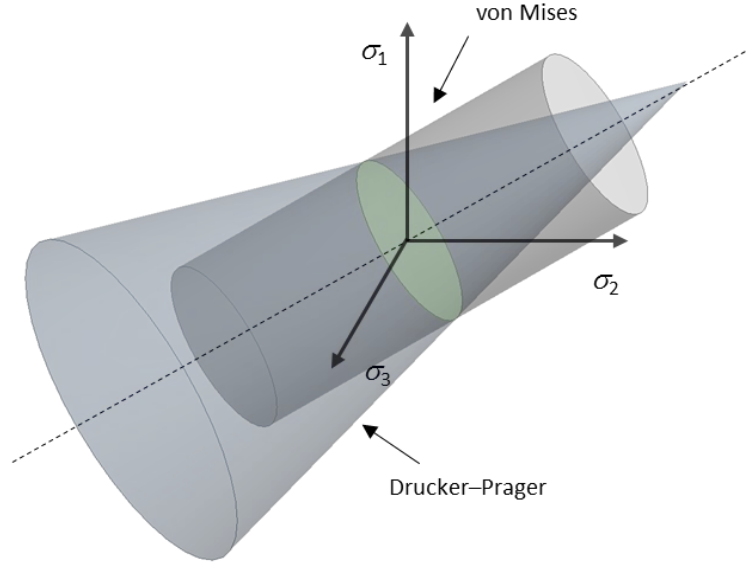

**Fig. 40:** Comparison between Drucker–Prager and von Mises yield surfaces

### 3.1.4  *Mohr–Coulomb failure criterion*

The Drucker–Prager yield criterion is closely related to the *Mohr–Coulomb model*, which is represented by a hexagonal cone inscribed inside the circular cone of the Drucker–Prager model. This is a direct analogy to the Tresca hexagonal prism inscribed inside the von Mises cylinder. According to the Coulomb–Mohr criterion, failure is supposed to occur on a given plane when a critical combination of shear and normal stress acts on this plane. At failure, the mathematical relationship between stresses is given by

$$\frac{|\tau| + \mu\sigma}{\sqrt{1+\mu^2}} = \tau_\mathrm{u} \tag{26}$$

where $\tau$ and $\sigma$ are the stresses acting on the fracture plane, while and $\tau_\mathrm{u}$ is the pure shear failure stress and $\mu$ is a material constant related to the angle formed by the plane of fracture with the plane of the maximum principal stress, and typically assumes values between 0.15 and 0.6.

The Mohr–Coulomb model is particularly suitable for reproducing the behaviour of brittle uneven materials in a predominantly compressive state.

### 3.1.5  *Stassi–d'Alia yield criterion*

A pressure dependency similar to the Drucker–Prager criterion is invoked by the *Stassi–d'Alia yield criterion*. Besides the distortion strain energy considered by von Mises, this criterion also takes into account a contribution of the hydrostatic pressure $P = -(\sigma_1 + \sigma_2 + \sigma_3)/3$ (the negative of the hydrostatic stress $\sigma_\mathrm{avg}$), so that the tensile equivalent stress for which yield starts to occur is given by the root of the equation

$$k \cdot \sigma_\mathrm{teqSA}^2 + 3(k-1)P \cdot \sigma_\mathrm{teqSA} - \sigma_\mathrm{eqVM}^2 = 0 \;, \tag{27}$$

where $k = -\dfrac{\sigma_{Yc}}{\sigma_{Yt}}$ is the ratio between the compressive and tensile yield strengths of the material.

The compressive equivalent stress is simply given by $\sigma_{ceq} = -k\sigma_{teqSA}$. It can easily be verified that the Stassi–d'Alia criterion reduces to the von Mises criterion when $k = 1$.

*3.1.6  Maximum normal stress theory*

The *maximum normal stress theory*, often referred to as the cut-off or *Rankine criterion*, is applicable to brittle materials. According to this theory, the material fails when the maximum principal stress of a given stress state reaches either the uniaxial tension strength $\sigma_{ut}$ or the uniaxial compression strength $\sigma_{uc}$. This criterion can be expressed mathematically as

$$\sigma_{uc} \leq \sigma_{max} \leq \sigma_{ut},$$
$$\sigma_{max} = \max\left(|\sigma_1|, |\sigma_2|, |\sigma_3|\right). \tag{28}$$

Note that no interaction between the principal stresses is considered. In the principal stress space, the Rankine criterion corresponds to a cube oriented according to the three principal axes that intersect with the stress axes at the values of $\sigma_{uc}$ and $\sigma_{ut}$. This model is therefore *not* independent of hydrostatic stress.

For brittle materials, $\sigma_{uc}$ is usually much larger than $\sigma_{ut}$; these materials commonly contain large numbers of randomly oriented microscopic cracks that cannot support significant tensile stresses, since these stresses tend to open these flaws and cause them to grow. If the dominant stresses are compressive, the planar flaws tend to have their opposite sides pressed together so that they have less effect on the failure behaviour. This explains the higher strengths in compression. Also, compressive failure occurs in planes aligned with planes of maximum shear.

The Rankine criterion gives reasonably accurate predictions of fracture in brittle materials as long as the normal stress having the largest absolute value is tensile, while agreement with data is much worse in compression. For such materials, the tensile ultimate strength is usually measured through flexural tests, since in a tensile test the material tends to break in correspondence with the testing machine grasps. However, flexural tests usually overestimate the stress to failure because the maximum tensile stress is only reached in one face of the bent specimen, elsewhere being smaller or even compressive. For these reasons, in recent years more advanced tests have been devised to measure the ultimate strength of brittle materials in a purely tensile state: as an example, the Hopkinson bar set-up can be configured to generate a plane tensile shock wave on the sample, also allowing the sensitivity of the material to the strain rate to be evaluated [36].

In practice, several brittle failure models consist of a combination of a Rankine criterion in tension and a more elaborate surface in compression (that allows interaction between principal stresses), such as the Mohr–Coulomb fracture criterion mentioned in Section 3.1.4: such a combination is represented for instance by the so-called *modified Mohr–Coulomb fracture criterion* (Fig. 41): the transition between tension dominated states, expressed by the Rankine criterion and the compressive dominated states, modelled by the Mohr–Coulomb criterion is usually found at ratios $\sigma_2/\sigma_1 \approx -1$, corresponding to the state of pure torsion.

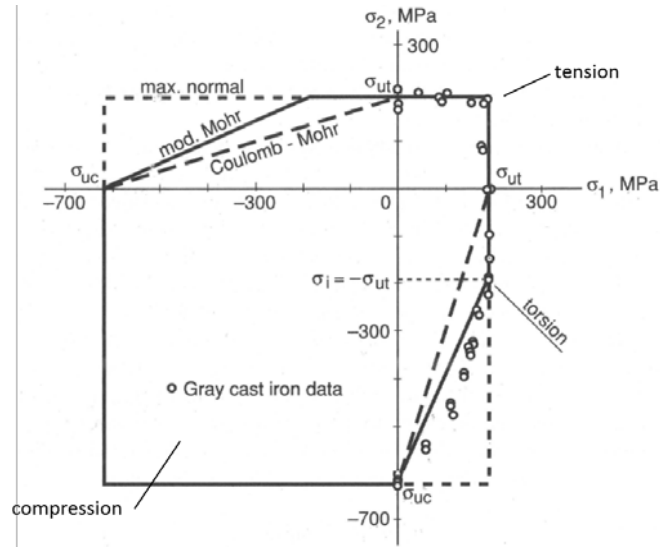

**Fig. 41:** Fracture data for grey cast iron compared with various failure criteria [37]

### 3.1.7 Hill criterion for orthotropic materials

All failure models presented so far are concerned with isotropic materials, thus they are represented by functions of principal stresses $\sigma_1$, $\sigma_2$, $\sigma_3$. This is not possible for anisotropic materials, since the orientation of the material does not permit the rotation of the stress tensor to the eigenvectors. As a result, failure criteria for orthotropic materials inevitably depend on all six components of the stress tensor.

The most commonly used orthotropic yield function is the *Hill criterion*, which is an extension of the isotropic von Mises model. It employs six independent material parameters that may be accessed experimentally from uniaxial tension and pure shear tests in the three material orientations. The criterion is of the form

$$F(\sigma_{11} - \sigma_{22})^2 + G(\sigma_{22} - \sigma_{33})^2 + H(\sigma_{33} - \sigma_{11})^2 + 2L\sigma_{12}^2 + 2M\sigma_{23}^2 + 2N\sigma_{31}^2 = 1 \ . \qquad (29)$$

Equation (29) reduces to the von Mises criterion, albeit in terms of the stress components instead of the principal stresses, simply by setting $F = G = H = 1/(2\sigma_Y^2)$ and $L = M = N = 3/(2\sigma_Y^2)$. Because orthotropic materials cannot be written as functions of the principal stresses, a graphic visualization of the yield locus is impossible.

The Hill criterion can be extended in the same fashion as the Drucker–Prager criterion to include differences in tensile and compressive strength. This results in an anisotropic Drucker–Prager model with three additional parameters accounting for the effect of hydrostatic pressure in each direction of material symmetry. Note that with increasing refinement of the models, the number of parameters increases. As a result, the applicability and accuracy of a constitutive model is often determined by the experimental accessibility of the parameters.

### 3.1.8 Criteria for non-linear materials: deformation to failure

The linear approximation is a powerful means of describing the stress–strain relationship in the elastic regime. For some materials, however, the $\sigma$–$\varepsilon$ relationship can depart appreciably from linearity. Examples are 'soft' materials, such as annealed copper or aluminium and magnesium alloys, and, most interestingly for BIDs, graphitic materials.

For deformation-driven problems, such as beam-induced energy deposition, considerable overestimation can be made when considering tension as the limiting factor (Fig. 42).

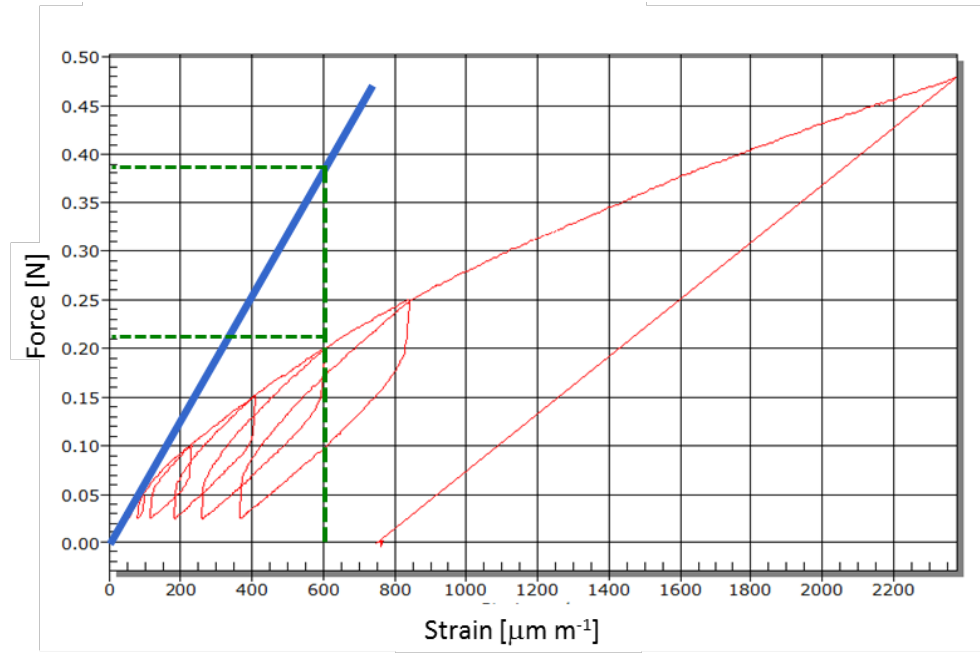

**Fig. 42:** Stress–strain test for a molybdenum–graphite grade. Departure from linearity is shown, as well as overestimation of tension stress in the case where a linear stress–strain relationship is assumed.

In the case of brittle behaviour, as in the example shown in Fig. 42, it is more appropriate to replace a fracture criterion based on ultimate strength, particularly in tension, such as the Rankine theory, with one based on *deformation to fracture*: failure is reached once the maximum principal strain reaches the value of the ultimate strain obtained from a uniaxial tension or bending test.

### *3.1.9    Summary*

A multitude of failure criteria exist for various applications. Some of the most prevalent isotropic and orthotropic rate-independent yield functions have been introduced (Table 5). Smooth failure surfaces are often suited to ductile materials, whereas most brittle materials exhibit a competition between different failure modes, each of which is represented by a separate patch of an overall non-smooth failure envelope.

Beam intercepting devices often make use of brittle materials for their active parts. In such cases, the use of a modified Mohr–Coulomb criterion should be considered, whereas if the material exhibits a strongly non-linear behaviour, the maximum normal strain criterion is better suited, at least in the tension region of the principal stresses space.

For ductile isotropic materials, either the von Mises or the Tresca–Guest criterion is usually chosen.

**Table 5:** Summary of relevant failure criteria, categorized according to the smoothness of the failure surfaces, dependence on hydrostatic pressure, and applicability to orthotropic materials.

|  | **Pressure-independent** | **Pressure-dependent** |
| --- | --- | --- |
| Isotropic | Huber–Hencky–von Mises *(smooth)* | Drucker–Prager *(smooth)* |
|  | Tresca–Guest *(non-smooth)* | Stassi–d'Alia *(smooth)* |
|  |  | Rankine *(non-smooth)* |
|  |  | Mohr–Coulomb *(non-smooth)* |
|  |  | Modified Mohr–Coulomb *(non-smooth)* |
|  |  | Maximum normal strain *(non-smooth)* |
| Orthotropic | Hill *(smooth)* | (Extended) Hill *(smooth)* |

## 3.2 Material selection: figures of merit

The choice of a particular material for BIDs, as much as for any other mechanical component, is driven by its performance against a large range of requirements. To general aspects, such as availability, manufacturing feasibility, costs, weight, delivery times, etc., one must add application-specific requirements, which, in the case of BIDs, typically include mechanical robustness, resistance to high temperatures, geometrical stability, cleaning efficiency, low contribution to RF impedance, and resistance to radiation.

To classify and rank potential materials against this large number of requirements, it is useful to introduce *figures of merit* which permit several material properties related to a specific requirement to be condensed into a single indicator: the higher the figure of merit, the better the material performance against that specific requirement.

A set of indices can be particularly helpful to orient material choice in the early phases of design; however, one must be aware of the fact that figures of merit rely on simplified, constant, linearized, temperature-independent material properties and on largely approximate extrapolations of certain factors, such as energy deposition. Hence, they should be used as indicative, comparative tools and not for quantitative assessment of material or component performance. Additionally, in the case of anisotropic materials, relevant properties are usually averaged over the three directions.

The most relevant figures of merit for the design of BIDs are related to:

- thermomechanical robustness;
- thermal stability;
- electrical conductivity;
- radiation resistance.

An index called the *thermomechanical robustness index* (TRI), is proposed to assess the material robustness against particle beam impacts. Given that thermal shock problems are, to a large extent, governed by the thermal deformation induced by a sudden temperature increase, it appears reasonable to base this index on the ratio between material *admissible strain* or *strain to failure* $\varepsilon_{\text{adm}}$ and the actual strain

$$\text{TRI} = \frac{\varepsilon_{\text{adm}}}{\varepsilon_{\text{ref}}} \cdot \left( \frac{T_{\text{m}}}{\Delta T_q} - 1 \right)^m , \tag{30a}$$

where $\varepsilon_{\text{ref}}$ and $\Delta T_q$ are the strain and the temperature increase generated by a reference energy deposition given in Eqs. (30c) and (30d), $T_{\text{m}}$ is the melting (or degradation) temperature and $m$ is a coefficient related to the material loss of strength with temperature increase.

Note that TRI tends to zero when the melting point is reached.

Since *strain to failure* values are hardly available for many materials while effective *failure strength* values, usually related to fracture or yielding for brittle or ductile materials, respectively, are much easier to obtain in the literature, it is convenient to express $\varepsilon_{\text{adm}}$ as a function of the failure strength $R_{\text{M}}$.

Incidentally, we observe that this assumption is conservative for non-linear materials, given that it involves a linear stress–strain relationship up to failure:

$$\varepsilon_{\text{adm}} = \frac{R_{\text{M}}}{\overline{E} \cdot (1-\nu)} . \tag{30b}$$

In Eq. (30b), we also implicitly assume that the stress distribution is the one encountered in a plane strain problem (see Eq. (7d)): this is in fact usually justified by the assumption that the beam impact occurs relatively close to the component surface, so that thermal expansion is not constrained in the direction normal to the surface. If a deeper impact is expected, a coefficient equal to $(1 - 2\nu)$ should be used instead.

The actual reference strain is expressed by

$$\varepsilon_{\text{ref}} = \bar{\alpha} \cdot \Delta T_q \ . \tag{30c}$$

The temperature increase $\Delta T_q$ can be assumed to be equal to a reference quasi-instantaneous *energy deposition* $q_d$, assumed to depend on the *geometric radiation length* $X_g$ and the material density $\rho$, divided by the *specific heat* $c_p$:

$$\Delta T_q = \frac{q_d}{c_p} = \frac{C_R \rho^n}{c_p X_g} \ . \tag{30d}$$

In these equations, $\bar{E}$ is the (*averaged*) *Young modulus*, $\nu$ the *Poisson ratio*, $\bar{\alpha}$ the (*averaged*) CTE, $C_R$ an arbitrary *scaling factor* and $n$ a *coefficient* expressing the influence of density on the *energy distribution* generated by the impact.

Equation (30d) implies that the energy deposited by a given number of particles, and therefore the material temperature increase, is related to the material density and to the geometric radiation length; it has been empirically observed that the coefficient $n$ for materials impacted by protons at several hundreds of GeV is ~0.2.

Combining Eqs. (30a–d), TRI can finally be written as

$$\text{TRI} = \frac{R_M c_p X_g}{\bar{E}(1-\nu)\bar{\alpha} C_R \rho^n} \cdot \left( \frac{T_m c_p X_g}{C_R \rho^n} - 1 \right)^m \ . \tag{30e}$$

The *thermal stability index* (TSI) provides an indication of the ability of the material to maintain the geometrical stability of the component under steady-state particle losses. This is particularly important for components such as collimators and long absorbers, which are required to interact with the halo of the particle, maintaining their longitudinal straightness to a fraction of a beam transverse sigma (Fig. 43).

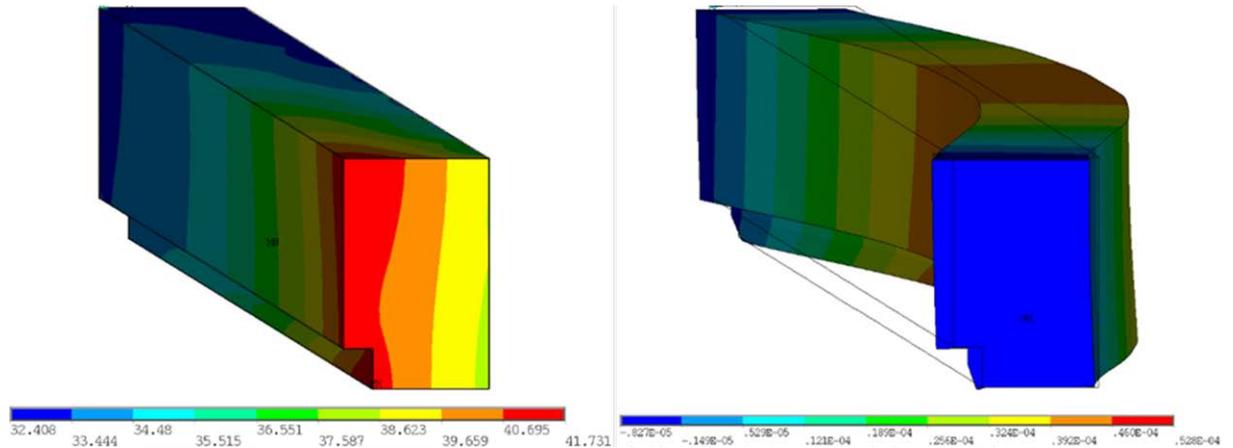

**Fig. 43:** Operating temperatures (in °C) (left) and thermally induced deflection (in m) (right) of a LHC secondary collimator jaw in steady-state conditions. The length of the jaw is 1 m, its deflection ≈40 µm.

The TSI is proportional to the radius of curvature of an elongated structure induced by a non-uniform temperature distribution; since we are particularly interested in deformations induced by grazing beams, we can assume a steady-state energy deposition in which all the heat is flowing from the surface exposed to the beam through the thickness of the BID. In this case, the radius of curvature of the component is given by

$$\rho_c = \frac{\bar{\lambda}}{\bar{\alpha}\dot{q}} \,, \tag{31a}$$

where $\dot{q}$ is the *heat flux* in W m$^{-2}$ and $\bar{\lambda}$ is the (*averaged*) *thermal conductivity*. As with TRI, we can express the steady-state flux of (deposited) heat as

$$\dot{q} = \frac{C_\text{S}\rho^n}{X_\text{g}} \,, \tag{31b}$$

where $C_\text{S}$ is a scaling factor.

The TSI is then given by

$$\text{TSI} = \frac{\bar{\lambda} X_\text{g}}{\bar{\alpha} C_\text{S}\rho^n} \,. \tag{31c}$$

Beam intercepting devices located in the accelerator ring, such as collimators and certain absorbers, are usually the machine components sitting closest to the circulating beam; therefore, their contribution to the accelerator global RF impedance is by far the highest. The part of the beam coupling impedance related to the resistive losses in the material surrounding the beam, the so-called *wall impedance*, is directly related to the material electrical resistivity. Therefore, maximizing the electrical conductivity of the materials mostly interacting with the beam can play a major role in minimizing the risk of impedance-induced beam instabilities.

A first approximation of the contribution to the total impedance at relatively high frequencies (above ~1 MHz) by resistive objects is given by the so-called classic thick-wall regime [38]. In this regime, the transverse wall impedance of a cylindrical beam pipe is approximately given by

$$Z_t(\omega) = (1+\text{j})\frac{LZ_0\mu_r}{2\pi b^3}\sqrt{\frac{2}{\mu_0\mu_r\gamma\omega}} \,, \tag{32a}$$

where $\omega$ is the frequency, j is the imaginary unit, $L$ is the length of the pipe, $Z_0$ is the free-space impedance, $b$ is the radius of the beam pipe, $\mu_r$ and $\mu_0$ are the relative and free-space permeability, respectively, and $\gamma$ is the *electrical conductivity*.

We can hence define a *RF impedance index* (RFI), minimizing the material contribution to the system wall impedance, as

$$\text{RFI} = \sqrt{\frac{\gamma}{\mu_r}} \,. \tag{32b}$$

Irradiation of materials by energetic particles causes microstructural defects, which translate into a degradation of the thermophysical properties. Radiation resistance is defined as the ability of the material to maintain its properties under and after irradiation. An analysis of the effects induced on materials by ionizing radiation goes beyond the scope of these lectures. A review of radiation effects on materials can be found in several sources, such as Ref. [5].

### 3.3  Novel materials for beam intercepting devices

As seen, the introduction in recent years of new and extremely energetic particle accelerators, such as the LHC, brought about the need for advanced cleaning and protection systems, to safely increase the energy and intensity of particle beams to unprecedented levels. This has greatly increased the requirements for materials exposed to accidental impact from highly energetic and intense particle beam pulses; on top of outstanding thermal shock resistance, materials for halo cleaning and machine protection devices are typically required to maximize the figures of merit defined in Section 3.2, such as electrical conductivity, geometrical stability, and resistance to radiation damage. These requirements are set to become even more compelling in consideration of the High-Luminosity upgrade of the LHC (HL-LHC), expected to increase the beam intensity and its stored energy by a factor of two [39]: as an example, carbon–carbon (C–C) composites used for primary and secondary collimators in the LHC may limit the HL-LHC performance because of C–C low electrical conductivity leading to beam instability at high intensities [40], while the tungsten alloy (Inermet180) used in LHC tertiary collimators has very low robustness in case of beam impacts even at relatively low intensities.

In view of these challenges, an extensive R&D program has been launched at CERN in recent years to explore and develop a number of novel materials aiming to combine the excellent properties of graphite or diamond, specifically their low density, high thermal conductivity. and low thermal expansion, with those of metals or transition metal based ceramics, possessing high mechanical strength and good electrical conductivity. The most promising materials so far identified (early 2015) are molybdenum carbide – graphite (MoGr) and copper–diamond (CuCD).

#### 3.3.1   *Molybdenum carbide – graphite*

Molybdenum carbide – graphite is a novel composite developed in a collaboration between CERN and an Italian enterprise [41]: it is produced from molybdenum and graphite powders by high-temperature *fast direct hot pressing*, a pressure-assisted sintering technique in which heating is obtained by the passage of an electrical current through moulds and powders [42].

Pure molybdenum has a very high melting point and a low CTE, as well as excellent mechanical strength and electrical conductivity, while graphitic materials feature low density, extremely high service temperatures, large damping properties (particularly useful in attenuating shock waves) and, provided graphite crystallite ordering is sufficiently extended and a high graphitization degree is attained, excellent thermal conductivity and a very low CTE, at least in the direction aligned with the graphite basal plane. At high temperatures, molybdenum reacts rapidly with carbon, forming stable carbides ($MoC_{1-x}$) which, in spite of their ceramic nature, retain a good electrical conductivity; in this respect, MoGr becomes a *ceramic–matrix composite*.

A broad range of compositions, powder types, and dimensions with processing temperatures ranging from 1700°C to 2600°C were developed: the best results so far were obtained for a sintering temperature of 2600°C.

The carbonaceous phase may be composed, in different grades, either of natural graphite flakes or of a mixture of natural graphite flakes and mesophase pitch-derived carbon fibres: these were selected to act as structural reinforcement and nucleation sites for enhanced graphitization, contributing to the improvement of thermal properties, thanks to their well-ordered graphitic structure (Fig. 44), and their mechanical strength.

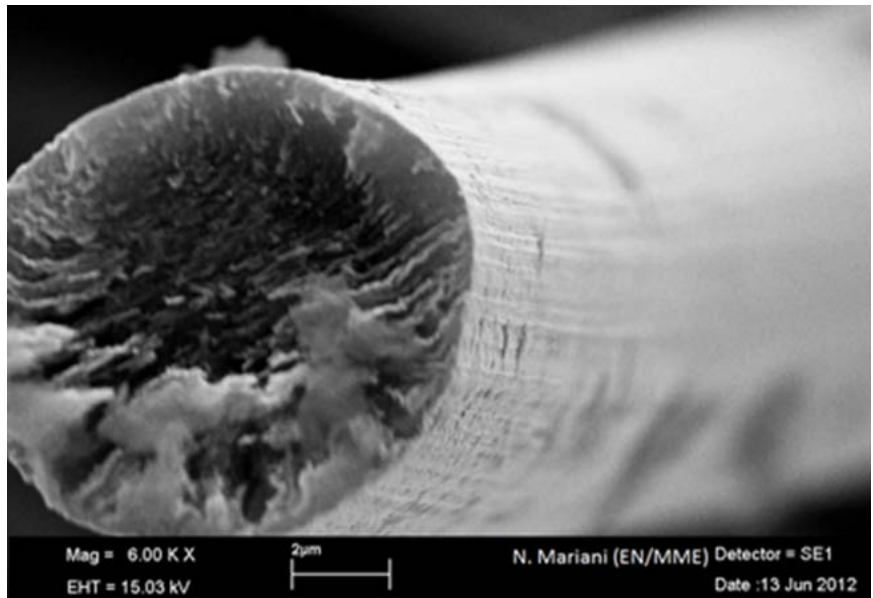
**Fig. 44:** Mesophase-pitch-derived carbon fibre (6000×)

High processing temperature grades are produced by *liquid-phase sintering* above the melting point of molybdenum carbide (2589°C). Scanning electron micrographs provide evidence of a very homogeneous microstructure with a regular distribution of small (5–10 µm) carbide particles and a high degree of graphitization of the carbonaceous phase (Fig. 45).

The high ordering and orientation of the graphitic phase is most likely catalysed by the presence of a carbide liquid phase at high temperatures; this lowers the required activation energy for the graphite arrangement and improves the diffusion rate of carbon atoms, with graphite crystallite growing through molten material as the graphitization process proceeds.

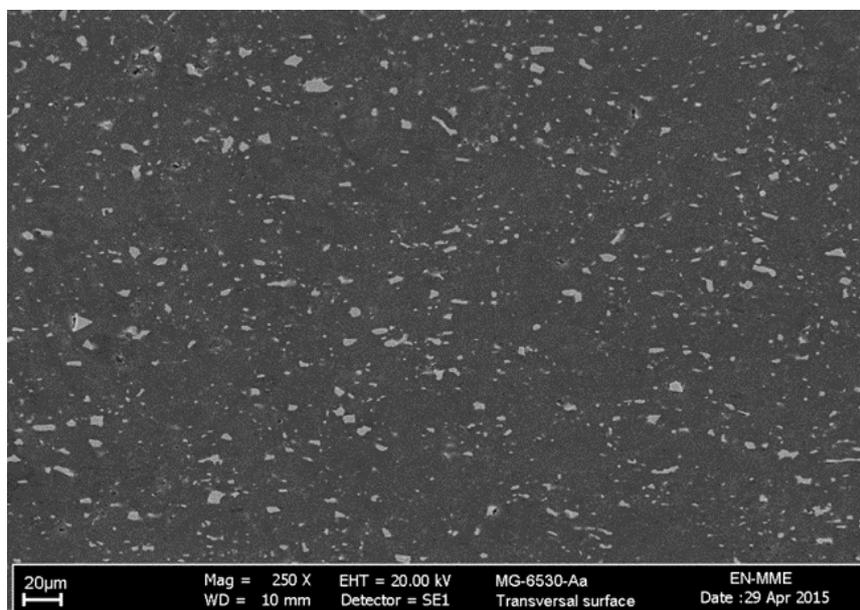
**Fig. 45:** Scanning electron micrograph of MoGr; note the finely dispersed carbide grains (250×)

To favour a liquid carbide infiltration and material compaction rate, a significant quantity of molten carbides is allowed to flow out of the mould during liquid-phase sintering, so that the final density of the material is reduced to ≈2.5 g cm$^{-3}$. Thanks to the extensive catalysed graphitization, MoGr has an electrical conductivity in the preferential direction (perpendicular to the pressing direction and parallel to the basal planes of graphite crystallites) of ~1 MS m$^{-1}$, one order of magnitude larger than that of isotropic graphite and of the C–C composites used for the LHC collimator. This property can be further increased by cladding or coating the external surface with pure molybdenum or other high electrical conductivity materials.

On top of its low density and good electrical conductivity, MoGr presents outstanding thermal properties, which are particularly useful in increasing the TSI and TRI: along the preferential direction, the thermal conductivity at RT is of the order of 700 Wm$^{-1}$K$^{-1}$ (Fig. 46), almost twice that of pure copper and a factor of four greater than that of C–C, while the CTE is $1.8 \times 10^{-6}$ K$^{-1}$ for temperatures spanning from RT to 2000°C.

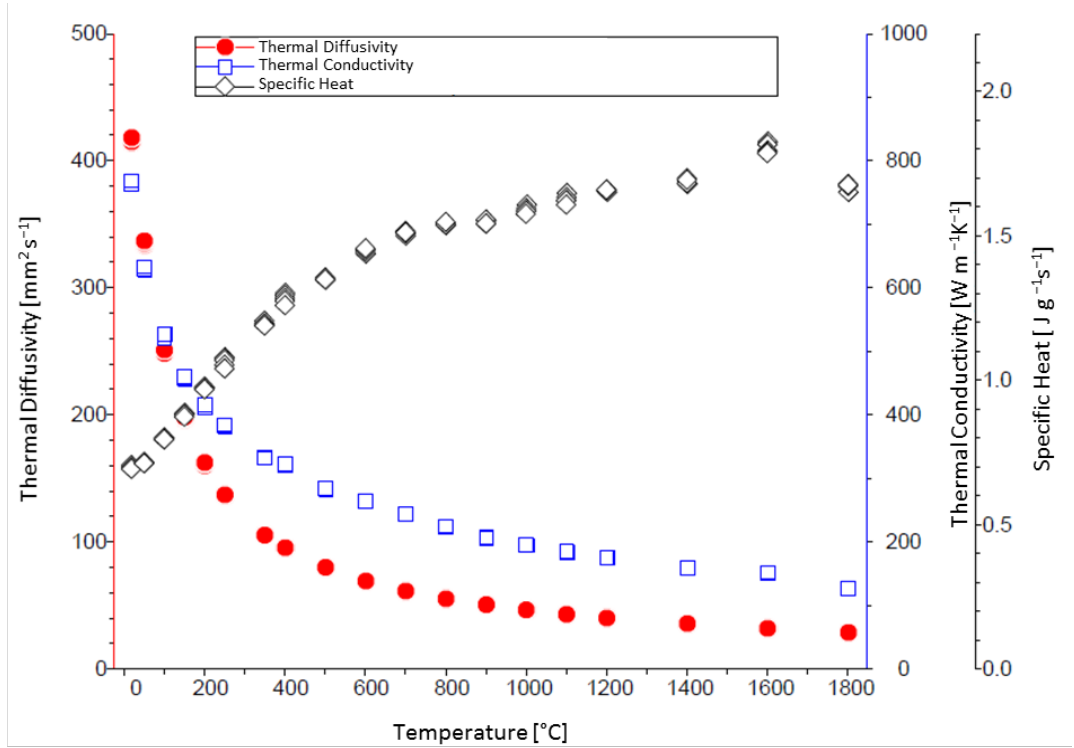

**Fig. 46:** Thermal conductivity, diffusivity, and specific heat of MoGr between RT and 1800°C

Relevant reference properties of MoGr are provided in Table 6: the subscript $a$ indicates the preferential direction parallel to graphite basal planes, while the subscript $c$ indicates the direction orthogonal to these planes.

**Table 6:** Selected properties of MoGr

| Property | Value |
|---|---|
| Density, $\rho$ | 2.5 g/cm$^3$ |
| CTE, $\alpha_a$ (RT to 1000°C) | $1.8 \times 10^{-6}$ K$^{-1}$ |
| CTE, $\alpha_c$ (RT to 1000°C) | $12 \times 10^{-6}$ K$^{-1}$ |
| Thermal conductivity, $\lambda_a$ (RT) | >700 Wm$^{-1}$K$^{-1}$ |
| Thermal conductivity, $\lambda_c$ (RT) | 85 Wm$^{-1}$K$^{-1}$ |
| Electrical conductivity, $\gamma_a$ (RT) (uncoated) | 1 MSm$^{-1}$ |
| Electrical conductivity, $\gamma_c$ (RT) | 0.3 MSm$^{-1}$ |
| Young modulus $E_a$ (flexural) (RT) | 53 GPa |
| Ultimate strength $R_m$ (flexural) (RT) | 85 MPa |

*3.3.2 Copper–diamond*

Copper–diamond is produced by RHP Technology (Austria) by *solid-state sintering*; the initial volumetric composition is 60% diamond, 39% copper, and 1% boron [41, 42]. Copper is chosen for its excellent thermal and electrical conductivity, along with its good ductility, while diamond is added to reduce the density and the CTE, while contributing to the thermal conductivity.

A higher proportion of diamond would not allow a good material compaction, which is also achieved through the use of diamonds of various sizes, to optimize the filling of interstitials. Unlike MoGr, the main issue for material adhesion is the low chemical affinity between the two main elements, which leads to a lack of bonding between copper and diamond: this would jeopardize not only the material's mechanical strength but also its thermal conductivity. Boron is added to offset such limitations, since this element promotes the formation of carbides at the diamond–copper interface, improving material internal bonding (Fig. 47).

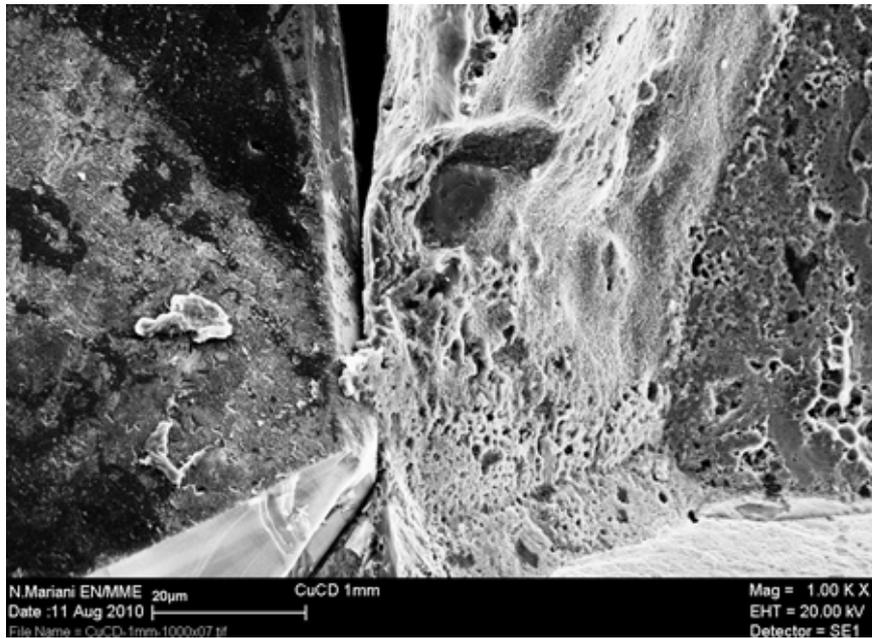

**Fig. 47:** High magnification scanning electron micrograph of the fracture surface of CuCD. Note the small boron carbide platelet, which is connecting the diamond grain to the detached copper matrix. 1000×.

Copper–diamond has very good thermal and electrical conductivity, and its CTE is reduced by a factor of 2–3, compared with pure copper (Table 7).

**Table 7:** Selected properties of CuCD

| | |
|---|---|
| Density, $\rho$ | 5.4 g cm$^{-3}$ |
| CTE, $\alpha$ (RT to 900°C) | 6–12 × 10$^{-6}$ K$^{-1}$ |
| Thermal conductivity, $\lambda$ (RT) | 490 W m$^{-1}$ K$^{-1}$ |
| Electrical conductivity, $\gamma$ (RT) | 12.6 MS m$^{-1}$ |
| Young modulus $E$ (flexural) (RT) | 220 GPa |
| Ultimate Strength $R_m$ (flexural) (RT) | 70 MPa |

However, density and CTE are higher than MoGr, and the industrialization of the material is rather difficult: while thin samples of constant section can be produced via water-jet cutting, more complicated shapes with precise tolerances can only be produced by applying a pure copper cladding on the outer surfaces, which may limit material performance in the case of accidental grazing impacts (Fig. 48).

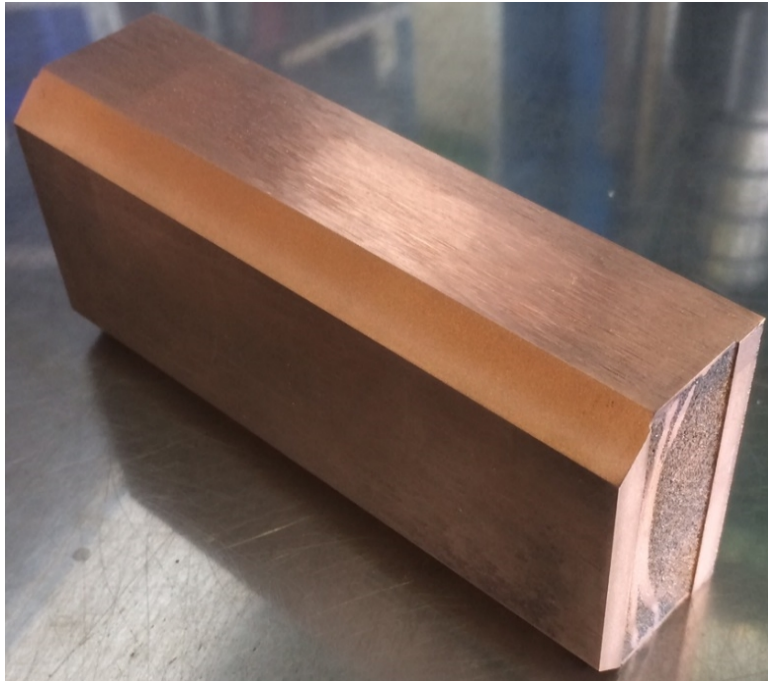

**Fig. 48:** CuCD block produced for a prototype jaw of HL-LHC secondary collimator [43]. Note the copper cladding on all functional surfaces.

### 3.4 Comparison of BID materials performances

A comparison of various materials of interest for BIDs, including the figures of merit defined in Section 3.2 is presented in Table 8. It can be seen from the table that no material perfectly meets all the requirements: material choice will therefore depend on which performance aspects must be favoured.

**Table 8:** Relevant properties and figures of merit for typical BID materials

|  | **Beryllium** | **Carbon–carbon** | **Graphite** | **MoGr** | **CuCD** | **Glidcop** | **Molybdenum** | **Tungsten heavy alloy** |
|---|---|---|---|---|---|---|---|---|
| $\rho$ [g cm$^{-3}$] | 1.84 | 1.65 | 1.9 | 2.50 | 5.4 | 8.90 | 10.22 | 18 |
| $Z$ | 4 | 6 | 6 | ≈6.5 | ≈11.4 | ≈29 | 42 | ≈70.8 |
| $X_g$ [cm] | 35 | 26 | 19 | 17 | 4.8 | 1.4 | 0.96 | 0.35 |
| $c_p$ [Jkg$^{-1}$K$^{-1}$] | 1925 | 780 | 760 | 750 | 420 | 391 | 251 | 150 |
| $\bar{\alpha}$ [10$^{-6}$ K$^{-1}$] | 18.4 | 4.1 | 5.5 | 5.0 | 7.8 | 20.5 | 5.3 | 6.8 |
| $\bar{\lambda}$ [Wm$^{-1}$K$^{-1}$] | 216 | 167 | 70 | 547 | 490 | 365 | 138 | 90.5 |
| $T_m$ [°C] | 1273 | 3650 | 3650 | 2589 | ≈1083 | 1083 | 2623 | ≈1400 |
| $\bar{E}$ [GPa] | 303 | 62.5 | 12 | 44 | 220 | 130 | 330 | 360 |
| $R_M$ [MPa] | 370 | 87 | 30 | 80 | 70 | 365 | 660 | 660 |
| $\Delta Tq$ [K] | 0.36 | 1.2 | 1.7 | 2.1 | 15.1 | 60.1 | 144 | 745 |
| TRI | 800 | 800–1200 | 800–1100 | 500–800 | 7 | 5 | 6.5 | 0.5 |
| TSI | 17 | 45 | 10 | 70 | 10 | 0.8 | 0.7 | 0.1 |
| RFI | 4.8 | 0.38 | 0.27 | 1 | 3.5 | 7.3 | 4.4 | 2.9 |

Beryllium performs well in practically all aspects: unfortunately, extensive use is severely limited by its toxicity.

In general, carbon-based materials feature excellent TRI and TSI, thanks to their low atomic number and density, reduced CTE, high degradation temperature, and high thermal conductivity; however, they are penalized by low electrical conductivity when RF impedance is a critical requirement. In such a case, MoGr is the most promising compromise, particularly if coated with higher-conductivity thin films, which would further improve the electrical conductivity by a factor of 10.

The poor performance of tungsten heavy alloys as to thermomechanical robustness should be noted: this is due to a combination of factors, including, in particular, the low melting temperature of the nickel–copper matrix that is used to bind the tungsten particles and to increase material ductility.

## 4  Experimental testing and validation

### 4.1  Introduction

Advanced numerical simulation codes are powerful tools that enable the analysis of extremely complex dynamic phenomena; however, to provide reliable results, they require sufficiently accurate constitutive models for all the conditions that materials might undergo during such events.

Unfortunately, constitutive material models, in particular, at the extreme conditions generated by high-energy beam impacts, are far from being readily available and experimentally validated; many constitutive models for existing materials were obtained through military R&D and are therefore classified. The situation is even more delicate for non-conventional alloys, compounds and composite materials presently used or likely to be used in state-of-the-art BIDs for very high-energy particle accelerators, for which experimental studies have never been carried out.

Moreover, numerical simulations cannot easily predict additional, far-reaching, consequences of beam accidents on nearby equipment, ultra-high vacuum performance, electronics, etc.

This is why only ad-hoc material tests can provide the correct inputs for numerical analyses, enabling the benchmarking and validation of simulation results on simple specimens as well as on full-scale, complex structures.

### 4.2  HiRadMat facility at CERN

A dedicated facility has been designed and commissioned at CERN to test materials and systems under high-intensity pulsed particle beams: HiRadMat (High Radiation to Materials) (Fig. 49) [12].

Given the high destructive power reached by the LHC, it has been decided that any new beam intercepting device must be tested, prior to its installation, for sufficient robustness to as realistic as possible conditions for future operation, to at least ensure that possible and unavoidable damage can be locally constrained, in order to prevent catastrophe (e.g. causing damage to nearby components, water leaks into vacuum from the cooling system, spreading of sputtered materials, or vacuum quality deterioration over long distances).

Previous tests of robustness and damage effects on BIDs and materials were performed in ad-hoc installations in the TT40 transfer beam line to LHC and CNGS in 2004 and 2006. The difficulty in performing such important tests on temporary installations and the potential impact on operating transfer lines were the main motivations for building HiRadMat, which was purposely designed to study beam shock impacts on materials and accelerator components.

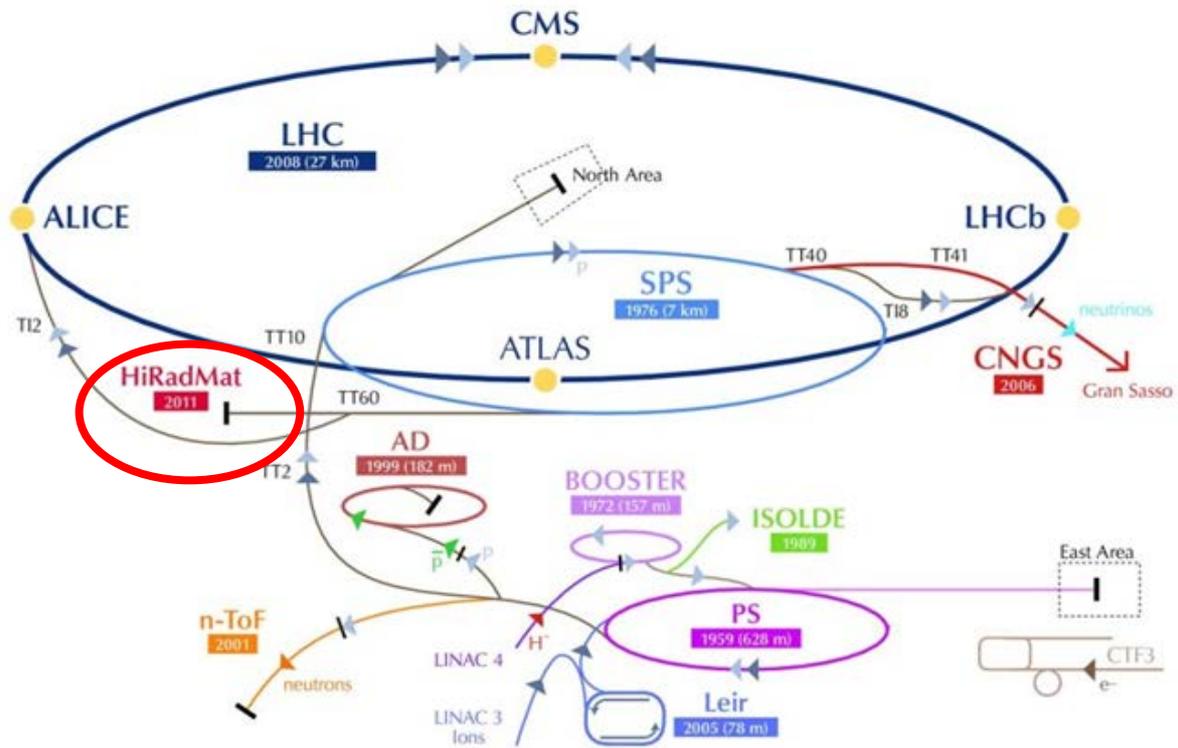

**Fig. 49:** Location of HiRadMat in CERN accelerator complex

HiRadMat uses an extracted primary proton or ion beam from the LHC SPS. The main beam parameters are listed in Table 9. The beam spot size at the focal point of the experiment can be varied from 0.5 to 2 mm$^2$, which, together with the variable beam intensity, offers sufficient flexibility to test materials at different deposited energy densities (Fig. 50).

**Table 9:** Key parameters for the HiRadMat beam

|  | **Protons** | **Ions (Pb$^{82+}$)** |
|---|---|---|
| Energy | 440 GeV | 173.5 GeV u$^{-1}$ |
| Bunch intensity (max) | $1.7 \times 10^{11}$ p$^+$ | $7 \times 10^9$ ions |
| No of bunches | 1 to 288 | 52 |
| Pulse intensity (max) | $4.9 \times 10^{13}$ p$^+$ | $3.6 \times 10^9$ ions |
| Pulse energy (max) | 3.4 MJ | 21 kJ |
| Bunch length | 11.24 cm | 11.24 cm |
| Bunch spacing | 25, 50, 75, 150 ns | 100 ns |
| Pulse length | 7.2 µs | 5.2 µs |

Beyond the needs of CERN, HiRadMat is open to other users and is also included in the EUCARD FP7 European Project as transnational access to facilitate its use by European teams.

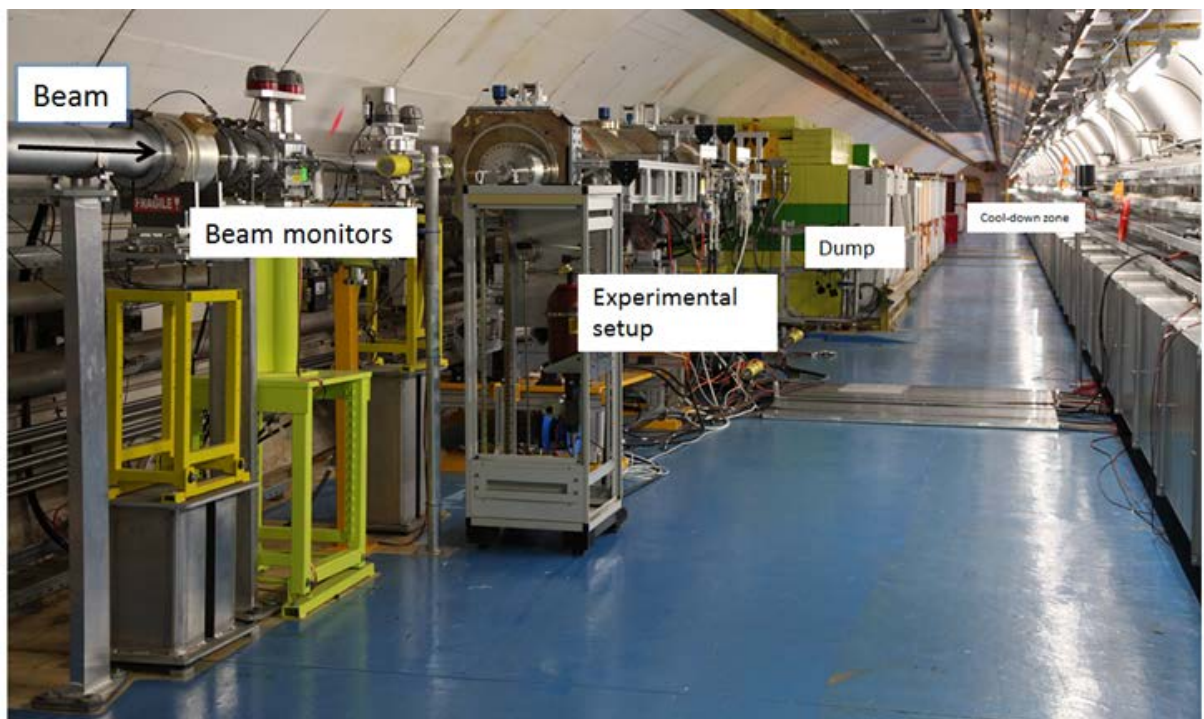

**Fig. 50:** HiRadMat facility experimental set-up

HiRadMat is not an irradiation facility, where large doses on equipment can be accumulated. It is rather a test area, designed to perform single experiments to evaluate the effect of high-intensity pulsed beams on materials or accelerator component assemblies in a controlled environment. The facility is designed for a maximum of $10^{16}$ protons per year, distributed among 10 experiments, each having a total of $10^{15}$ protons or about 100 high-intensity pulses. This limit allows reasonable cool-down times for the irradiated objects (from a few months to a year) before they can be analysed in specialized facilities.

Nine experiments were performed in the facility in 2012 before LHC Long Shutdown 1; three of them are described next.

### 4.3 HiRadMat 12 experiment

In normal operating modes, each of the 362 MJ LHC beams is safely extracted into a 700 m long transfer line, its energy density diluted and finally absorbed in large graphite blocks. However, several failure modes could abnormally deflect the beam into a graphite absorber, septum magnets, or superconducting magnets. It is therefore extremely important to predict the consequences of such events.

Extensive simulation studies of the full impact of the ultra-relativistic proton beam generated by the LHC on solid targets of different materials were carried out: they predicted that the energy deposited in the target by the protons in the first 10 bunches and their hadronic shower would lead to the phenomenon of hydrodynamic tunnelling described in Section 2.4.4. The strongly heated material would undergo phase transitions that include liquefaction, evaporation, and even conversion into weakly ionized strongly coupled plasma. The high temperature in the absorption zone would induce high pressures in the core, generating a radially outgoing shock wave, which causes substantial density depletion at the axis. As a consequence, the protons that are delivered in subsequent bunches would penetrate the target much more deeply. For example, the range of a hadronic shower of 7 TeV protons in solid carbon, which is about 3 m, would be extended to around 25 m for the full LHC beam with 2808 bunches because of hydrodynamic tunnelling [44].

This phenomenon therefore has very important implications for machine protection system design. To check the validity of these theoretical considerations, especially the existence of the hydrodynamic tunnelling, a dedicated experiment was performed at the HiRadMat facility using the SPS proton beam [45].

The set-up consisted of three targets of 15 copper cylinders each, spaced by 1 cm to allow visual inspection after irradiation. Each cylinder had a radius of 4 cm and length of 10 cm. Figure 51 shows the targets before the installation. The targets were mounted on a movable table that could be moved to four different positions: target 1, target 2, target 3, and an off-beam position. The set-up was equipped with pCVD diamond particle detectors, PT100 temperature sensors, strain gauges and secondary electron emission particle detectors to obtain additional information during the beam interaction.

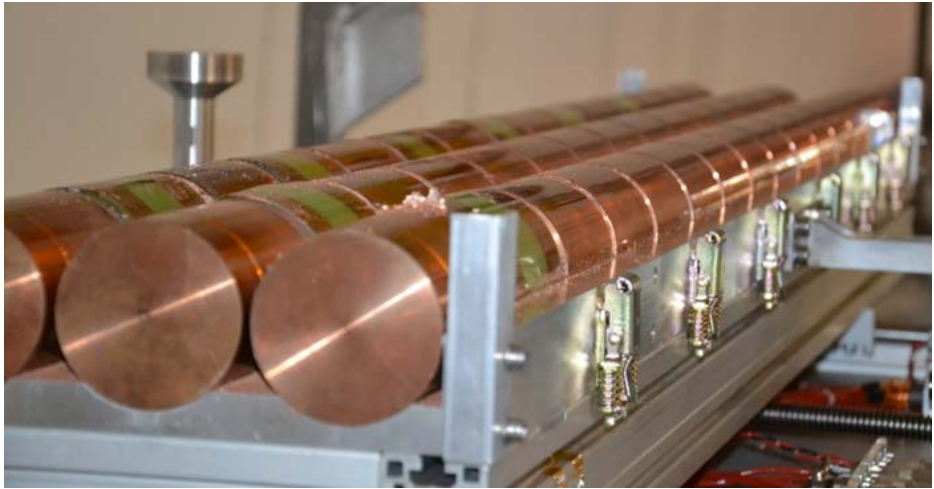

**Fig. 51:** HRMT-12 set-up; each individual target is made of 15 copper cylinders

For all the experiments, the proton energy was 440 GeV, bunch intensity $1.5 \times 10^{11}$ protons, bunch length 0.5 ns, and bunch separation 50 ns. Target 1 was irradiated with 144 bunches with a beam sigma of 2 mm. Target 2 was irradiated with 108 bunches, whereas target 3 was irradiated with 144 bunches; in both these cases, the beam had a much smaller focal spot size, characterized by $\sigma = 0.2$ mm. The beam parameters used in these three experiments are summarized in Table 10.

The target was opened for visual inspection after 8 months of cool-down. Droplets and splashes of molten and evaporated copper were found on the copper cylinders, on the aluminium housing at the position of the gaps between cylinders, and in the front aluminium caps.

**Table 10:** Experimental beam parameters used in the three experiments

| Target | Number of bunches | Beam $\sigma$ [mm] | Beam energy [MJ] | Expectation |
|---|---|---|---|---|
| 1 | 144 | 2.0 | 1.52 | Some tunnelling |
| 2 | 108 | 0.2 | 1.14 | Moderate tunnelling |
| 3 | 144 | 0.2 | 1.52 | Significant tunnelling |

Figure 52 shows the aluminium cover that was placed on top of the target assembly. After the beam impact, molten or evaporated material is projected outwards and deposited on the cover. The traces of the projected copper between the 10 cm long cylinders are clearly visible. It can be seen that in the experiment using 144 bunches and a beam focal spot of $\sigma = 2.0$ mm (bottom picture), the splash of molten copper occurs up to the gap between the fifth and sixth cylinders. This means that the material melted or evaporated over a length of $55 \pm 5$ cm. In the second experiment, with 108 bunches and a beam focal spot of $\sigma = 0.2$ mm (middle picture), the melting and evaporation zone extends to the eighth cylinder, indicating a damage length of $75 \pm 5$ cm. In the experiment with 144 bunches and a beam focal spot of $\sigma = 0.2$ mm (top picture), the melting and evaporation zone extends to the ninth cylinder, a length of $85 \pm 5$ cm.

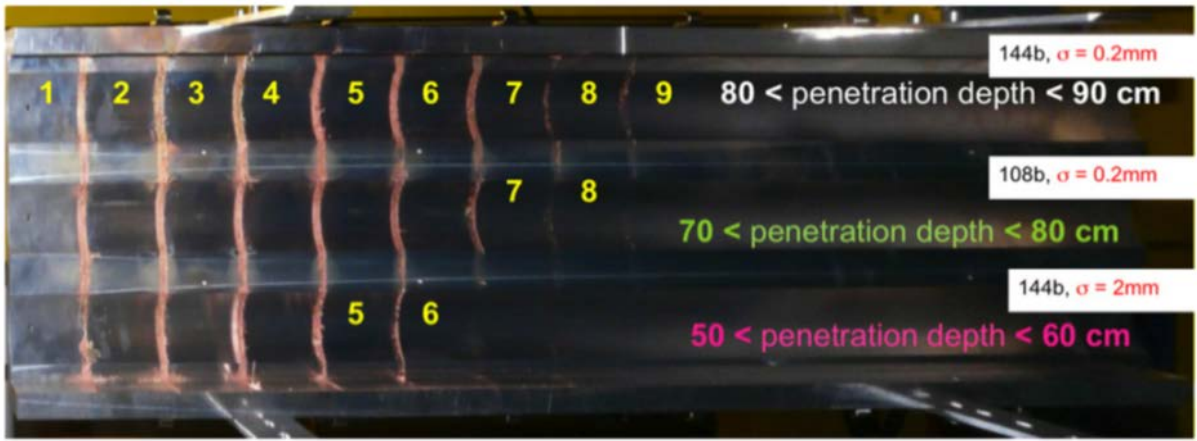

**Fig. 52**: Top cover of the experimental set-up after the irradiation. Traces of projected copper between the 10 cm long cylinders of the targets indicate the length of the melting and evaporation zone [45].

Detailed numerical simulations were carried out running the energy deposition code FLUKA and the two-dimensional hydrodynamic code BIG2 iteratively, using a time step of 700 ns, corresponding to the time during which the target density changes by about 15% at the target centre because of hydrodynamic processes. A semi-empirical multi–phase EOS was chosen to model different phases of the copper target during and after the irradiation.

In Fig. 53, density and temperature are plotted along the axis at $t = 5800$ ns, for the 108 bunches of case 2. It can be seen that the flat part of the temperature curve that represents the melting region lies within $L = 75$ and 80 cm, which is equivalent to the second half of the eighth cylinder. The temperature curve also shows that the material along the axis up to 75 cm is liquefied or even evaporated, depending on the temperature. The liquefied material escapes from the left face of cylinder number 8 and collides with the molten or gaseous material ejected from the right face of cylinder number 7. As a result of this collision, the material is splashed vertically and is deposited at the inner surface of the target cover above the gap between cylinders 7 and 8. The simulations are therefore in full agreement with experimental observations. Similar conclusions can be drawn for cases 1 and 3.

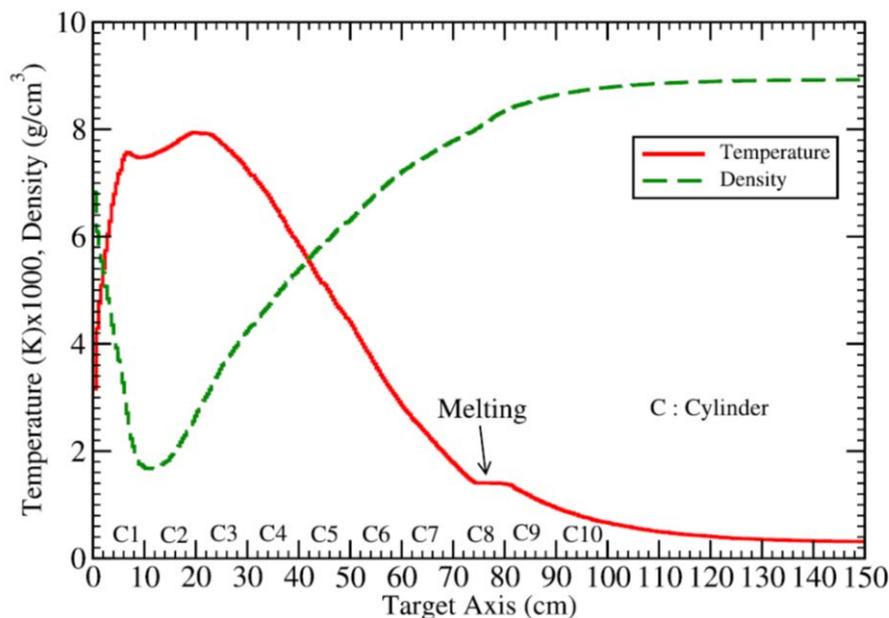

**Fig. 53:** Temperature and density on axis of target 2 after 5800 ns (case 2, 108 bunches)

Figure 54 the physical state of the target up to a radius of 0.5 cm. It can be seen that different parts of the beam-heated region lie in different phases of high energy density matter. These include gas, two-phase liquid–gas, liquid, and melting states. This suggests that the HiRadMat facility is very much suited to the important research area of high energy density physics.

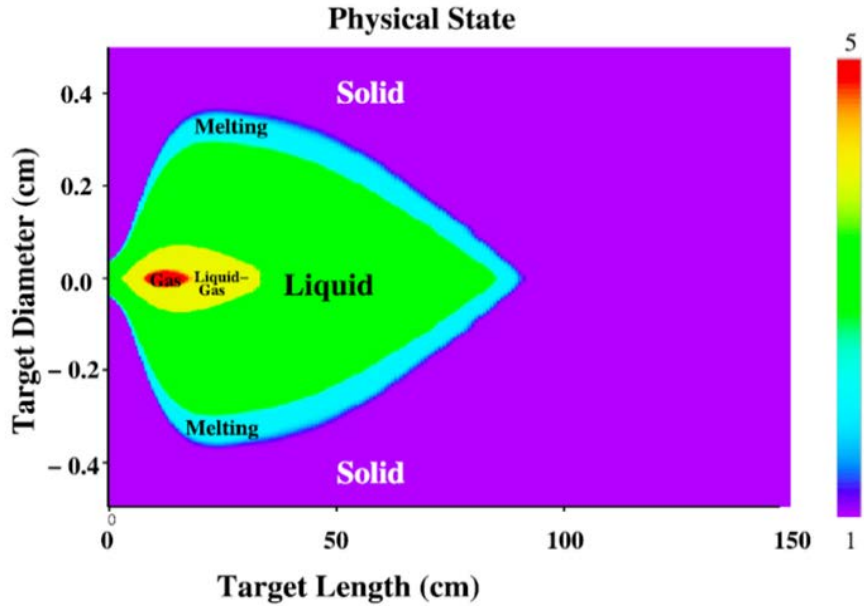

**Fig. 54:** Physical state of target 3 material after 7850 ns (case 3, 144 bunches) [45]

### 4.4 HiRadMat 09 experiment

As anticipated in Section 2.4.3, a thorough numerical analysis of a tertiary collimator of the LHC was completed to simulate the effects of several asynchronous beam abort cases with different values of beam emittance, energy, and intensity. This computation relied on advanced simulations performed with the wave propagation code Autodyn, applied to a multicomponent three-dimensional model [33].

The most important issue of these simulations concerned the reliability of constitutive models of relevant materials, especially at extreme conditions of temperature, pressure, and energy induced by the beam impact. To probe and evaluate such models, two experiments were performed in the HiRadMat facility in 2012. The first experiment, known as HRMT09, entailed the destructive test of a complete tertiary collimator, to assess not only the mechanical damage provoked to the structure but also other consequences of the beam accident, such as degradation of vacuum pressure in the beam line, contamination of the inner walls of the vacuum vessel, and impacts on collimator dismounting procedure.

The aim of the experiment was to verify the robustness and performance integrity of a fully assembled tertiary collimator under direct beam impact [46]. Three different tests were performed, with different beam intensities and different goals (Fig. 55).

- Test 1: investigate the effects of asynchronous beam dump under an impact equivalent to 1 LHC bunch at 7 TeV [47].
- Test 2: identify the onset of plastic damage on blocks made of tungsten heavy alloy.
- Test 3: reproduce a destructive scenario, inducing severe damage on the collimator jaw (damage on the collimator equivalent to four bunches at 5 TeV).

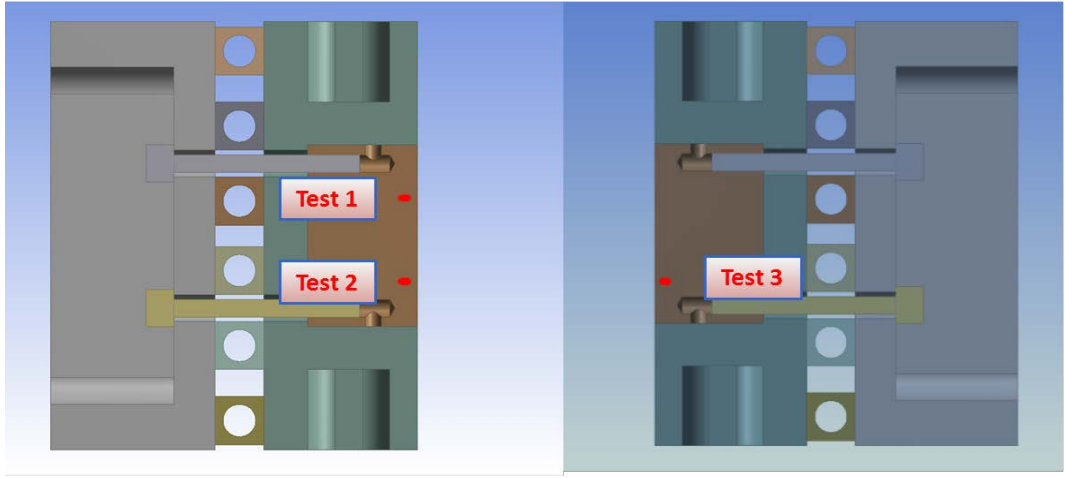

**Fig. 55:** Schematic diagram of three tests performed on tertiary collimator during HRMT09 experiment

For each test, the intensity and emittance of the SPS pulse was calculated so that the mechanical damage on the jaw would be equivalent to the one induced by the LHC (Table 11). For example, a SPS pulse with $3.36 \times 10^{12}$ protons is expected to produce a mechanical damage on the jaw equivalent to one LHC nominal bunch at 7 TeV.

A visual inspection performed a few months after the irradiation revealed the effects anticipated in Fig. 5. The damage provoked by tests 1 and 3 is clearly visible; an impressive quantity of tungsten alloy was ejected, partly stuck on the opposite jaw, partly fallen on the tank bottom or towards entrance and exit flanges.

The observation also highlighted other possible issues.

- Contamination of bellows, tank, and vacuum chambers, owing to activated tungsten fragments; procedures for maintenance, intervention, and replacement must take this into account.
- Ejected particles may affect the correct functionality of movable parts (RF fingers sliding on upper and lower rails).
- Degradation of ultra-high vacuum along the beam line.

**Table 11:** Beam parameters and impact positions of tests performed during HRMT09

|  | Test 1 | Test 2 | Test 3 |
|---|---|---|---|
| Beam energy | 440 GeV | 440 GeV | 440 GeV |
| Pulse intensity | $3.36 \times 10^{12}$ p | $1.04 \times 10^{12}$ p | $9.34 \times 10^{12}$ p |
| No of bunches | 24 | 6 | 72 |
| Bunch spacing | 50 ns | 50 ns | 50 ns |
| Beam size [$\sigma_x \times \sigma_y$] | 0.53 mm × 0.36 mm | 0.53 mm × 0.36 mm | 0.53 mm × 0.36 mm |
| Impact location | Left jaw, +10 mm | Left jaw, −8.3 mm | Right jaw, −8.3 mm |
| Impact depth | 2 mm | 2 mm | 2 mm |
| Jaw half-gap | 14 mm | 14 mm | 14 mm |

A qualitative comparison of visible damaged areas with Autodyn simulations is provided in Figs. 56 to 58.

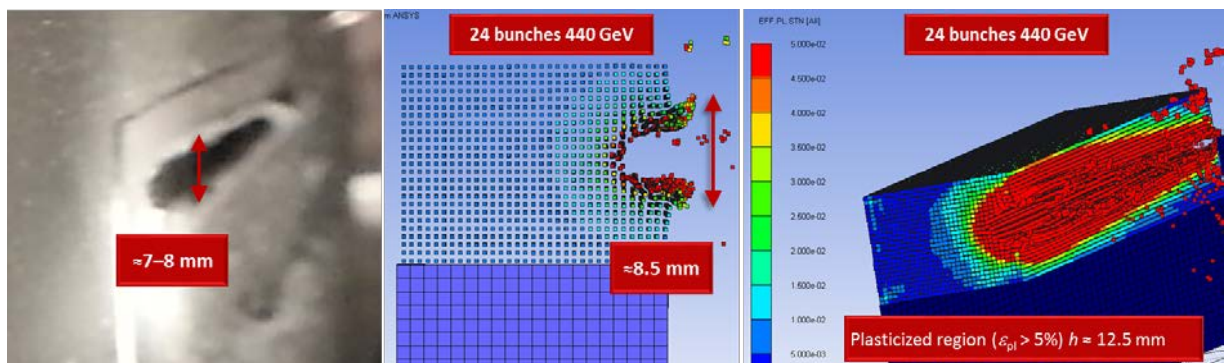

**Fig. 56:** Comparison between actual damage and numerical simulation for test 1 beam impact

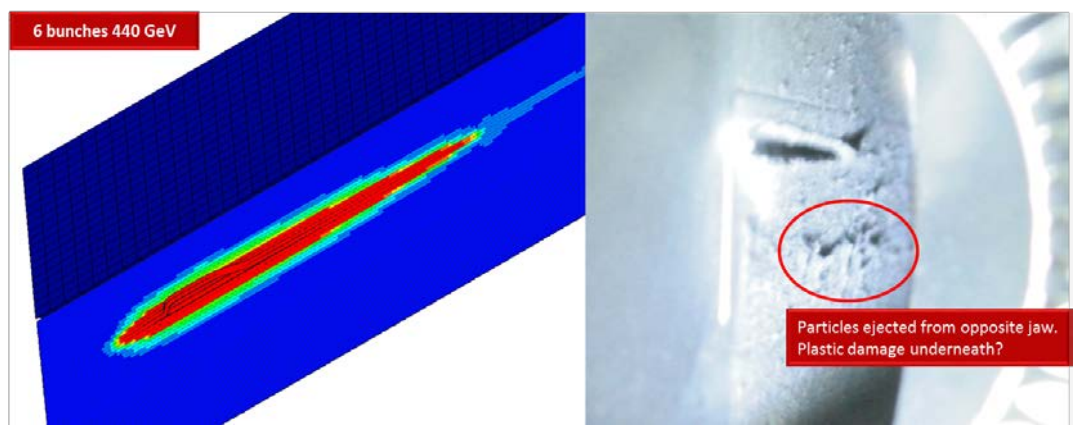

**Fig. 57:** Comparison between numerical analysis and actual damage for test 2 beam impact. Simulations predict a rather extended plasticized region, but only a tiny groove that might have been covered by the particles ejected from the opposite jaw during test 3.

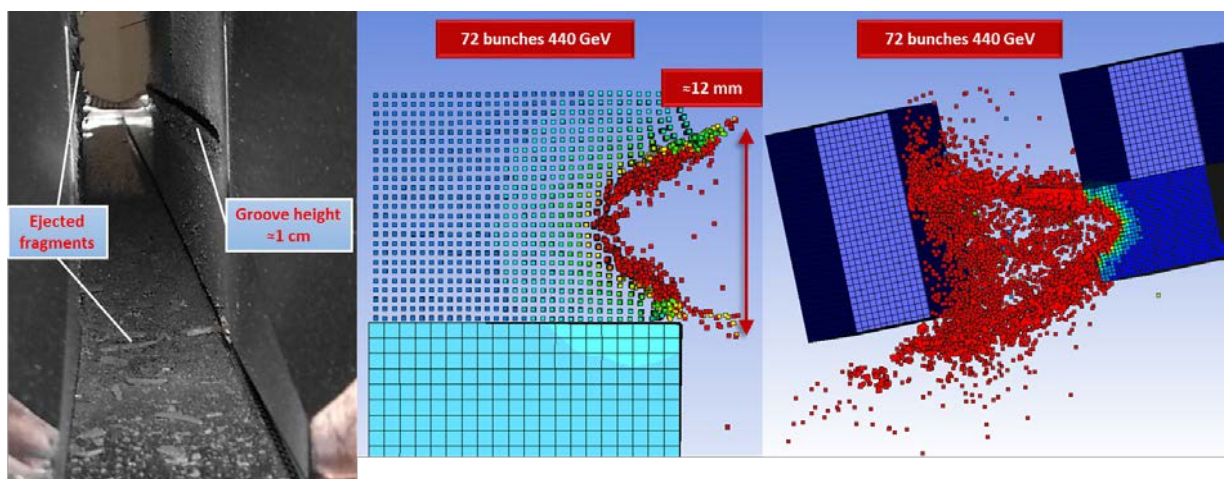

**Fig. 58:** Comparison between actual damage and numerical simulation for test 3 beam impact. Note the spray of ejected particles, reaching velocities close to 1 km s$^{-1}$.

Simulations of test 1 and test 3 show very good agreement with visual inspections, while it is impossible to visualize the plastic deformation produced by test 2. The zone is, in fact, covered with particles ejected from the opposite jaw during test 3, which reached a velocity of about 1 km s$^{-1}$ according to simulations; however, no signs of a significant groove are visible, in line with simulations.

## 4.5 HiRadMat 14 experiment

The main goal of the HRMT14 experiment was to derive new material constitutive models collecting, mostly in real time, experimental data from different acquisition devices: strain gauges, a laser Doppler vibrometer, a high-speed video camera, and temperature and vacuum probes [48].

The material sample holder constituted a vacuum vessel and a specimen housing featuring 12 material sample tiers arranged in two arrays of six (Fig. 59).

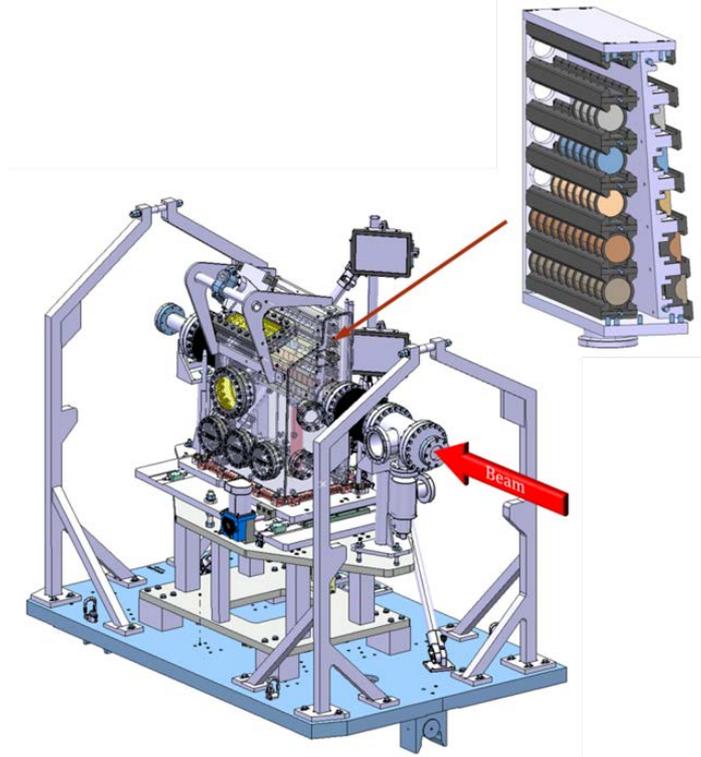

**Fig. 59:** General assembly of the HRMT-14 test-bench

Each tier hosted specimens made of materials currently used for collimators, such as tungsten heavy alloy (Inermet 180), Glidcop® AL-15 LOX (dispersion-strengthened copper), and molybdenum, as well as novel materials under development, i.e. molybdenum–copper–diamond (MoCuCD), copper–diamond (CuCD), and molybdenum carbide–graphite (MoGr) composites.

Two different specimen shapes were chosen for each tested material: cylindrical discs (type 1) for medium-intensity tests, to measure axially symmetrical shock waves, and semicircular prisms (type 2) for high-intensity tests, to allow extreme surface phenomena (melting, material explosion, debris projections, etc.) to be visualized and imaged (Fig. 60).

Part of the instrumentation was installed directly on the specimens; resistive strain gauges measured the strain produced on samples by shock wave propagation, to benchmark time-dependent simulations (Fig. 61). Temperature sensors, vacuum pressure gauges, and microphones were also installed inside or in the vicinity of the tank. Optical devices (a laser Doppler vibrometer and a high-speed camera) were installed remotely in a concrete bunker, to protect them from the effects of radiation. The laser Doppler vibrometer measured the radial velocity on the outer surface of one cylindrical sample per tier. The high-speed camera filmed the particle projection produced by high-energy impacts on type 2 specimens; the lighting necessary for the acquisition was provided by a battery of radiation-hard xenon flashes mounted atop the tank.

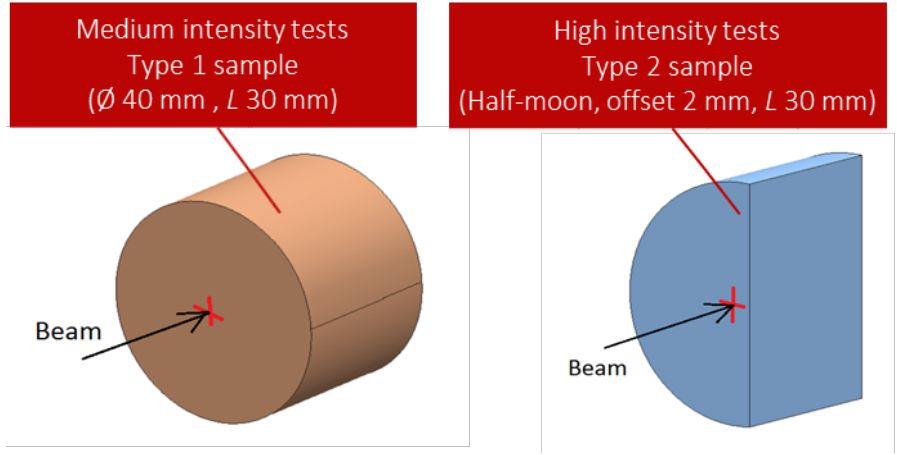

**Fig. 60:** Material specimen shapes for medium-intensity (type 1, left) and high-intensity (type 2, right) tests

Table 12 and Table 13 report the characteristic values of the most intense pulses shot on medium-intensity (type 1 specimens) and high-intensity (type 2 specimens) tiers respectively.

**Table 12:** Beam parameters for most intense pulses shot on each material on medium-intensity tiers

| Type 1 Specimens | Bunches (maximum) per pulse | Delivered protons | Beam size ($\sigma_x \times \sigma_y$) [mm × mm] | Pulse energy [MJ] |
|---|---|---|---|---|
| **Inermet 180** | 24 | $2.70 \times 10^{12}$ | $1.4 \times 2$ | 0.19 |
| **Molybdenum** | 72 | $4.75 \times 10^{12}$ | $1.35 \times 1.25$ | 0.33 |
| **Glidcop** | 72 | $4.66 \times 10^{12}$ | $1.35 \times 1.25$ | 0.33 |
| **MoCuCD** | 72 | $7.62 \times 10^{12}$ | $1.8 \times 1.8$ | 0.54 |
| **CuCD** | 72 | $7.57 \times 10^{12}$ | $1.8 \times 1.8$ | 0.53 |
| **MoGr** | 72 | $7.82 \times 10^{12}$ | $1.8 \times 1.8$ | 0.55 |

**Table 13:** Beam parameters for most intense pulses shot on each material on high-intensity tiers

| Type 2 Specimens | Bunches (maximum) per pulse | Delivered protons | Beam size ($\sigma_x \times \sigma_y$) [mm × mm] | Pulse energy [MJ] |
|---|---|---|---|---|
| **Inermet 180** | 72 | $9.05 \times 10^{12}$ | $2 \times 2$ | 0.64 |
| **Molybdenum** | 144 | $1.95 \times 10^{13}$ | $2 \times 2$ | 1.37 |
| **Glidcop** | 72 | $9.03 \times 10^{12}$ | $1.9 \times 1.9$ | 0.64 |
| **MoCuCD** | 144 | $1.96 \times 10^{13}$ | $2 \times 2$ | 1.38 |
| **CuCD** | 144 | $1.95 \times 10^{13}$ | $2 \times 2$ | 1.37 |
| **MoGr** | 144 | $1.95 \times 10^{13}$ | $2 \times 2$ | 1.37 |

Strain gauges measured axial and hoop strains on the external surface of type 1 samples, while a laser Doppler vibrometer measured the radial velocity. Experimental data were then compared with the results of numerical simulations (Fig. 61).

A strong electromagnetic disturbance, induced by the particle beam, perturbed the strain gauge measurements during the first few microseconds after the impact, covering the first deformation peak. However, this effect rapidly disappeared, enabling the remainder of the phenomenon to be recorded. Measured and simulated signals are in good accordance, especially during the first reflections of the shock wave.

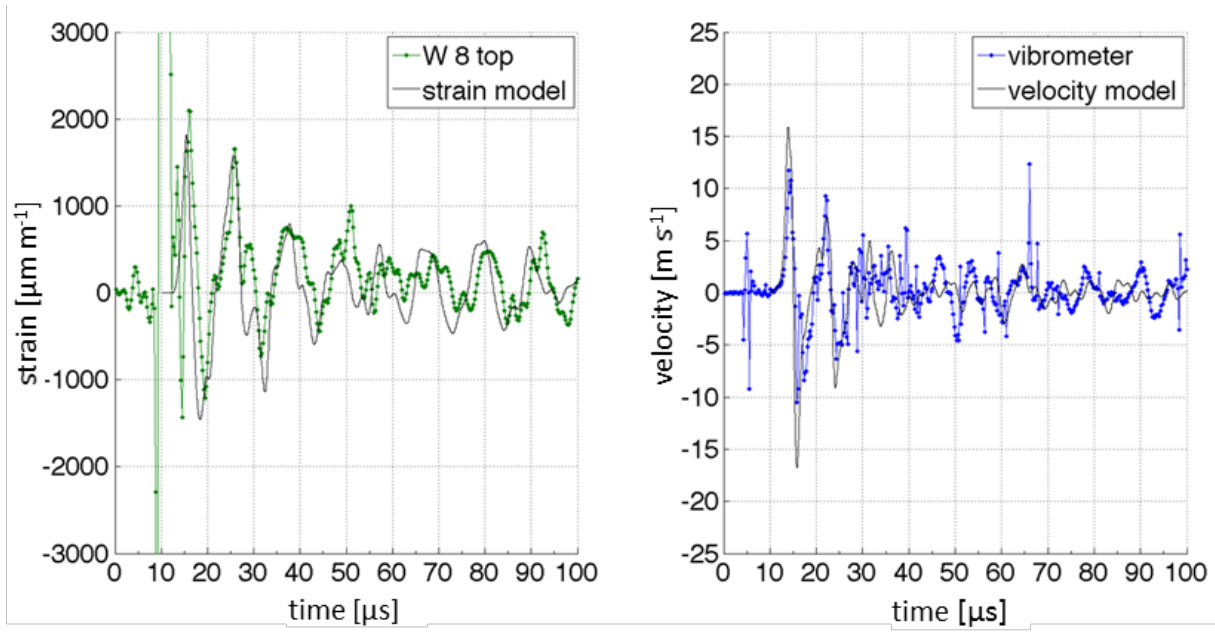

**Fig. 61:** Comparison between measurements (dotted lines) and simulations (continuous lines) of the impact of $2.7 \times 10^{12}$ p ($\sigma \approx 1.4$ mm) on Inermet type 1 sample (slot no 8) at $r = 20$ mm, $L = 15$ mm; axial strain (left) and radial velocity (right).

The high-speed camera and flash systems enabled images of the impact of a hadron beam on solid targets and of the effects induced to be recorded for the first time. The most remarkable phenomena occurred during beam impact on type 2 specimens made of Inermet, the material with the highest stopping power.

As shown in Fig. 62, a large quantity of hot material was ejected at high velocity from the two most loaded Inermet 180 specimens; the high temperatures reached are confirmed by the intense light emitted by the fragments over a few hundred microseconds.

Both the front shape and velocity of the ejected particles are consistent with data acquired using the high-speed camera (Fig. 63), even considering the differences in beam size between the real ($\sigma = 1.9$ mm) and simulated ($\sigma = 2.5$ mm) scenarios.

Smoothed-particle hydrodynamics simulation results are consistent with the ejected particle front shape and velocity acquired by the high-speed camera (Fig. 64), even considering the differences in beam size between real and simulated scenarios. The velocity of the fragment front has been estimated by measuring the displacement between two successive frames and is $\approx 275$ m s$^{-1}$, well matching the simulated velocity of 316 m s$^{-1}$.

The excellent fit between numerical results and experimental measurements confirms the reliability of the simulation techniques and provides a positive indication of the validity of the equation of state and strength model for Glidcop. The match between captured pictures of the Inermet explosion and SPH simulations is also good. Similar analyses will be performed in the near future on molybdenum, for which constitutive models exist, although they are less well established than for copper and tungsten.

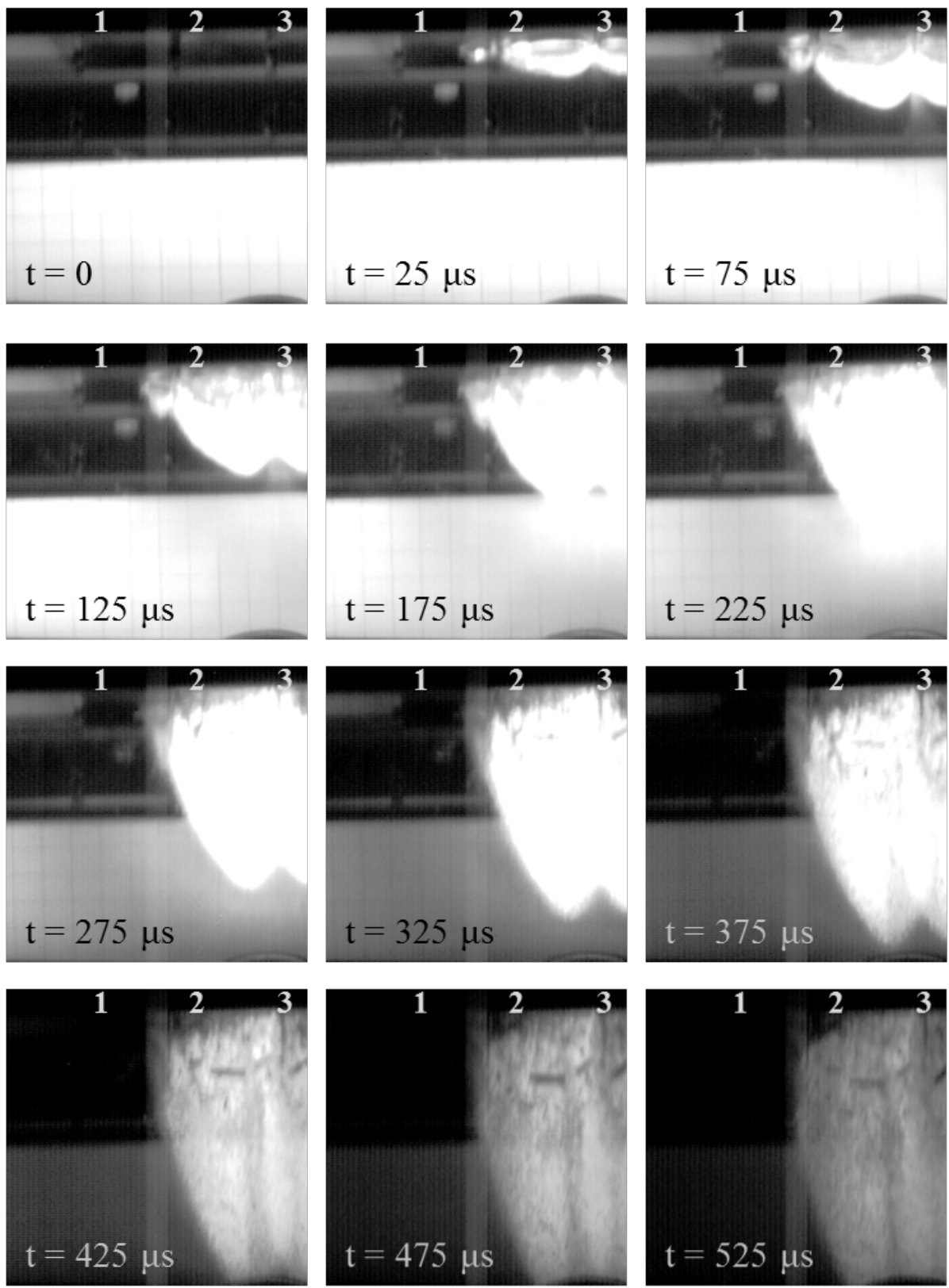

**Fig. 62:** Image sequence of the impact on Inermet of a 72 bunch SPS proton pulse. The beam is coming from the left; three Inermet samples are partially visible (numbered 1 to 3).

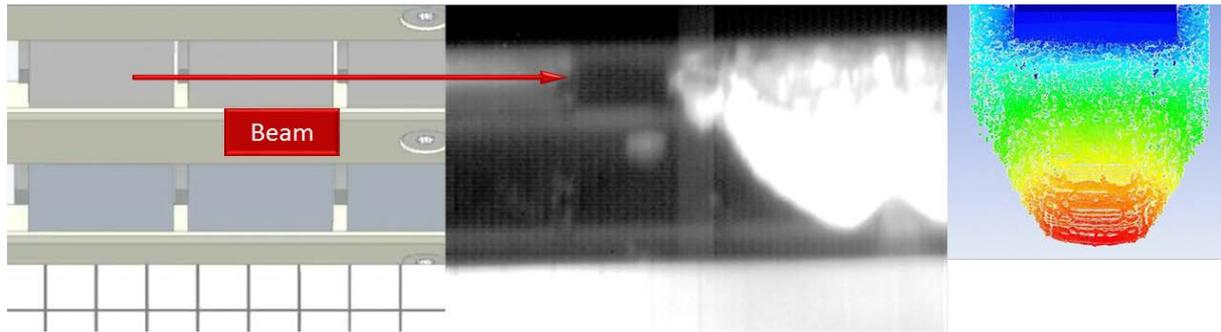

**Fig. 63:** Comparison between SPH simulation and acquired image 125 µs after the impact. Orientation of the Inermet samples and direction of the beam are provided for convenience. Calculated maximum velocity of the fragment front is 316 m s$^{-1}$.

Post-irradiation observations of specimens that underwent high-intensity tests (Fig. 65) confirmed that CuCD and MoGr resisted the impact of 144 bunches of the SPS, with CuCD showing coloration of the surface and a possible slight superficial deformation. Three MoGr grades were tested with densities ranging from 3.8 to 5.3 g cm$^{-3}$: apart from an older grade of higher density, which has since been abandoned, the lighter MoGr grades showed no sign of degradation after visual inspection and non-destructive testing. It is worth noting that since 2012 newer grades of MoGr have been developed with even lower density (down to 2.5 g cm$^{-3}$) and better thermophysical properties.

All higher-$Z$ materials were damaged, to a variable degree of severity. MoCuCD experienced a catastrophic brittle failure and has been since abandoned; Glidcop suffered ejection of molten material at the point of impact (2 mm below the surface) of 72 bunch pulses, although, thanks to its ductility, the specimens' surfaces were largely deformed but not fractured.

Molybdenum exhibited somehow surprising behaviour: under a 72 bunch pulse (hitting the centre), the three last and most loaded specimens did not show evident signs of damage (although a later, more accurate, inspection found some small cracks), while the second sample in the series, which was less loaded, revealed a deep crack extending across most of the specimen. A second pulse at double the intensity (144 bunches) was delivered 10 mm apart from the first: in this case, a groove was produced on the most loaded specimens (although less extended than on Inermet samples at half the intensity), while cracks were induced, particularly on the third samples, several millimetres away from the point of impact; this behaviour may be explained by the temperature increase induced by the energy deposition: when the so-called ductile-to-brittle transition temperature is exceeded, materials shift from a highly brittle to a ductile behaviour. The ductile-to-brittle transition temperature in pure molybdenum is typically several tens of degrees above RT; therefore, up to a certain point, beam-induced heating may have had a beneficial effect in increasing ductility to a level that effectively countered the higher induced stresses.

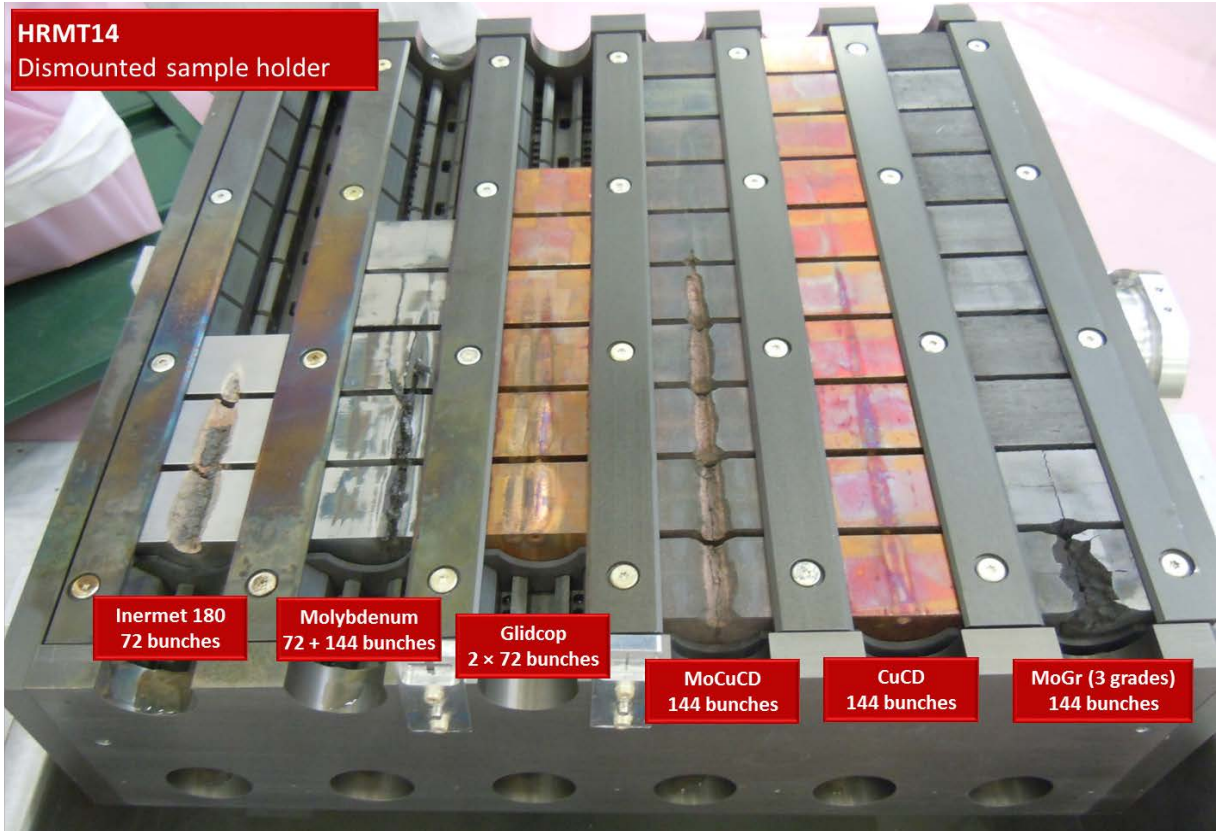

**Fig. 64:** Post-irradiation observation of HRMT-14 sample holder. Beam arrived from the top. Note that the two last specimens in MoGr were from an obsolete grade.

Inermet experienced a brittle failure, with no signs of plastic deformation on the brim of the damaged area and on the flat surface. The low melting point of copper and nickel probably played an important role in determining the extent of damaged zone. The simulated damage extension is consistent with experimental observations (Fig. 65).

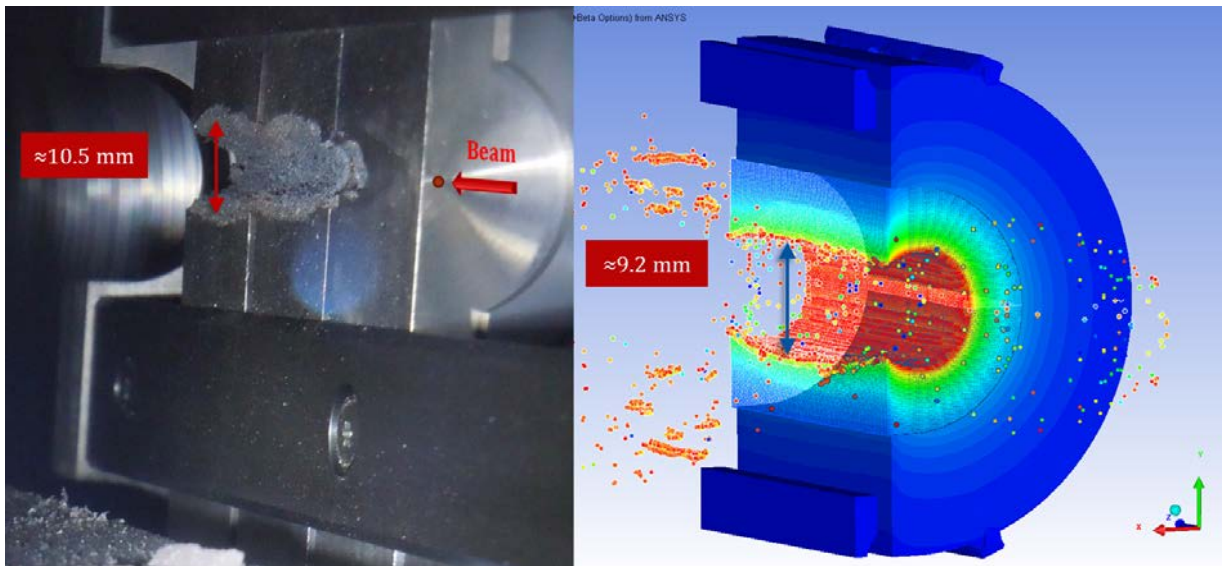

**Fig. 65:** Post-irradiation observation of Inermet 180 samples (left) and simulated failure (right)

## Acknowledgments

I would like to heartily thank Federico Carra, Alessandro Dallocchio, and Marco Garlaschè (CERN, EN/MME) to whom I am indebted for their support in preparing and proofreading the lectures and these proceedings and, even more, for the stimulating discussions and exchanges which, through the years, allowed me to put together the material presented here.

I would also like to express my gratitude to Paolo Gradassi, Linus Mettler, and Jorge Guardia Valenzuela (CERN, EN/MME) for their important contributions to these proceedings.